\documentclass[a4paper,11pt]{article}
\pdfoutput=1 

\usepackage{jheppub} 

\usepackage[T1]{fontenc} 
\graphicspath{{Figures/}} 
\usepackage{braket}
\usepackage{mathtools}
\usepackage{placeins}

\title{\boldmath Quantum Spread Complexity in Neutrino Oscillations}

\author[a]{Khushboo Dixit,}
\author[b,c,e]{S. Shajidul Haque,}
\author[a,d,e]{Soebur Razzaque}

\affiliation[a]{Centre for Astro-Particle Physics (CAPP) and Department of Physics, University of Johannesburg, PO Box 524, Auckland Park 2006, South Africa}
\affiliation[b]{High Energy Physics, Cosmology \& Astrophysics Theory Group and \\
	The Laboratory for Quantum Gravity \& Strings, Department of Mathematics \& Applied Mathematics, University of Cape Town, Cape Town, South Africa}
\affiliation[c]{Department of Mathematics and Natural Sciences, Brac University, Dhaka, Bangladesh}
\affiliation[d]{Department of Physics, The George Washington University, Washington, DC 20052, USA}
\affiliation[e]{National Institute for Theoretical and Computational Sciences (NITheCS), Private Bag X1, Matieland, South Africa}

\emailAdd{kdixit@uj.ac.za}
\emailAdd{shajid.haque@uct.ac.za}
\emailAdd{srazzaque@uj.ac.za}

\abstract{Quantum information theory has recently emerged as a flourishing area of research and quantum complexity, one of its powerful measures, is being applied for investigating complex systems in many areas of physics. Its application to practical physical situations, however, is still few and far between. Neutrino flavor oscillation is a widely studied physical phenomena with far reaching consequences in understanding the standard model of particle physics and to search for physics beyond it. Oscillation arises because of mixing between the flavor and mass eigenstates, and their evolution over time. It is an inherent quantum system for which flavor transitions are traditionally studied with probabilistic measures. We have applied quantum complexity formalism as an alternate measure to study neutrino oscillations. In particular, quantum spread complexity revealed additional information on the violation of charge-parity symmetry in the neutrino sector. Our results indicate that complexity favors the maximum violation of charge-parity, hinted recently by experimental data.}

\begin{document} 
\maketitle
\flushbottom

\section{Introduction}

In recent years, quantum complexity, a widely recognized measure in information theory, has found application in various branches of physics, encompassing quantum many-body systems, quantum field theory, and even cosmology. 
The interest in quantum complexity stemmed from the study of anti-de Sitter/conformal field theory (AdS/CFT) duality, also known as the gauge/gravity duality. Complexity is considered a useful probe \cite{Susskind:2014moa} to investigate the physics behind the horizon of an eternal AdS black hole, employing proposals such as “complexity = volume” and “complexity = action” \cite{Susskind:2014rva, Stanford:2014jda, Brown:2015bva, Brown:2015lvg}.

From the standpoint of the dual quantum (field) theory, complexity has emerged as a valuable tool for characterizing quantum chaos \cite{Ali:2019zcj, Bhattacharyya:2019txx, Bhattacharyya:2020iic, Balasubramanian:2019wgd, Bhattacharyya:2020art, Balasubramanian_2021}, detecting quantum phase transitions \cite{Ali:2018aon}, quantum decoherence \cite{Bhattacharyya:2022rhm, Bhattacharyya:2021fii}, and more. For example, recent studies \cite{Bhattacharyya:2020rpy, Bhattacharyya:2020kgu, Haque:2021hyw} have delved into the cosmological perturbation model and the evolution of the universe, utilizing Nielsen's approach \cite{Nielsen_2006, nielsen2006geometric, dowling2006geometry, Jefferson:2017sdb, Ali:2018fcz} to complexity. Interestingly, in reference \cite{Bhattacharyya:2020kgu}, it was discovered that de Sitter space, which offers the most popular model for inflation, exhibits the highest rate of complexity growth among expanding backgrounds that satisfy the null energy condition. It would be intriguing to investigate whether this maximization of complexity occurs in other natural processes of evolution.

In our work, we will use a more recent approach to measuring complexity, known as spread complexity \cite{Balasubramanian:2022tpr, haque2022krylov}, to understand the evolution of neutrino flavor states. Spread complexity offers a clear definition that is valid in arbitrary quantum systems and is relatively straightforward to compute. It has already demonstrated its usefulness in diagnosing quantum chaos \cite{Balasubramanian:2022tpr} and quantum phase transitions \cite{Caputa_2023}. In this paper we will apply this information theoretic tool to gain insight about neutrino oscillations. Specifically, we will investigate if spread complexity can be used as an alternative to the oscillation probabilities for different flavors of neutrinos. 

The phenomena of neutrino oscillations are due to mixing of the flavor eigenstates $\nu_\alpha$ ($\alpha = e, \mu, \tau$ for three generations) in the mass eigenstates $\nu_i$ of masses $m_i$ ($i = 1, 2, 3$ for three generations). The former are associated with weak interactions --- neutrinos with definite flavor are created in charge current interactions --- while the latter are associated with the propagation of massive neutrinos governed by a Hamiltonian. The flavor states are superposition of the mass states and vice versa. The proportions of mass states in a neutrino with definite flavor $\nu_\alpha$ change while propagation from the creation point and can be identified as a different flavor $\nu_\beta$ in a detector at a distance. This is the essence of neutrino oscillations. Detection of these phenomena, first by the solar neutrino experiments \cite{Kamiokande-II:1991pyu, Cleveland:1998nv, Ahmad:2002jz}, and subsequently by the atmospheric \cite{ Fukuda:1994mc, Fukuda:1998mi} and reactor neutrino experiments \cite{Abe:2011fz, DayaBay:2018yms, RENO:2018dro} provide the first signal for physics beyond the standard model. Mixing of the flavor states in mass states, and subsequently probabilities for oscillations between flavors, is governed by the well known Pontecorvo-Maki-Nakagawa-Sakata (PMNS) matrix \cite{Maki:1962mu, Bilenky:1978nj}, which is a $3\times3$ unitary matrix admitting one Charge-Parity (CP) violating Dirac phase ($\delta$), and is typically parameterized by three mixing angles $\theta_{12}, \theta_{23}$ and $\theta_{13}$.

The neutrino oscillations are driven by two independent mass-squared differences $\Delta m_{21}^2 \equiv m_2^2 - m_1^2$ and $\Delta m_{31}^2 \equiv m_3^2 - m_1^2$ for three neutrino masses. The absolute mass scale does not affect oscillations, however, the hierarchy of masses, whether $m_3 > m_2 > m_1$ (normal hierarchy) or $m_2 > m_1 > m_3$ (inverted hierarchy), is unknown.\footnote{From the solar neutrino experiments, it is known that $m_2 > m_1$.} The CP phase $\delta$ is also unknown, apart from a hint from the T2K experiment at $\delta\sim -2.14$ radian \cite{T2K:2021xwb}, which however is in tension with the NOvA experiment excluding this value~\cite{NOvA:2021nfi}. The angles $\theta_{12}$ and $\theta_{13}$ are known with good accuracy, while $\theta_{23}$ is not. Neutrino experiments measure events, typically from $\nu_e ({\bar\nu}_e)$ and/or $\nu_\mu ({\bar\nu}_\mu)$ induced interactions in a detector, given a flux of neutrinos produced by an accelerator or a reactor. The measured events are fitted with simulations based on oscillation probabilities by varying the mixing parameters and mass-squared differences.

Moreover, neutrinos participate in weak interactions only and hence they have little chance to experience the effects such as decoherence during their travel to a distant detector. It makes these particles efficient candidates to be utilized to perform several tasks related to quantum information \& computation. In this line, many aspects of quantumness embedded in the neutrino system have been analyzed thoroughly in previous studies \cite{Blasone:2007wp, Blasone:2007vw, Gangopadhyay:2013aha, Alok:2014gya, Banerjee:2015mha, Formaggio:2016cuh, Fu:2017hky, Dixit:2018kev, Naikoo:2017fos, Richter:2017toa, Song:2018bma, Wang:2020vdm, Ettefaghi:2020otb, Ravari:2022yfd, Li:2022mus}. For example, an indirect test of $Leggett-Garg$ (LG) inequalities, which can verify the temporal quantum correlations, has been performed using the oscillation data coming from the MINOS and Daya-Bay experiments \cite{Formaggio:2016cuh, Fu:2017hky}. Furthermore, several measures of entanglement, spatial and temporal correlations have been studied for neutrinos and these measures have also been found to provide important pieces of information regarding open issues in the neutrino sector as discussed above.  For example, in refs. \cite{Dixit:2018kev, Naikoo:2017fos} it has been discussed that the test of Bell-type and LG inequalities can indicate the specific choice of neutrino mass ordering. LG inequalities have also been shown to discriminate between the Dirac and Majorana nature of neutrinos \cite{Richter:2017toa}. Some measures of quantumness have also been seen to be sensitive to the new physics effects due to non-standard neutrino-matter interactions \cite{Dixit:2019swl,Sarkar:2020vob,Shafaq:2020sqo,Yadav:2022grk}. Recently, a trade-off relation between the local coherence of individual subsystems as well as of bipartitions and the non-local coherence shared among the given subsystems has also been studied for neutrino oscillations  \cite{Bittencourt:2023asd}, which is an interesting way to look for complementarity relations among divergent features of quantum correlations.  

In this paper, we compute the spread complexities of neutrinos of a particular flavor oscillating to other flavors after propagation. We show that the cost function, which is automatically minimized in the Krylov basis used in our computations, gives an alternate description of the neutrino flavor oscillations and is sensitive to the oscillation parameters. In particular, we explore the CP phase value, mass hierarchy and $\theta_{23}$ value predicted by the spread complexity. 

In Section~\ref{Dynamics} we discuss the dynamics of neutrino oscillations and mixing in two- and three-flavor scenarios. In Section~\ref{Complexity} we introduce a basic description of spread complexity and cost function, and in Section~\ref{Complex_oscillation} we apply it to neutrino oscillations. We show our results from numerical calculations and discuss them in Section~5, summarize our findings in Section~6 and conclude our study in Section~7.

\section{Dynamics of Neutrino Oscillations} \label{Dynamics}
Here we discuss the evolution of neutrino flavor states both in the case of two flavor approximation and the complete three flavor neutrino oscillations scenario. In the neutrino-system, the flavor states are not the mass eigenstates, and in fact, the flavor states mix via a unitary matrix $U$ to generate mass eigenstates as given below
\begin{equation}
	\ket{\nu_{\alpha}} = U^\ast \ket{\nu_i} \,,
	\label{eq:mass-to-flavor}
\end{equation}   
where $\ket{\nu_{\alpha}}$ and $\ket{\nu_i}$ are column vectors with neutrino flavor and mass eigenstates as their components, respectively. Here, we discuss the time evolution of neutrino flavor states for both two and three flavor oscillation scenarios.

\subsection{Two-flavor Neutrino Oscillations}
Evolution of the flavor states is represented by Schr\"odinger equation as \footnote{We have used natural units: $\hslash = c = 1$ throughout the paper.}
\begin{equation*}\label{schrodEq}
	i \frac{\partial}{\partial t} \begin{pmatrix}
		\ket{\nu_e (t)}\\
		\ket{\nu_{\mu} (t)}
	\end{pmatrix} = H_f \begin{pmatrix}
		\ket{\nu_e (t)}\\
		\ket{\nu_{\mu} (t)}
	\end{pmatrix}
\end{equation*}
where $H_f = U H_m U^{-1}$, $U$ is a $2\times 2$ mixing matrix and $H_m$ is the Hamiltonian (diagonal) that governs the time evolution of neutrino mass eigenstate. The forms of $H_m$ and $U$ are given below
\begin{equation*}
	H_m = \begin{pmatrix}
		E_1  &0\\
		0  &E_2
	\end{pmatrix}, ~~~~~U = \begin{pmatrix}
		\cos\theta  &\sin\theta\\
		-\sin\theta  &\cos\theta
	\end{pmatrix}.
\end{equation*}
The Hamiltonian in flavor basis can be expressed as
\begin{equation}\label{FlavHamilt}
	H_f = \begin{pmatrix}
		E_1 \cos^2 \theta + E_2 \sin^2 \theta    & (E_2-E_1)\sin\theta \cos\theta\\
		(E_2-E_1)\sin\theta \cos\theta  & E_1 \sin^2 \theta + E_2 \cos^2 \theta
	\end{pmatrix} \,.
\end{equation}
Finally, we have a system of coupled differential equations to solve in the neutrino flavor-basis, {\it i.e.,}
\begin{align*}
	\frac{\partial }{\partial t}\ket{\nu_e(t)} = -i&(E_1 \cos^2 \theta + E_2 \sin^2 \theta) \ket{\nu_e(t)} 
	-i (E_2-E_1)\sin\theta \cos\theta \ket{\nu_{\mu}(t)} \\
	\frac{\partial }{\partial t}\ket{\nu_{\mu}(t)} = -i&(E_2-E_1)\sin\theta \cos\theta \ket{\nu_e(t)}  
	-i (E_1 \cos^2 \theta + E_2 \sin^2 \theta) \ket{\nu_{\mu}(t)} \,.
\end{align*}
Let us consider an $M$ matrix defined as
\begin{equation*}
	M = -i\begin{pmatrix}
		E_1 \cos^2 \theta + E_2 \sin^2 \theta   &(E_2-E_1)\sin\theta \cos\theta\\
		(E_2-E_1)\sin\theta \cos\theta       &E_1 \cos^2 \theta + E_2 \sin^2 \theta
	\end{pmatrix}
\end{equation*}
that has eigenvalues $\lambda_1 = -i E_1$ and $\lambda_2 = -i E_2$ with corresponding eigenvectors as $(-\cot\theta, 1)^T$, and $(\tan\theta, 1)^T$, respectively. It implies that we can write 
\begin{equation*}
	\begin{pmatrix}
		\ket{\nu_e(t)}\\
		\ket{\nu_\mu(t)}
	\end{pmatrix} = \begin{pmatrix}
		-\cot\theta  & \tan\theta\\
		1                      &1
	\end{pmatrix} \begin{pmatrix}
		c e^{-i E_1 t}\\
		d e^{-i E_2 t}
	\end{pmatrix}.
\end{equation*}
Then, we proceed to get the time evolved neutrino flavor states as 
\begin{eqnarray}\label{timeEvol1}
	\ket{\nu_e(t)} &=& c~ (-\cot\theta) e^{-i E_1 t} + d~ (\tan\theta) e^{-i E_2 t}\nonumber\\
	\ket{\nu_{\mu}(t)} &=& c~ e^{-i E_1 t} + d~ e^{-i E_2 t}.
\end{eqnarray}
where $c$ and $d$ are constants whose values we can obtain by applying the initial conditions (at $t=0$) and can be expressed as
\begin{eqnarray*}
	c = -\sin\theta \cos\theta \ket{\nu_e(0)} + \sin^2\theta \ket{\nu_{\mu}(0)}\nonumber\\
	d = \sin\theta \cos\theta \ket{\nu_e(0)} + \cos^2\theta \ket{\nu_{\mu}(0)}.~
\end{eqnarray*}
Therefore, Eq.~(\ref{timeEvol1}) takes the form
\begin{align}\label{timeEvol}
	\ket{\nu_e(t)} = &(\cos^2\theta e^{-i E_1 t} + \sin^2\theta e^{-i E_2 t}) \ket{\nu_e(0)}
	+ \sin\theta \cos\theta (e^{-i E_2 t} - e^{-i E_1 t}) \ket{\nu_{\mu}(0)}\nonumber  \\ 
	\ket{\nu_{\mu}(t)} = &\sin\theta \cos\theta (e^{-i E_2 t} - e^{-i E_1 t}) \ket{\nu_e(0)}
	+ (\sin^2\theta e^{-i E_1 t} + \cos^2\theta e^{-i E_2 t}) \ket{\nu_{\mu}(0)}.
\end{align}
We can see that the time evolved flavor states are now superpositions of the initial flavor states at time $t=0$, hence, their coefficients can be used to obtain the survival and oscillation probabilities for each flavor, after propagation over a distance $L$, as
\begin{equation*}
	P_{\alpha \alpha} = 1 - \sin^2 2\theta \sin^2\left((E_2-E_1)L/2\right) \,.
\end{equation*}

\subsection{Three-flavor Neutrino Oscillations}
It is straightforward to obtain the time evolution of the flavor states in case of three flavor oscillations. In this case, the Schr\"odinger equation takes the following form 
\begin{equation}\label{schrodEq}
	i \frac{\partial}{\partial t} \begin{pmatrix}
		\ket{\nu_e(t)}\\
		\ket{\nu_{\mu}(t)}\\
		\ket{\nu_{\tau}(t)}
	\end{pmatrix} = H_f \begin{pmatrix}
		\ket{\nu_e(t)}\\
		\ket{\nu_{\mu}(t)}\\
		\ket{\nu_{\tau}(t)}
	\end{pmatrix} \,,
\end{equation}
where $H_f = U H_m U^{-1}$ and $H_m = diag (E_1, E_2, E_3)$ is the Hamiltonian of neutrino energies $E_i$, with $i = 1, 2, 3$. For relativistic neutrinos of momentum $p$, $E_i = \sqrt{p^2 + m_i^2} \simeq p + m_i^2/2E$. Therefore, $E_j - E_i \simeq (m_j^2 - m_i^2)/2E = \Delta m_{ji}^2/2E$ and the Hamiltonian, after subtracting $E_1$ and removing the identity term that do not affect oscillations, can be written as
\begin{equation*}
	H_m = \frac{1}{2E} \begin{pmatrix}
		0  & 0 & 0 \\
		0  & \Delta m_{21}^2 & 0 \\
		0  & 0 & \Delta m_{31}^2
	\end{pmatrix} \,.
\end{equation*} 
In the three-flavor case, $U$ is a $3\times 3$ unitary matrix, called the PMNS mixing matrix \cite{Maki:1962mu, Bilenky:1978nj}. It is parametrized by three angles and a complex phase and is of the form \cite{ParticleDataGroup:2022pth}
	\begin{equation}\label{PMNS}
		\centering
		U = \begin{pmatrix}
			U_{e1}   &   U_{e2}   &   U_{e3} \\
			U_{\mu 1} &  U_{\mu 2} &  U_{\mu 3} \\
			U_{\tau 1} & U_{\tau 2} & U_{\tau 3}
		\end{pmatrix} = \begin{pmatrix}
			c_{12} c_{13} & s_{12} c_{13} & s_{13} e^{-i \delta} \\ - s_{12}c_{23} - c_{12} s_{23}s_{13} e^{i\delta} & c_{12}c_{23}-s_{12}s_{23}s_{13} e^{i\delta} & s_{23}c_{13} \\ s_{13}s_{23} - c_{12}c_{23}s_{13} e^{i\delta} & -c_{12}s_{23}-s_{12}c_{23}s_{13} e^{i\delta} & c_{23}c_{13}\end{pmatrix} \,.
	\end{equation}
\noindent
Here, $c_{ij} = \cos\theta_{ij}$, $s_{ij} = \sin\theta_{ij}$ with mixing angles $\theta_{ij}$ and $\delta$ is the $CP$-violating Dirac phase. There are, therefore, six parameters in three-flavor oscillations: two mass-square differences ($\Delta m_{21}^2$ and $\Delta m_{31}^2$), three mixing angles ($\theta_{12}$, $\theta_{13}$ and $\theta_{23}$) and one CP phase ($\delta$). In case the neutrinos are Majorana particles, there are two additional complex phases in the mixing matrix, which however do not affect oscillations.  

Hence, for three flavor oscillation scenario, after solving the set of three coupled differential equations, we get the time-evolved flavor states of neutrinos as 
\begin{eqnarray}
	\ket{\nu_e(t)} &=& A_{ee}(t) \ket{\nu_e(0)} + A_{e\mu}(t) \ket{\nu_{\mu}(0)} + A_{e\tau}(t) \ket{\nu_{\tau}(0)} \nonumber \\
	\ket{\nu_{\mu}(t)} &=& A_{\mu e}(t) \ket{\nu_e(0)} + A_{\mu \mu}(t) \ket{\nu_{\mu}(0)} + A_{\mu \tau}(t) \ket{\nu_{\tau}(0)} \nonumber \\
	\ket{\nu_{\tau}(t)} &=& A_{\tau e}(t) \ket{\nu_e(0)} + A_{\tau \mu}(t) \ket{\nu_{\mu}(0)} + A_{\tau \tau}(t) \ket{\nu_{\tau}(0)} \,.
	\label{eq:time-evolved-kets}
\end{eqnarray}
The explicit expressions of the amplitudes $A_{\alpha \beta}(t)$ with $\alpha, \beta = e, \mu, \tau$ for standard vacuum oscillations are given in the Appendix. 
It is straightforward to follow the dynamics of antineutrino oscillations by applying the change $\delta \rightarrow -\delta$ in the amplitudes $A_{\alpha\beta}$ obtained for neutrinos. Hence, the parameter $\delta$ can induce $CP$-violation in neutrino sector that is measured in terms of $\Delta CP$ as
\begin{equation*}
	\Delta CP = P_{\alpha \beta} - P_{\bar{\alpha} \bar{\beta}}.
\end{equation*}
Here, $\Delta CP$ becomes maximum for $\delta = \pm 90^o$.

\section{Spread Complexity and Cost Function} \label{Complexity}
We will be interested in the complexity of some general quantum state $|\psi(t)\rangle$. The evolution of this state can be obtained from the Schr\"odinger equation as 
\begin{equation*}
	i \frac{\partial}{\partial t} \ket{\psi(t)} = H \ket{\psi(t)}.
\end{equation*}
The solution 
gives the time evolution of the state $|\psi\rangle$ as follows
\begin{equation}
	\ket{\psi(t)} = e^{-i H t}\ket{\psi(0)},
	\label{eq:ketpsi-t}
\end{equation}
where $|\psi(0)\rangle$ is the initial state at $t=0$. The spread complexity can be defined as the spread of $\ket{\psi(t)}$ in the Hilbert space relative to $|\psi(0)\rangle$, where the former, often referred to as ``target state'', and the latter, often referred to as ``reference state'', are connected by unitary transformations \cite{Balasubramanian:2022tpr, Caputa:2022eye}.

We expand Eq.~(\ref{eq:ketpsi-t}) in series and write 
\begin{equation*}
	\ket{\psi(t)} = \sum_{n=0}^{\infty}\frac{(-i t)^n}{n!} H^n \ket{\psi(0)} = \sum_{n=0}^{\infty}\frac{(-i t)^n}{n!} \ket{\psi_n},
\end{equation*}
where, $\ket{\psi_n} = H^n \ket{\psi(0)}$. Hence, we can see that the time evolved state $\ket{\psi(t)}$ is represented as a superposition of infinite $\ket{\psi_n}$ states. However, in this representation, the $\ket{\psi_n}$ states are not necessarily orthonormal. Hence, we use Gram-Schmidt procedure to obtain an ordered orthonormal basis from these $\ket{\psi_n}$ states. 
We have the following forms of $\psi_n$ states as
\begin{eqnarray*}
	\ket{\psi_0} &=& H^0 \ket{\psi(0)}\\
	\ket{\psi_1} &=& H^1 \ket{\psi(0)}\\
	\ket{\psi_2} &=& H^2 \ket{\psi(0)}
\end{eqnarray*}
and so on. These states \{$\ket{\psi_0}$, $\ket{\psi_1}$, $\ket{\psi_2}$, $\dots$\} are not orthonormalized yet. Following the Gram-Schmidt procedure we subtract the component of $\ket{\psi_n}$ (parallel to the previous state $\ket{\psi_{(n-1)}}$) from the given state $\ket{\psi_n}$. Hence, we have
\begin{eqnarray*}
	\ket{K_0} &=&\ket{\psi_0},\\
	\ket{K_1} &=& \ket{\psi_1} - \frac{\langle K_0|\psi_1\rangle}{\langle K_0|K_0\rangle} \ket{K_0},\\
	\ket{K_2} &=& \ket{\psi_2} - \frac{\langle K_0|\psi_2\rangle}{\langle K_0|K_0\rangle} \ket{K_0} - \frac{\langle K_1|\psi_2\rangle}{\langle K_1|K_1\rangle} \ket{K_1},
\end{eqnarray*} 
and so on. These orthonormal set of vectors form the Krylov basis \cite{Balasubramanian:2022tpr}. 

The extent of spread of the evolved state $\ket{\psi(t)}$ in the Hilbert space depends on how complex the time evolution is. A cost function is defined as a measure of this complexity from a minimum of all possible basis choices \cite{Balasubramanian:2022tpr}. Therefore, this cost function is an immediate candidate for measuring the spread complexity. 
More explicitly, for a time evolved state $\ket{\psi(t)}$ and the Krylov basis defined as \{$\ket{K_n}$\}, the cost function can be defined as 
\begin{equation}\label{cost}
\mathcal{\chi} = \sum C_n|\langle K_n|\psi(t)\rangle|^2 = \sum C_n P_{K_n}, 	
\end{equation}
Here $C_n$ is a real increasing number, and a convenient choice is $C_n = n = 0,1,2, \dots$ \cite{Balasubramanian:2022tpr,Caputa_2023}. $P_{K_n}$ is the probability of $\psi(t)$ being in one of the Krylov basis states. The complexity is minimized for this cost function, constructed using the Krylov basis. 

One should appreciate the simplicity of this measure to capture the complexity embedded in a given system. The above choice of $n$ also justifies the fact that the time evolution will be more complex as the number of Krylov states increases. In other words, the increasing weight $n$ implies that the cost of a wavefunction increases if it spreads deeper into the basis. Since the Gram-Schmidt procedure provides an ordered basis, the weight or the contribution of the last Krylov state will be the most. However, the overall complexity may depend on parameters guiding the dynamics of the system evolution.

\section{Complexity for Neutrino Oscillations} \label{Complex_oscillation}


In this section, we will delve into the study of spread complexity within the context of the neutrino system undergoing flavor oscillations. The motivation for this investigation stems from the probabilistic structure of the spread complexity measure, as described in Eq.~(\ref{cost}), prompting us to make a natural comparison between spread complexity and the conventional flavor oscillation probabilities.
Our primary goal is to explore how spread complexity can serve as an alternative measure to various transition (and survival) probabilities in the realm of neutrino oscillations. By doing so, we will display that spread complexity can provide new insight in enhancing our understanding of neutrino oscillations.

We will initially focus on the two-flavor oscillation scenario and subsequently extend our analysis to the three-flavor case. As mentioned earlier, in the context of spread complexity, we begin with a specific flavor state and evolve it into a superposition state involving all flavors. Since the weight factor for the reference state is zero according to Eq. (\ref{cost}), the reference state does not really contribute to complexities. Hence, in the case of two-flavor oscillations, we can directly compare the spread complexity with the transition probabilities between two flavors.

In the case of three-flavor oscillations, however, evolved states become superposition states comprising all flavors. Consequently, a natural comparison for complexity would be with unity minus the survival probability of a given flavor.
For instance, we can compare the spread complexity $\chi_e$ with $1-P_{ee}$, which is directly applicable to neutrino oscillation experiments. This approach allows us to directly compare the information obtained from complexity with experimental results.

Since the transition probabilities in the three-flavor case involve two distinct flavor states, such as electron (initial) to muon (final), a direct comparison between spread complexity $\chi_e$ (with electron as the initial state) and the final evolved state (a mixed state) is not feasible. Nonetheless, we will separately compare both $P_{e\mu}$ and $P_{e\tau}$ with $\chi_e$, and likewise for other flavors. This analysis aims to determine if the information extracted from these transition amplitudes is comparable to the information obtained solely from spread complexity.

\subsection{Complexity for Two-flavor Neutrino Oscillations \label{twoflavor}}
Spread complexity measures how an initial state is spread in the Hilbert space by a unitary evolution. Here, we will consider the spreading of both the $\ket{\nu_e}$ and $\ket{\nu_{\mu}}$ initial states. As we will see shortly, there are exactly two non-zero Krylov states, which are the same as the flavor states. 
\subsubsection{Initial electron-neutrino ($\nu_e$) state}
For the initial state $\ket{\nu_e}$, we will consider
$\ket{\nu_e(0)} = (1,0)^{T}$.
Then we get the basis $\ket{\psi_n}$, $(n=0,1,2,\dots)$ as 
\begin{eqnarray*}
	\ket{\psi_0} &=& \ket{\nu_e(0)}\\
	\ket{\psi_1} &=& H_f \ket{\nu_e(0)}\\
	\ket{\psi_2} &=& H_f^2 \ket{\nu_e(0)}\\
	\ket{\psi_3} &=& H_f^3 \ket{\nu_e(0)}
\end{eqnarray*}
and so on. The Hamiltonian $H_f$ is defined in Eq.~(\ref{FlavHamilt}). It turns out that the Krylov basis for this two-flavor oscillations scenario is $\{\ket{K_n}\} = \{\ket{K_0},\ket{K_1}\}$ where, $\ket{K_0}=(1,0)^T$ and $\ket{K_1}=(0,1)^T$, {\it i.e.,} $\ket{K_n}= \{\ket{\nu_e},\ket{\nu_{\mu}}\}$. Hence, for the initial $\ket{\nu_e}$ flavor the complexity takes the form
\begin{align}\label{twoflavorchie}
	\chi_e &= 0 \times |\langle K_0|\nu_e(t)\rangle|^2 + 1 \times |\langle K_1|\nu_e(t)\rangle|^2 \nonumber \\
	&= 0 \times |\langle \nu_e|\nu_e(t)\rangle|^2 + 1 \times |\langle \nu_{\mu}|\nu_e(t)\rangle|^2 \nonumber \\
	&= |\langle \nu_{\mu}|\nu_e(t)\rangle|^2 \nonumber \\
	&= P_{e\mu},
\end{align}
which is the $\nu_e\to \nu_\mu$ transition probability. The time evolved state $\ket{\nu_e(t)}$ is defined in Eq.~(\ref{timeEvol}).
\subsubsection{Initial muon-neutrino ($\nu_{\mu}$) state}
Similarly, if the initial state is $\ket{\nu_{\mu}}$, then we can start by considering $\ket{K_0} = (0, 1)^T$ 
and find out that $\ket{K_1} = (1, 0)^T$ 
{\it i.e.,} the Krylov basis is now $\{\ket{K_n}\}= \{\ket{K_0}, \ket{K_1}\} = \{\ket{\nu_{\mu}},\ket{\nu_e}\}$. Then, in this case, the complexity can be calculated as
\begin{align}\label{twoflavorchimu}
	\chi_{\mu} &= 0 \times |\langle K_0|\nu_{\mu}(t)\rangle|^2 + 1 \times |\langle K_1|\nu_{\mu}(t)\rangle|^2 \nonumber \\
	&= 0 \times |\langle \nu_{\mu}|\nu_{\mu}(t)\rangle|^2 + 1 \times |\langle \nu_e|\nu_{\mu}(t)\rangle|^2 \nonumber \\
	&= |\langle \nu_e|\nu_{\mu}(t)\rangle|^2 \nonumber \\
	&= P_{\mu e}
\end{align}
Again, the time evolved state $\ket{\nu_{\mu}(t)}$ is defined in Eq.~(\ref{timeEvol}).

Hence, we see that in the case of two-flavor oscillations the complexity comes out to be equal to the flavor transition probabilities $P_{e\mu}$ (in case of initial $\ket{\nu_e}$) and $P_{\mu e}$ (in case of initial $\ket{\nu_{\mu}}$). It means the complexity is also higher if the probability of transition from one flavor to the other is higher.   
Also, since $P_{e\mu} = P_{\mu e}$ in the case of standard vacuum two-flavor neutrino oscillations, the complexity embedded in this system comes out to be same for both cases of initial flavor, {\it i.e.,} in this case the complexity of the system doesn't depend on the initial flavor of neutrino.\footnote{In the case of Majorana neutrinos we have the Krylov basis-states as $\begin{pmatrix}
		1\\
		0
	\end{pmatrix}$ and $e^{-i \phi}\begin{pmatrix}
		0\\
		1
	\end{pmatrix}$, however, the resultant expression of complexity is true for either case of Dirac or Majorana neutrinos.} In summary, complexity does not reveal additional information compared to probability in the two-flavor neutrino oscillation scenario. 
\subsection{Complexity for three-flavor neutrino oscillations \label{threeflavor}}
In this case, we have three choices of initial states as $\ket{\nu_e}$, $\ket{\nu_{\mu}}$ and $\ket{\nu_{\tau}}$. 
These states can be represented as $\ket{\nu_e}=(1, 0, 0)^T$, $\ket{\nu_{\mu}}=(0, 1, 0)^T$ and $\ket{\nu_{\tau}}=(0, 0, 1)^T$. We follow the same procedure as in the two-flavor case in order to construct the Krylov basis. As we will see shortly, there are exactly three non-zero Krylov states, however, the Krylov states are not equivalent to the flavor states of neutrino in the three-flavor oscillations. Below we provide the forms of Krylov basis for each initial state. 

\subsubsection{Initial electron-neutrino ($\nu_e$) state}
We start by considering
\begin{equation*}
	\ket{K_0} \equiv \ket{\nu_e} = (1,0,0)^T
\end{equation*}
then, other states spanning the Krylov basis take the form as 
\begin{equation*}
	\ket{K_1} = N_{1e}(0,a_1,a_2)^T ~~{\rm and}~~ \ket{K_2} = N_{2e}(0,b_1,b_2)^T, 
\end{equation*}
where,
\begin{align*}
	a_1 =& \left(\frac{\Delta m^2_{21}}{2E}\right) U_{e2}^\ast U_{\mu 2} + \left(\frac{\Delta m^2_{31}}{2E}\right) U_{e3}^\ast U_{\mu 3}\\  
	&= \left(\frac{\Delta m^2_{21}}{2E}\right) \sin \theta_{12} \cos \theta_{12} \cos \theta_{23}
	+ e^{i \delta} \sin \theta_{13} \sin \theta_{23} \left(\left(\frac{\Delta m^2_{31}}{2E}\right) - \left(\frac{\Delta m^2_{21}}{2E}\right) \sin^2 \theta_{12} \right),
\end{align*}
\begin{align*}
	a_2 =& \left(\frac{\Delta m^2_{21}}{2E}\right) U_{e2}^\ast U_{\tau 2} + \left(\frac{\Delta m^2_{31}}{2E}\right) U_{e3}^\ast U_{\tau 3}\\ 
	&= -\left(\frac{\Delta m^2_{21}}{2E}\right) \sin \theta_{12} \cos \theta_{12} \sin \theta_{23}
	+ e^{i \delta} \sin \theta_{13} \cos \theta_{23} \left(\left(\frac{\Delta m^2_{31}}{2E}\right) - \left(\frac{\Delta m^2_{21}}{2E}\right) \sin^2 \theta_{12} \right),
\end{align*}
\begin{align*}
	b_1 =& \left(\frac{\Delta m^2_{21}}{2E}\right)\left(\frac{\Delta m^2_{21}}{2E} - A_e\right) U_{e2}^\ast U_{\mu 2}
	+ \left(\frac{\Delta m^2_{31}}{2E}\right)\left(\frac{\Delta m^2_{31}}{2E} - A_e\right) U_{e3}^\ast U_{\mu 3}\\ 
	&= \sin \theta_{13} \cos \theta_{23} \left(\left(\frac{\Delta m^2_{31}}{2E}\right) - \left(\frac{\Delta m^2_{21}}{2E}\right) \sin^2 \theta_{12} \right)
	- e^{i \delta} \left(\frac{\Delta m^2_{21}}{2E}\right) \sin \theta_{12} \cos \theta_{12} \sin \theta_{23},
\end{align*}
\begin{align*}
	b_2 =& \left(\frac{\Delta m^2_{21}}{2E}\right)\left(\frac{\Delta m^2_{21}}{2E} - A_e\right) U_{e2}^\ast U_{\tau 2}
	+ \left(\frac{\Delta m^2_{31}}{2E}\right)\left(\frac{\Delta m^2_{31}}{2E} - A_e\right) U_{e3}^\ast U_{\tau 3}\\ 
	&= -\sin \theta_{13} \sin \theta_{23} \left(\left(\frac{\Delta m^2_{31}}{2E}\right) - \left(\frac{\Delta m^2_{21}}{2E}\right) \sin^2 \theta_{12} \right)
	- e^{i \delta} \left(\frac{\Delta m^2_{21}}{2E}\right) \sin \theta_{12} \cos \theta_{12} \cos \theta_{23},
\end{align*}
The variables $N_{1\alpha}$, $N_{2\alpha}$ and $A_{\alpha}$ ($\alpha = e, \mu, \tau$) are expressed at the end of this subsection. Then using Eq.~(\ref{eq:time-evolved-kets}) for the time-evolved flavor states and Eq.~(\ref{cost}) we calculate the complexity as
\begin{align}
	\chi_e = & P_{e\mu}(t) \left[ N_{1e}^2|a_1|^2 + 2 N_{2e}^2|b_1|^2) \right] 
	 + P_{e\tau}(t) \left[ N_{1e}^2|a_2|^2 + 2 N_{2e}^2 |b_2|^2 \right] \nonumber \\
	& + 2 \Re\left[ N_{1e}^2 a_1^\ast a_2 A_{e\mu}(t) A_{e\tau}(t)^\ast \right] 
	 + 4 \Re \left[ N_{2e}^2 b_1^\ast b_2 A_{e\mu}(t) A_{e\tau}(t)^\ast \right].
	\label{eq:nue_cost}
\end{align}
Here $\Re$ refers to the real part of the argument and the probabilities $P_{\alpha\beta} (t) = |A_{\alpha\beta}(t)|^2$. Note that the probability for $\nu_e$ to oscillate to other flavors is $1 - P_{ee} = P_{e\mu} + P_{e\tau}$, and differs from the complexity $\chi_e$, which has additional terms.

\subsubsection{Initial muon-neutrino ($\nu_{\mu}$) state}
Similarly, if we start by considering 
\begin{equation*}
	\ket{K_0} \equiv \ket{\nu_{\mu}} = (0,1,0)^T
\end{equation*}
then we get 
\begin{equation*}
	\ket{K_1} = N_{1\mu}(c_1,0,c_2)^T ~~{\rm and}~~ \ket{K_2} = N_{2\mu}(d_1,0,d_2)^T,
\end{equation*}
where,
\begin{align*}
	c_1 =& \left(\frac{\Delta m^2_{21}}{2E}\right) U_{\mu 2}^\ast U_{e2} + \left(\frac{\Delta m^2_{31}}{2E}\right) U_{\mu 3}^\ast U_{e3},\\
	c_2 =& \left(\frac{\Delta m^2_{21}}{2E}\right) U_{\mu 2}^\ast U_{\tau 2} + \left(\frac{\Delta m^2_{31}}{2E}\right) U_{\mu 3}^\ast U_{\tau 3},\\
	d_1 =& \left(\frac{\Delta m^2_{21}}{2E}\right)\left(\frac{\Delta m^2_{21}}{2E} - A_{\mu}\right) U_{\mu 2}^\ast U_{e2}
	+ \left(\frac{\Delta m^2_{31}}{2E}\right)\left(\frac{\Delta m^2_{31}}{2E} - A_{\mu}\right) U_{\mu 3}^\ast U_{e3},\\
	d_2 =& \left(\frac{\Delta m^2_{21}}{2E}\right)\left(\frac{\Delta m^2_{21}}{2E} - A_{\mu}\right) U_{\mu 2}^\ast U_{\tau 2}
	+ \left(\frac{\Delta m^2_{31}}{2E}\right)\left(\frac{\Delta m^2_{31}}{2E} - A_{\mu}\right) U_{\mu 3}^\ast U_{\tau 3}.
\end{align*}
Then following the same procedure as in the $\nu_e$ case we calculate the complexity for the $\nu_\mu$ case as 
\begin{align}\label{threeflavorchimu}
	\chi_{\mu} = &~P_{\mu e}(t) \left[ N_{1\mu}^2|c_1|^2 + 2 N_{2\mu}^2|d_1|^2\right] 
	+ P_{\mu \tau}(t) \left[ N_{1\mu}^2|c_2|^2 + 2 N_{2\mu}^2 |d_2|^2\right] \nonumber\\
	& + 2 \Re\left[ N_{1\mu}^2 c_1^\ast c_2 A_{\mu e}(t) A_{\mu \tau}(t)^\ast \right] 
	+ 4 \Re\left[ N_{2\mu}^2 d_1^\ast d_2 A_{\mu e}(t) A_{\mu \tau}(t)^\ast \right].
\end{align}

\subsubsection{Initial tau-neutrino ($\nu_{\tau}$) state}
Again, we start with the initial flavor state 
\begin{equation*}
	\ket{K_0} \equiv \ket{\nu_{\tau}} = (0,0,1)^T,
\end{equation*}
and obtain the two other Krylov states as
\begin{equation*}
	\ket{K_1} = N_{1\tau} (e_1,e_2,0)^T ~~{\rm and}~~ \ket{K_2} = N_{2\tau} (f_1,f_2,0)^T,
\end{equation*}
where,
\begin{align*}
	e_1 =& \left(\frac{\Delta m^2_{21}}{2E}\right) U_{\tau 2}^\ast U_{e2} + \left(\frac{\Delta m^2_{31}}{2E}\right) U_{\tau 3}^\ast U_{e3},\\
	e_2 =& \left(\frac{\Delta m^2_{21}}{2E}\right) U_{\tau 2}^\ast U_{\mu 2} + \left(\frac{\Delta m^2_{31}}{2E}\right) U_{\tau 3}^\ast U_{\mu 3},\\
	f_1 =& \left(\frac{\Delta m^2_{21}}{2E}\right)\left(\frac{\Delta m^2_{21}}{2E} - A_{\tau}\right) U_{\tau 2}^\ast U_{e2} 
	+ \left(\frac{\Delta m^2_{31}}{2E}\right)\left(\frac{\Delta m^2_{31}}{2E} - A_{\tau}\right) U_{\tau 3}^\ast U_{e3},\\
	f_2 =& \left(\frac{\Delta m^2_{21}}{2E}\right)\left(\frac{\Delta m^2_{21}}{2E} - A_{\tau}\right) U_{\tau 2}^\ast U_{\mu 2} 
	+ \left(\frac{\Delta m^2_{31}}{2E}\right)\left(\frac{\Delta m^2_{31}}{2E} - A_{\tau}\right) U_{\tau 3}^\ast U_{\mu 3}.
\end{align*}
The complexity in this case is given by
\begin{align}\label{threeflavorchitau}
	\chi_{\tau} = &~P_{\tau e}(t) \left[ N_{1\tau}^2|e_1|^2 + 2 N_{2\tau}^2|f_1|^2 \right] 
	+ P_{\tau \mu}(t) \left[ N_{1\tau}^2|e_2|^2 + 2 N_{2\tau}^2 |f_2|^2\right] \nonumber\\
	& + 2 \Re\left[ N_{1\tau}^2 e_1^\ast e_2 A_{\tau e}(t) A_{\tau \mu}(t)^\ast\right] 
	+ 4 \Re\left[ N_{2\tau}^2 f_1^\ast f_2 A_{\tau e}(t) A_{\tau \mu}(t)^\ast\right].
\end{align}

Here, we give analytical expressions for 
constants used in previous discussions for initial neutrino flavor $\nu_{\alpha}$.
\begin{align*}
	A_{\alpha} =& \frac{1}{2E}\left[\left(\Delta m^2_{21}\right)^3 |U_{\alpha 2}|^2 (1-|U_{\alpha 2}|^2) + \left(\Delta m^2_{31}\right)^3 |U_{\alpha 3}|^2 (1-|U_{\alpha 3}|^2)
		- \left(\Delta m^2_{21}\right)\left(\Delta m^2_{31}\right)\right.\\
		&\left. |U_{\alpha 2}|^2 |U_{\alpha 3}|^2 \left(\Delta m^2_{21}+\Delta m^2_{31}\right)\right] \left[\left(\Delta m^2_{21}\right)^2 |U_{\alpha 2}|^2 (1-|U_{\alpha 2}|^2) + \left(\Delta m^2_{31}\right)^2 |U_{\alpha 3}|^2\right.\\ &\left.(1-|U_{\alpha 3}|^2) - 2\left(\Delta m^2_{21}\right)\left(\Delta m^2_{31}\right) |U_{\alpha 2}|^2 |U_{\alpha 3}|^2\right]^{-1},
\end{align*}
and normalization constants
\begin{align*}
	N_{1\alpha} = &\left(\left(\frac{\Delta m^2_{21}}{2E}\right)^2 |U_{\alpha 2}|^2 (1-|U_{\alpha 2}|^2
+ \left(\frac{\Delta m^2_{31}}{2E}\right)^2 |U_{\alpha 3}|^2 (1-|U_{\alpha 3}|^2)\right.\\	
&\left.- 2 \left(\frac{\Delta m^2_{21}}{2E}\right)\left(\frac{\Delta m^2_{31}}{2E}\right) |U_{\alpha 2}|^2 |U_{\alpha 3}|^2\right)^{-1/2},
\end{align*}
\begin{equation*}
	\begin{split}
		N_{2\alpha} &= \left( \left(\frac{\Delta m^2_{21}}{2E}\right)^2 \left(\frac{\Delta m^2_{21}}{2E} - A_{\alpha}\right)^2 |U_{\alpha 2}|^2 (1-|U_{\alpha 2}|^2)\right. \\
		&\left.+ \left(\frac{\Delta m^2_{31}}{2E}\right)^2 \left(\frac{\Delta m^2_{31}}{2E} - A_{\alpha}\right)^2 |U_{\alpha 3}|^2 (1-|U_{\alpha 3}|^2)\right. \\
		&\left.- 2\left(\frac{\Delta m^2_{21}}{2E}\right) \left(\frac{\Delta m^2_{31}}{2E}\right) \left(\frac{\Delta m^2_{21}}{2E} - A_{\alpha}\right)
		\left(\frac{\Delta m^2_{31}}{2E} - A_{\alpha}\right) |U_{\alpha 2}|^2 |U_{\alpha 3}|^2\right)^{-1/2}.
	\end{split}
\end{equation*}
Explicit expressions of vacuum oscillation amplitudes $A_{\alpha\beta}(t)$ are given in the appendix.

\subsection{Matter effects on the complexity of neutrino system}
Neutrinos can also travel through a medium that may induce a matter potential  
due to coherent forward-scattering of electron neutrinos ($\nu_e$) with electrons contained inside that matter \cite{msw1, Mikheev:1986gs}. In that case, the Hamiltonian in flavor basis has an extra matter potential term. For a constant matter density this extra term $V=\pm \sqrt{2}G_f N_e$ is added to the vacuum Hamiltonian as
\begin{equation*}
	H_f = U H_m U^{-1} + V ~diag(1,0,0).
\end{equation*}
Here, $G_f$ and $N_e$ are the Fermi constant and electron number density in matter, respectively. The "+" and "-" signs of the potential correspond to neutrinos and antineutrinos, respectively.

In the case of constant matter density, the initial two Krylov states come out to be the same as those in the case of vacuum oscillations, {\it i.e.,}
\begin{eqnarray}
	\ket{K_0}_{\alpha}^{matter} = \ket{K_0}_{\alpha}^{vacuum}~\\
	\ket{K_1}_{\alpha}^{matter} = \ket{K_1}_{\alpha}^{vacuum},
\end{eqnarray}
where $\alpha$ represents the flavor of neutrino at the time of production. However, $\ket{K_2}$ contains the effects of constant matter density. The expression of the $\ket{K_2}$ state for the initial $\nu_e$ flavor is as follows
\begin{equation*}
	\ket{K_2}_e = N_{2e}^m (0, b_1^m, b_2^m)^T
\end{equation*}
where,
\begin{align*}
	b_1^m &= \left(\frac{\Delta m^2_{21}}{2E}\right)\left(\frac{\Delta m^2_{21}}{2E} + V - B_e\right) U_{e2}^\ast U_{\mu 2}
	+ \left(\frac{\Delta m^2_{31}}{2E}\right)\left(\frac{\Delta m^2_{31}}{2E} + V - B_e\right) U_{e3}^\ast U_{\mu 3},\\
	b_2^m &= \left(\frac{\Delta m^2_{21}}{2E}\right)\left(\frac{\Delta m^2_{21}}{2E} + V - B_e\right) U_{e2}^\ast U_{\tau 2}
	+ \left(\frac{\Delta m^2_{31}}{2E}\right)\left(\frac{\Delta m^2_{31}}{2E} + V - B_e\right) U_{e3}^\ast U_{\tau 3}.
\end{align*}
The superscript $m$ here stands for matter effects. Similarly, for the initial $\nu_\mu$ flavor
\begin{equation*}
	\ket{K_2}_{\mu} = N_{2\mu}^m (d_1^m, 0, d_2^m)^T,
\end{equation*}
where, 
\begin{align*}
	d_1^m &= \left(\frac{\Delta m^2_{21}}{2E}\right)\left(\frac{\Delta m^2_{21}}{2E} + V - B_{\mu}\right) U_{e2} U_{\mu 2}^\ast 
	+ \left(\frac{\Delta m^2_{31}}{2E}\right)\left(\frac{\Delta m^2_{31}}{2E} + V - B_{\mu}\right) U_{e3} U_{\mu 3}^\ast\\
	d_2^m &= \left(\frac{\Delta m^2_{21}}{2E}\right)\left(\frac{\Delta m^2_{21}}{2E} - B_{\mu}\right) U_{\mu 2}^\ast U_{\tau 2}
	+ \left(\frac{\Delta m^2_{31}}{2E}\right)\left(\frac{\Delta m^2_{31}}{2E} - B_{\mu}\right) U_{\mu 3}^\ast U_{\tau 3},
\end{align*}
and for the initial $\nu_{\tau}$ flavor
\begin{equation*}
	\ket{K_2}_{\tau} = N_{2\tau}^m (f_1^m, f_2^m, 0)^T
\end{equation*}
where,
\begin{align*}
	f_1^m &= \left(\frac{\Delta m^2_{21}}{2E}\right)\left(\frac{\Delta m^2_{21}}{2E} + V - B_{\tau}\right) U_{e2} U_{\tau 2}^\ast 
	+ \left(\frac{\Delta m^2_{31}}{2E}\right)\left(\frac{\Delta m^2_{31}}{2E} + V - B_{\tau}\right) U_{e3} U_{\tau 3}^\ast,\\
	f_2^m &= \left(\frac{\Delta m^2_{21}}{2E}\right)\left(\frac{\Delta m^2_{21}}{2E} - B_{\tau}\right) U_{\mu 2} U_{\tau 2}^\ast 
	+ \left(\frac{\Delta m^2_{31}}{2E}\right)\left(\frac{\Delta m^2_{31}}{2E} - B_{\tau}\right) U_{\mu 3} U_{\tau 3}^\ast.
\end{align*}
\vspace{0.9cm}
The constant $B_e$ is represented as
\begin{align*}
	B_e =& \left[\left(\Delta m^2_{21}\right)^2 \left(\Delta m^2_{21} + 2EV\right) |U_{e2}|^2 (1-|U_{e 2}|^2) + \left(\Delta m^2_{31}\right)^2 \left(\Delta m^2_{31} + 2EV\right) |U_{e3}|^2\right.\\
	&\left. (1-|U_{e3}|^2) - \left(\Delta m^2_{21}\right)\left(\Delta m^2_{31}\right) |U_{e2}|^2 |U_{e3}|^2 \left((\Delta m^2_{21} + 2EV)+(\Delta m^2_{31} + 2EV)\right)\right]\\
	& \left[2E\left[\left(\Delta m^2_{21}\right)^2 |U_{e2}|^2 (1-|U_{e2}|^2) + \left(\Delta m^2_{31}\right)^2 |U_{e3}|^2 (1-|U_{e3}|^2)\right.\right.\\
	&\left.\left.~~~~~~~~~~~~ - 2\left(\Delta m^2_{21}\right)\left(\Delta m^2_{31}\right) |U_{e2}|^2 |U_{e3}|^2\right]\right]^{-1}.
\end{align*}
For initial $\nu_{\mu}$ and $\nu_{\tau}$ state the constant $B_\alpha$ is
	\begin{align*}
		B_{\alpha} =& \left[\left(\Delta m^2_{21}\right)^3 |U_{\alpha 2}|^2 (1-|U_{\alpha 2}|^2) + \left(\Delta m^2_{31}\right)^3 |U_{\alpha 3}|^2 (1-|U_{\alpha 3}|^2) - \left(\Delta m^2_{21}\right)\left(\Delta m^2_{31}\right)\right.\\
		&\left. |U_{\alpha 2}|^2 |U_{\alpha 3}|^2
		 \left(\Delta m^2_{21}+\Delta m^2_{31}\right)+ 2EV\left(\left(\Delta m^2_{21}\right)^2 |U_{e2}|^2 |U_{\alpha 2}|^2 + \left(\Delta m^2_{31}\right)^2 |U_{e3}|^2 |U_{\alpha 3}|^2\right.\right.\\
		 &\left.\left.+ 2\left(\Delta m^2_{21}\right) \left(\Delta m^2_{31}\right)\Re(U_{e2}^\ast U_{\alpha 2}U_{e3}U_{\alpha 3}^\ast)\right)\right]  \left[2E\left[\left(\Delta m^2_{21}\right)^2 |U_{\alpha 2}|^2 (1-|U_{\alpha 2}|^2) + \left(\Delta m^2_{31}\right)^2\right.\right.\\
		 &\left.\left. |U_{\alpha 3}|^2 (1-|U_{\alpha 3}|^2) - 2\left(\Delta m^2_{21}\right)\left(\Delta m^2_{31}\right) |U_{\alpha 2}|^2 |U_{\alpha 3}|^2\right]\right]^{-1},
	\end{align*}
	where $\alpha=\mu,\tau$. The normalization factor $N_{1\alpha}$ remains the same in matter as in vacuum but the normalization factor $N_{2\alpha}$ is modified as given below.
	\begin{equation*}
		\begin{split}
			N_{2e}^m =& \left( \left(\frac{\Delta m^2_{21}}{2E}\right)^2 |U_{e2}|^2 (1-|U_{e2}|^2)\left[\left(\frac{\Delta m^2_{21}}{2E} + V - B_{e}\right)^2 \right]\right. \\
			&\left.
			+ \left(\frac{\Delta m^2_{31}}{2E}\right)^2 |U_{e3}|^2 (1-|U_{e3}|^2)\left[\left(\frac{\Delta m^2_{31}}{2E} + V - B_{e}\right)^2 \right]\right. \\
			&\left.- 2\left(\frac{\Delta m^2_{21}}{2E}\right) \left(\frac{\Delta m^2_{31}}{2E}\right) \left(\frac{\Delta m^2_{21}}{2E} + V - B_{e}\right) \left(\frac{\Delta m^2_{31}}{2E} + V - B_{e}\right)|U_{e2}|^2 |U_{e3}|^2 \right)^{-1/2},
		\end{split}
	\end{equation*}
	\begin{equation*}
		\begin{split}
			N_{2\mu}^m =& \left(\left(\frac{\Delta m^2_{21}}{2E}\right)^2 |U_{\mu 2}|^2\left[\left(\frac{\Delta m^2_{21}}{2E} + V - B_{\mu}\right)^2 |U_{e2}|^2 + \left(\frac{\Delta m^2_{21}}{2E} - B_{\mu}\right)^2 |U_{\tau 2}|^2\right]\right. \\
			&\left.+ \left(\frac{\Delta m^2_{31}}{2E}\right)^2 |U_{\mu 3}|^2\left[\left(\frac{\Delta m^2_{31}}{2E} + V - B_{\mu}\right)^2 |U_{e3}|^2 + \left(\frac{\Delta m^2_{31}}{2E} - B_{\mu}\right)^2 |U_{\tau 3}|^2\right]\right. \\
			&\left.+ 2\left(\frac{\Delta m^2_{21}}{2E}\right) \left(\frac{\Delta m^2_{31}}{2E}\right) \left[\left(\frac{\Delta m^2_{21}}{2E} + V - B_{\mu}\right) \left(\frac{\Delta m^2_{31}}{2E} + V - B_{\mu}\right) \Re(U_{\mu 2}^\ast U_{e2}U_{\mu 3}U_{e3}^\ast) \right.\right. \\
			&\left.\left.+ \left(\frac{\Delta m^2_{21}}{2E} - B_{\mu}\right) \left(\frac{\Delta m^2_{31}}{2E} - B_{\mu}\right) \Re(U_{\mu 2}^\ast U_{\tau 2}U_{\mu 3}U_{\tau 3}^\ast)\right]\right)^{-1/2},
		\end{split}
	\end{equation*}
	\begin{equation*}
		\begin{split}
			N_{2\tau}^m =& \left(\left(\frac{\Delta m^2_{21}}{2E}\right)^2 |U_{\tau 2}|^2\left[\left(\frac{\Delta m^2_{21}}{2E} + V - B_{\tau}\right)^2 |U_{e2}|^2 + \left(\frac{\Delta m^2_{21}}{2E} - B_{\tau}\right)^2 |U_{\mu 2}|^2\right]\right. \\
			&\left.+ \left(\frac{\Delta m^2_{31}}{2E}\right)^2 |U_{\tau 3}|^2\left[\left(\frac{\Delta m^2_{31}}{2E} + V - B_{\tau}\right)^2 |U_{e3}|^2 + \left(\frac{\Delta m^2_{31}}{2E} - B_{\tau}\right)^2 |U_{\mu 3}|^2\right]\right. \\
			&\left.+ 2\left(\frac{\Delta m^2_{21}}{2E}\right) \left(\frac{\Delta m^2_{31}}{2E}\right) \left[\left(\frac{\Delta m^2_{21}}{2E} + V - B_{\tau}\right) \left(\frac{\Delta m^2_{31}}{2E} + V - B_{\tau}\right) \Re(U_{\tau 2}^\ast U_{e2}U_{\tau 3}U_{e3}^\ast) \right.\right. \\
			&\left.\left.+ \left(\frac{\Delta m^2_{21}}{2E} - B_{\tau}\right) \left(\frac{\Delta m^2_{31}}{2E} - B_{\tau}\right) \Re(U_{\tau 2}^\ast U_{\mu 2}U_{\tau 3}U_{\mu 3}^\ast)\right]\right)^{-1/2}.
		\end{split}
	\end{equation*}
\section{Results} \label{Results}

\begin{figure*}[t] 
	\centering
	\begin{tabular}{cc}
		\includegraphics[width=.42\textwidth]{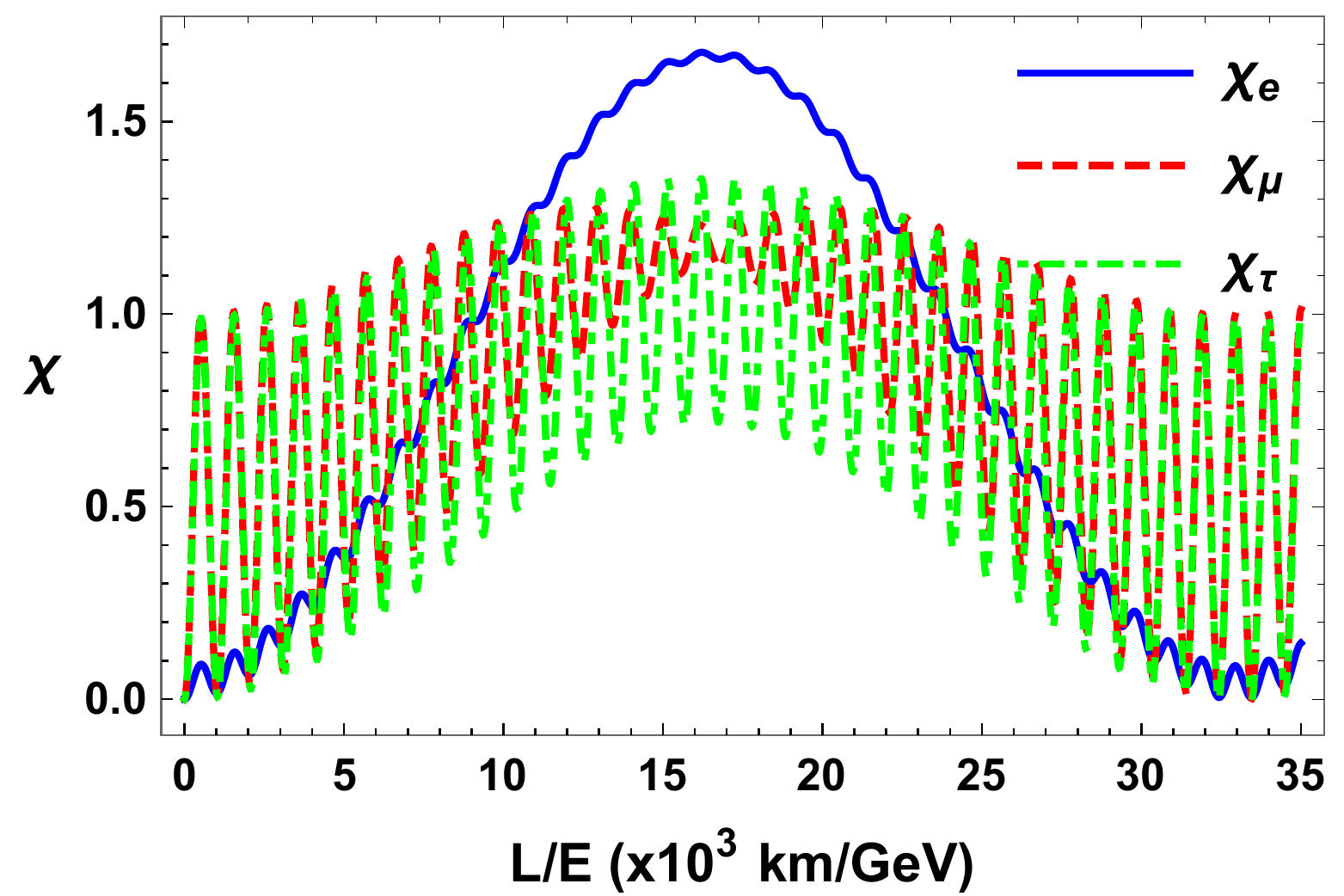}\ \ \ \ \ \ \ \ \
		\includegraphics[width=.42\textwidth]{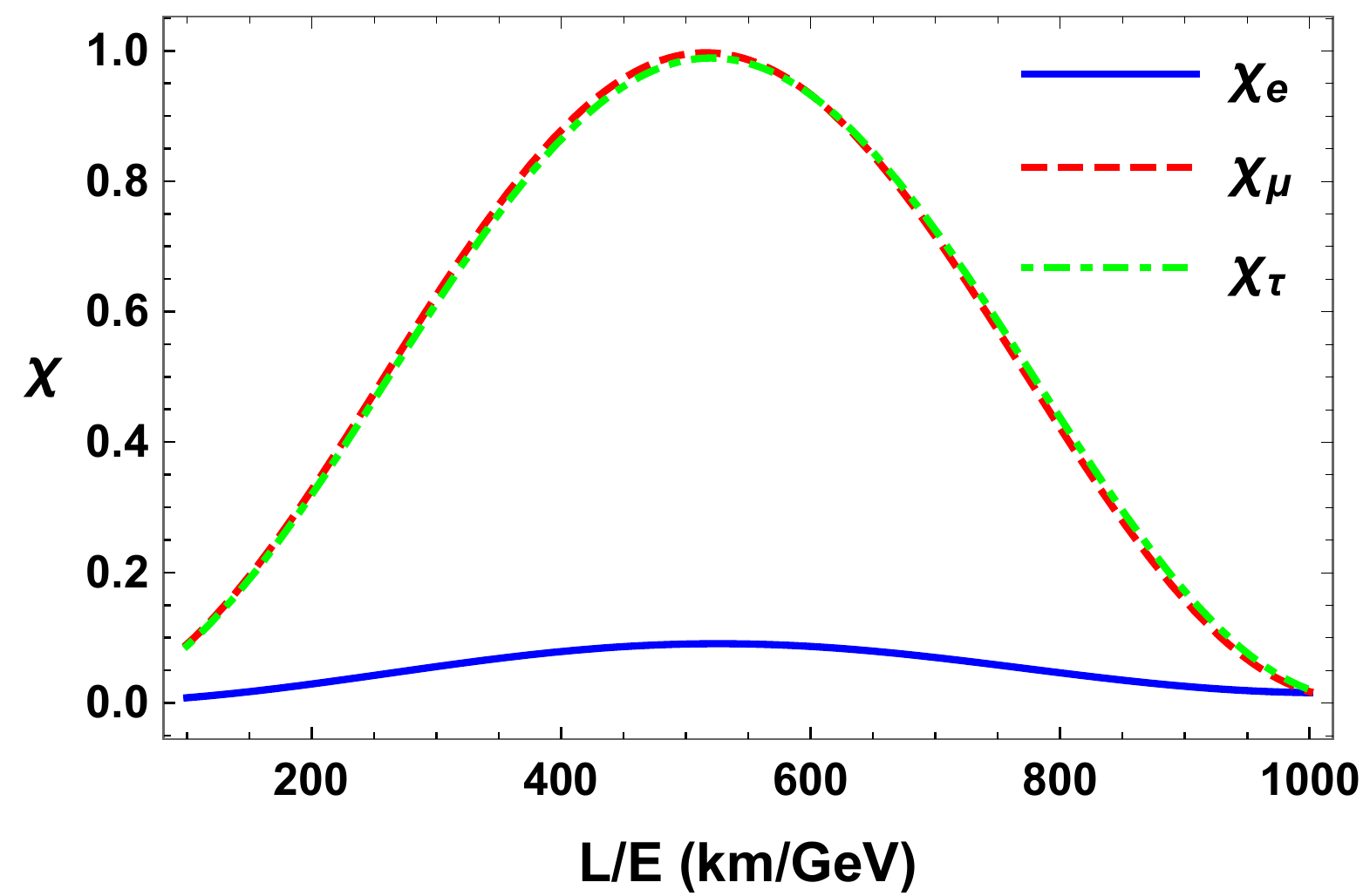}
	\end{tabular}
	\caption{(color online) Complexity plotted with respect to the distance $L$ over energy $E$ traveled by neutrinos in vacuum and in case if the initial flavor is $\nu_e$ (blue solid line), $\nu_{\mu}$ (red dashed line) and $\nu_{\tau}$ (green dot-dashed line) for $CP$-violating phase $\delta = 0^o$. All other parameters are set at their best-fit values.}
	\label{Chi_L}
\end{figure*}

In this section, we explore the effects of oscillation parameters on the complexity of the three-flavor neutrino oscillation system using numerical calculations. To obtain all the plots, we have considered the best-fit values of the oscillation parameters from reference \cite{ParticleDataGroup:2022pth} as $\theta_{12} = 33.64^o$, $\theta_{13} = 8.53^o$, $\theta_{23} = 47.63^o$ and $\Delta m^2_{21} = 7.53\times 10^{-5}$~eV$^2$. For normal hierarchy, we have used $\Delta m^2_{31} = 2.528\times 10^{-3}$~eV$^2$ and $\Delta m^2_{31} = -2.46\times 10^{-3}$~eV$^2$ for inverted hierarchy. 

\begin{figure*}[t] 
	\centering
	\begin{tabular}{cc}
		\includegraphics[width=.32\textwidth]{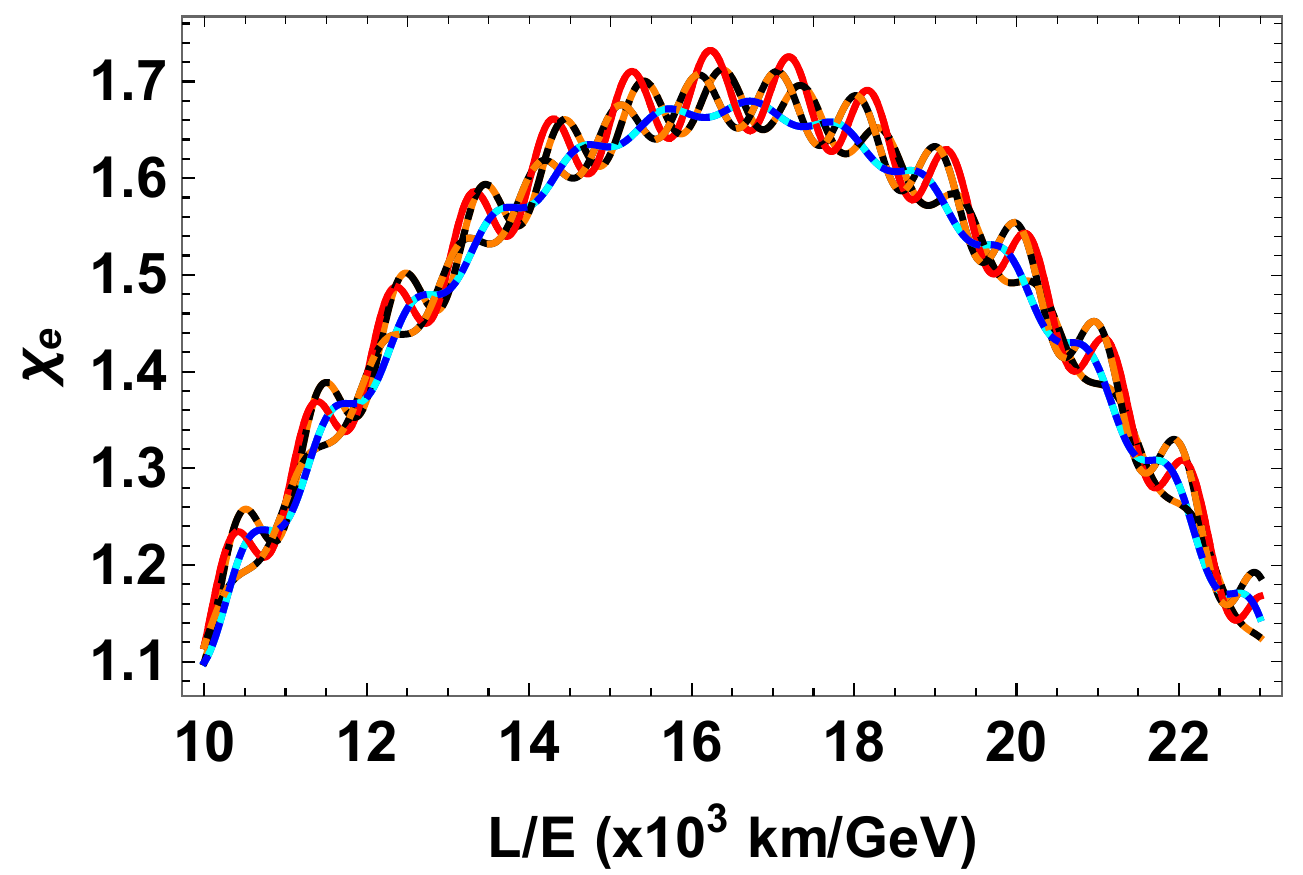}
		\includegraphics[width=.32\textwidth]{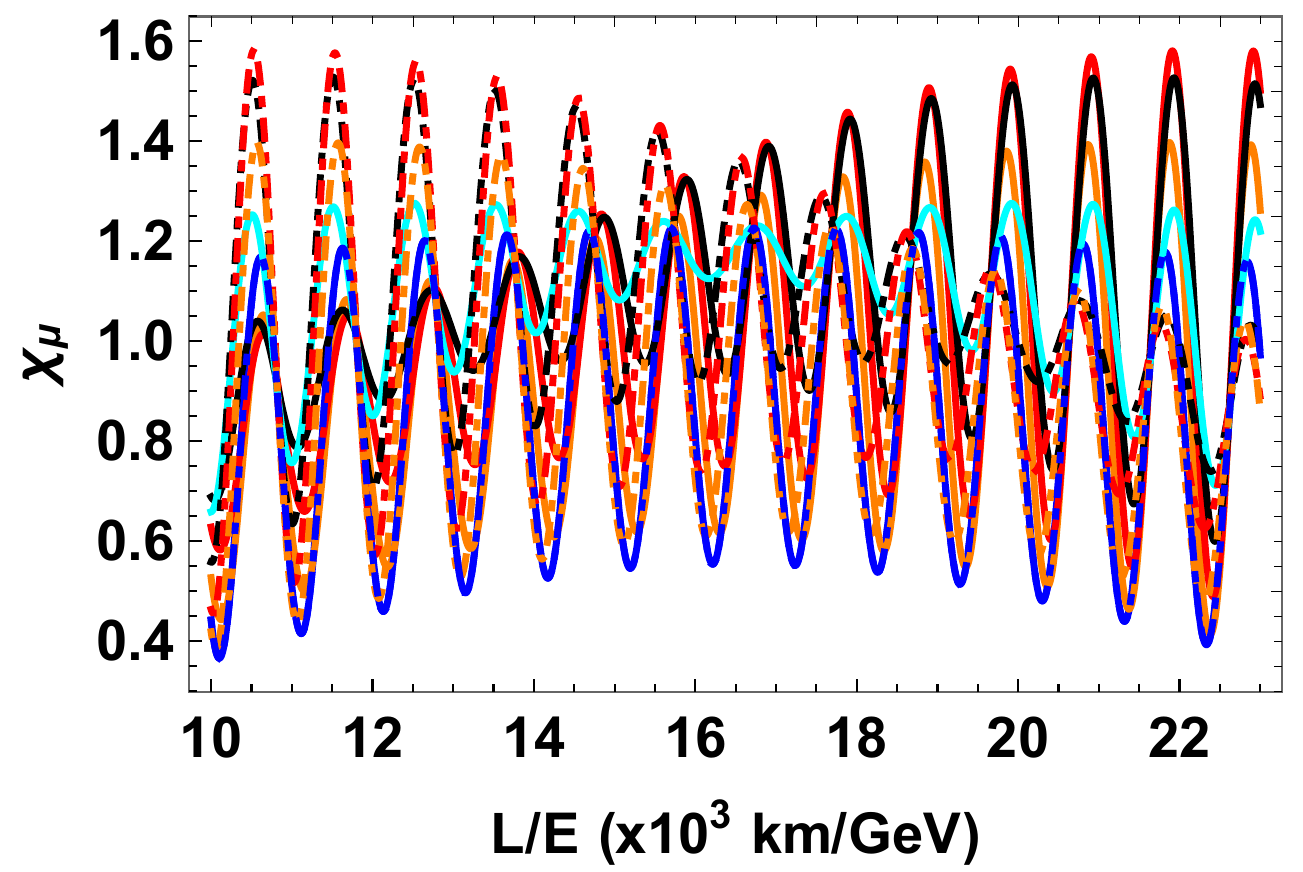}
		\includegraphics[width=.32\textwidth]{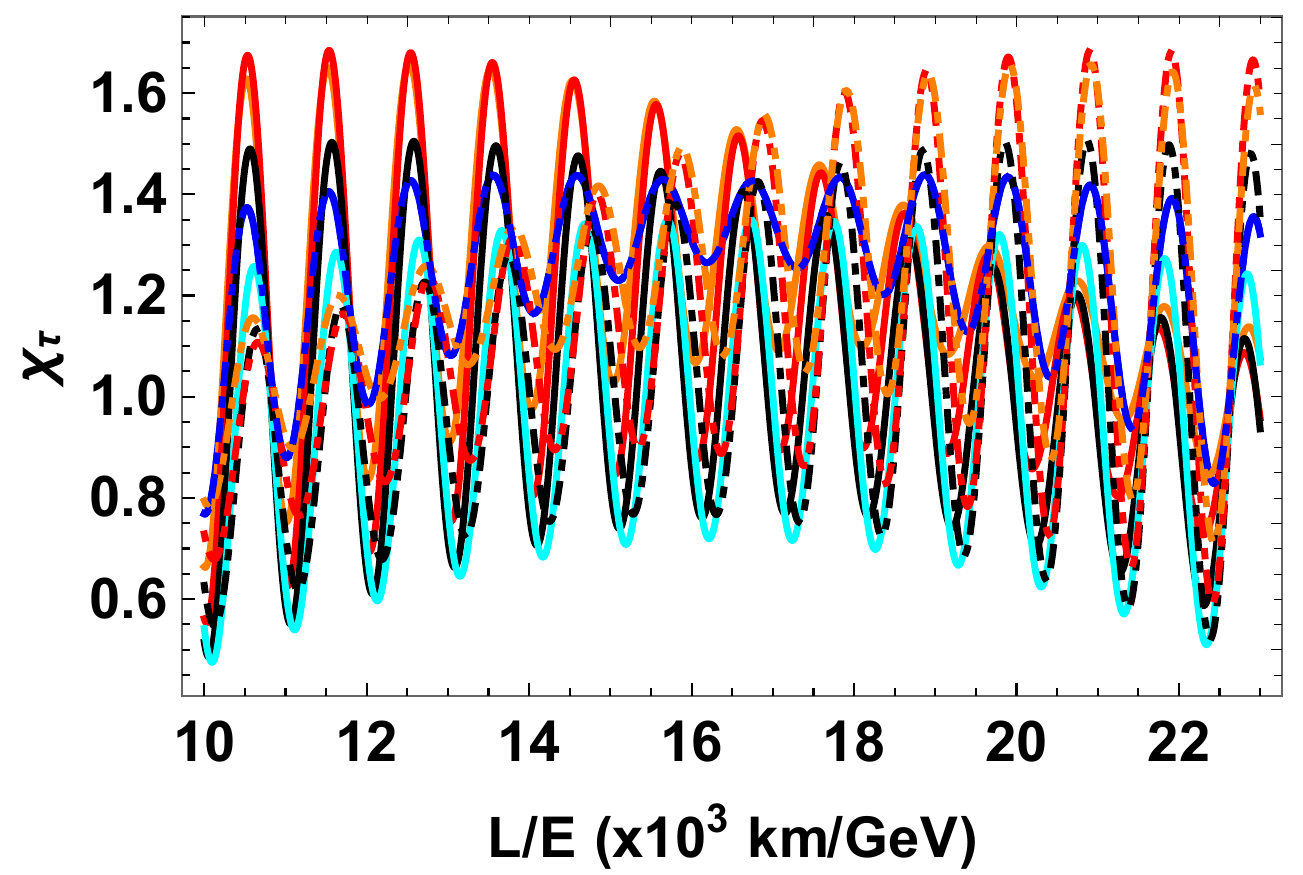} \\
		\includegraphics[width=.32\textwidth]{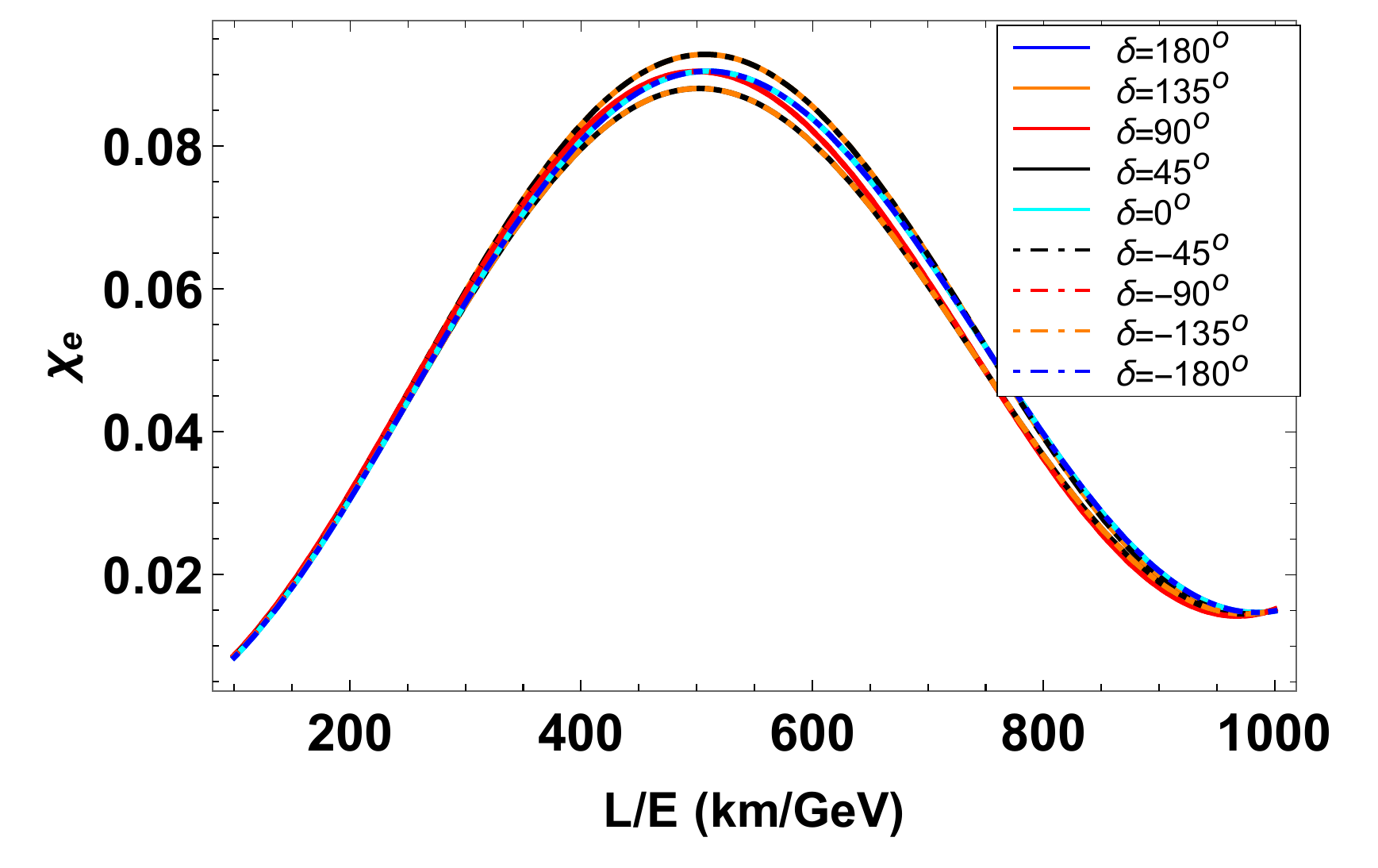}
		\includegraphics[width=.32\textwidth]{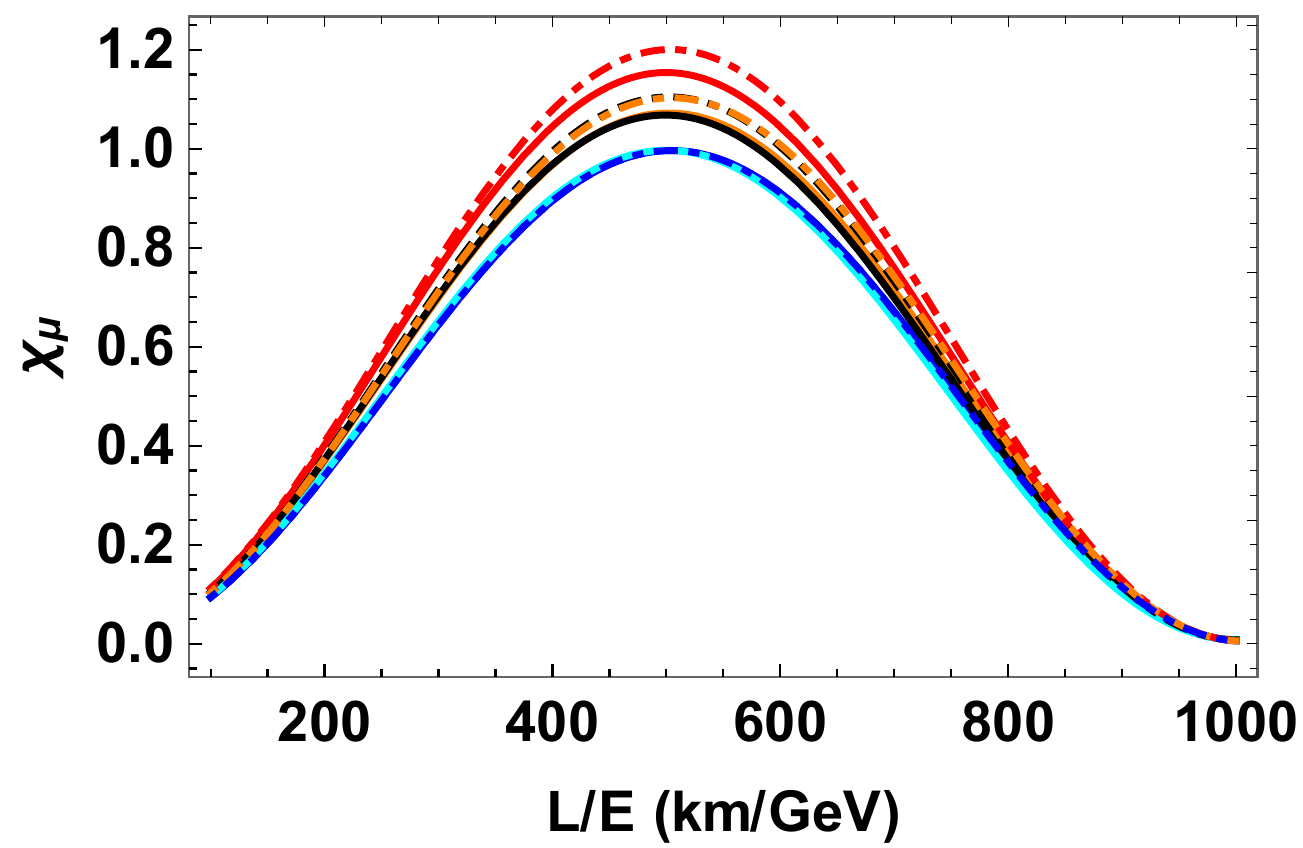}
		\includegraphics[width=.32\textwidth]{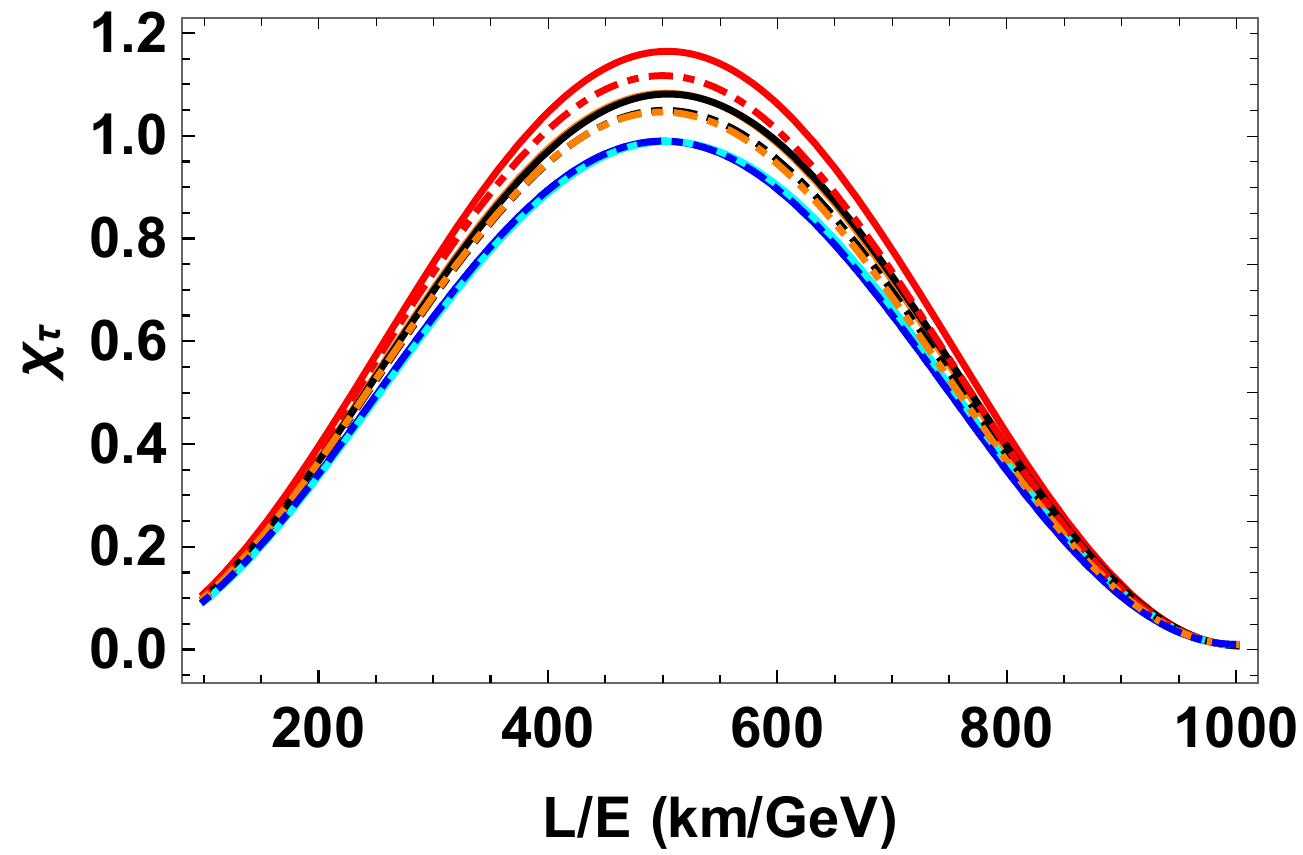}\\
		\includegraphics[width=.32\textwidth]{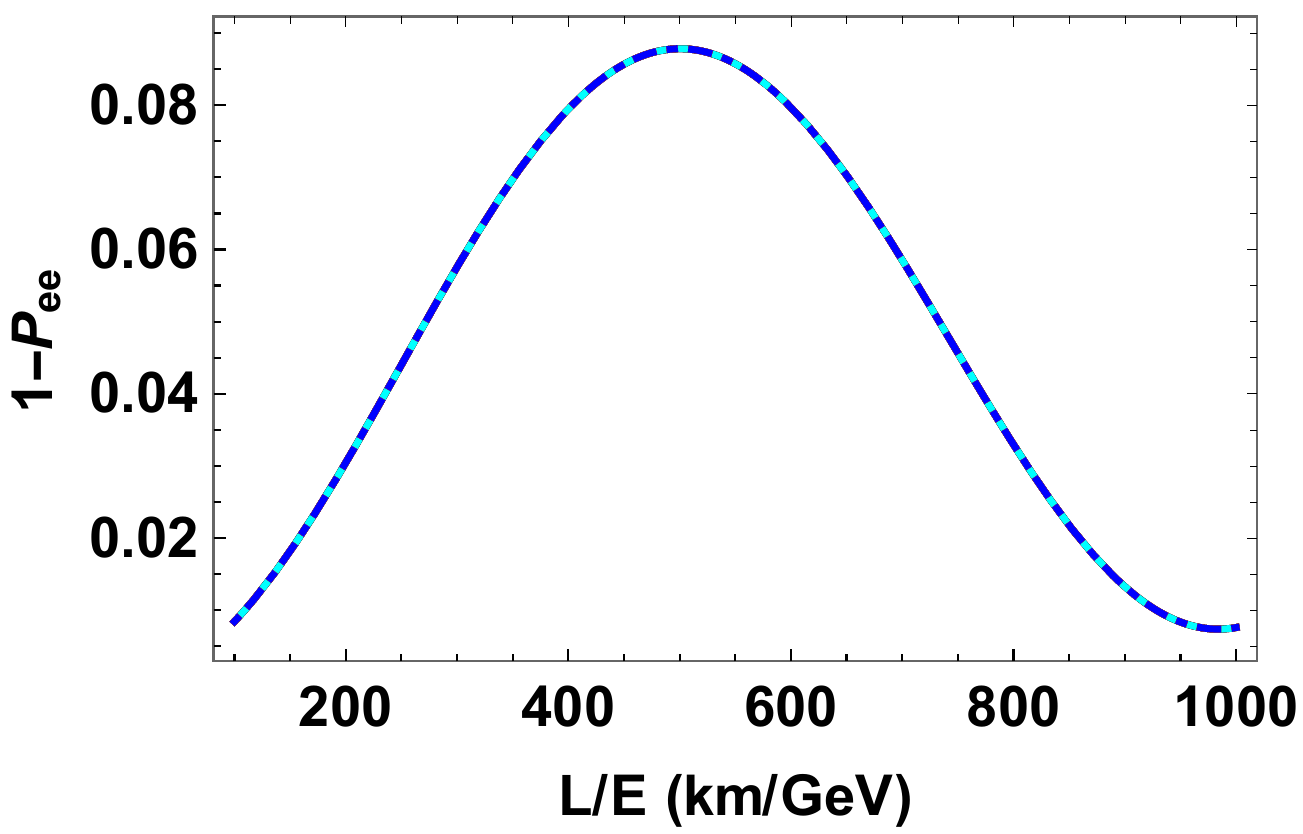}
		\includegraphics[width=.32\textwidth]{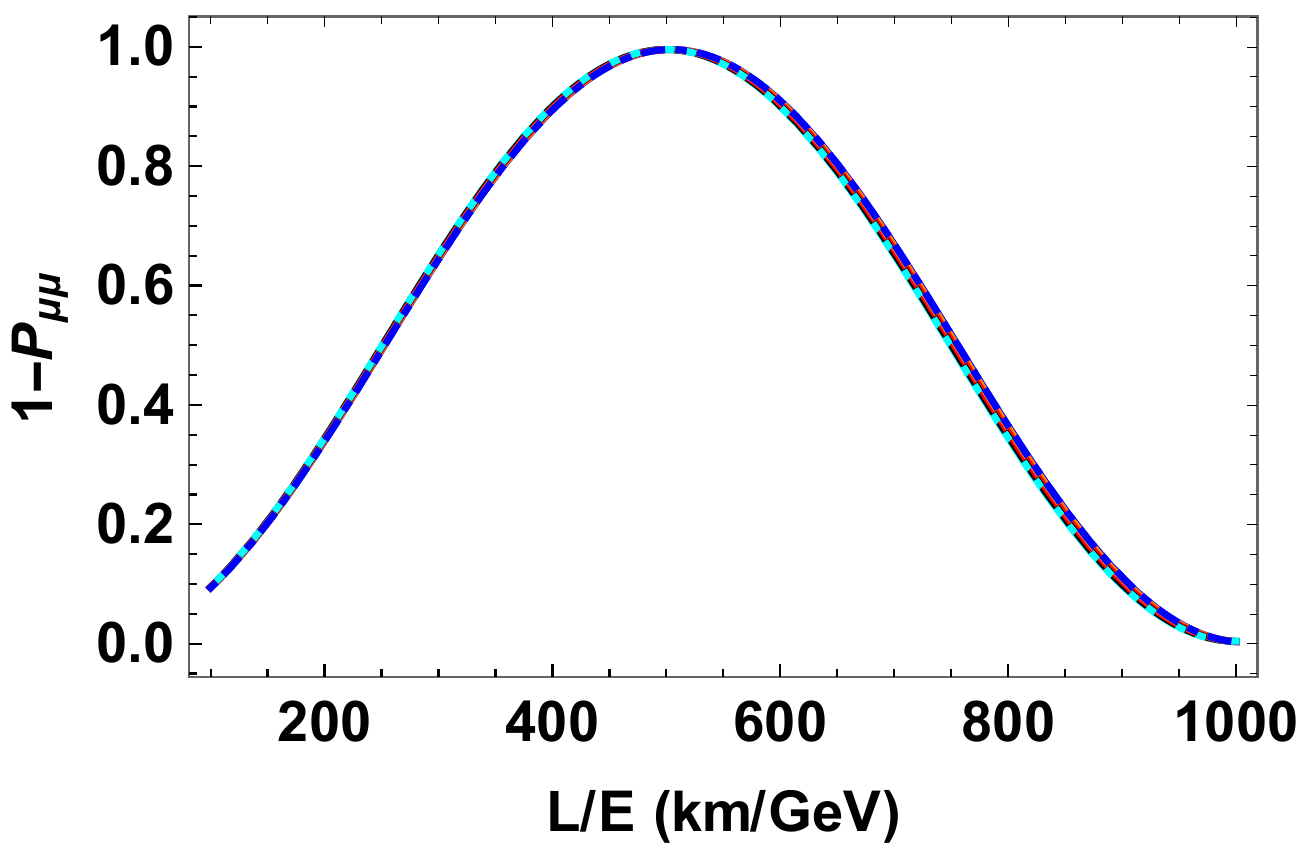}
		\includegraphics[width=.32\textwidth]{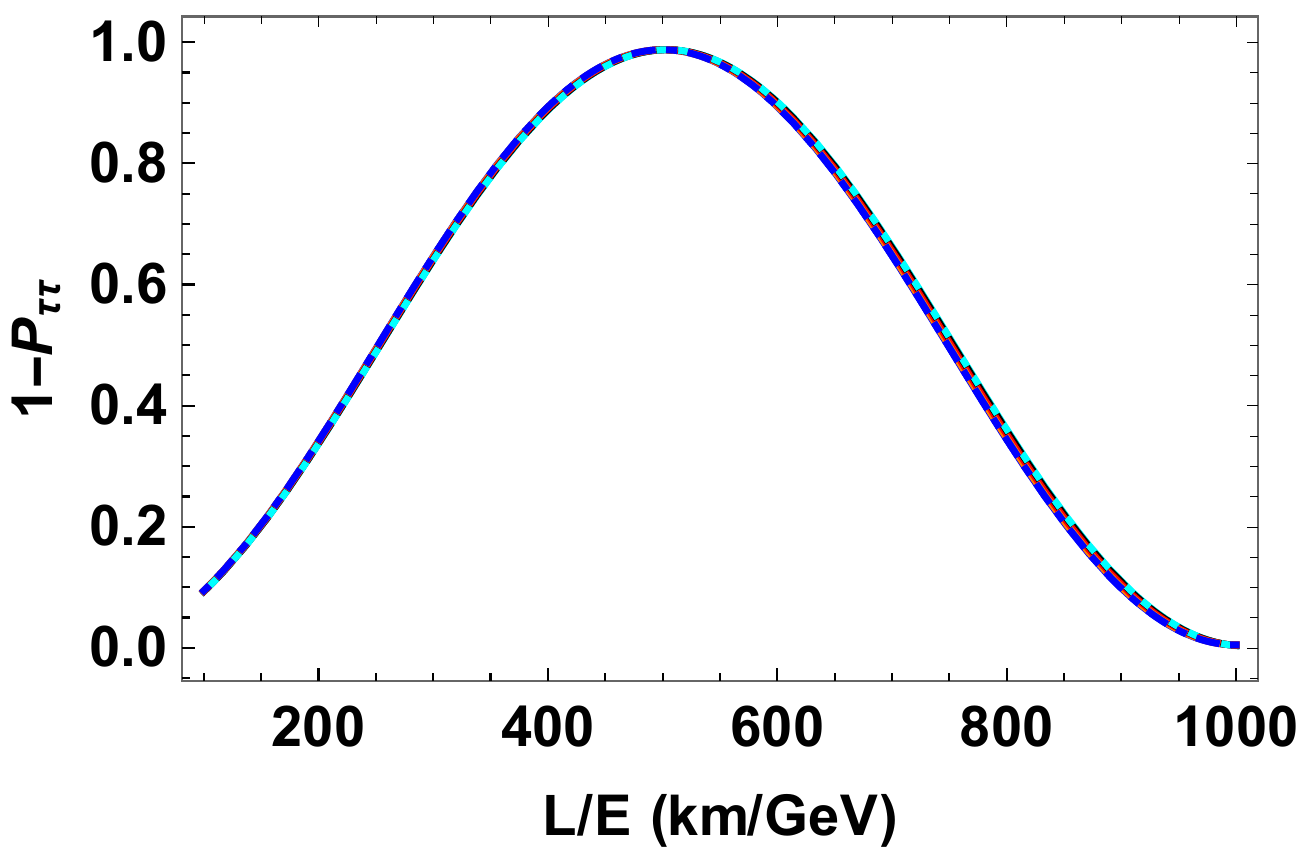}\\
	\end{tabular}
	\caption{(color online) Complexity for large $L/E$ range (upper panels), small $L/E$ range (middle panels) and 1-$P_{\alpha \alpha}$ (lower panels) with respect to  $L/E$ for neutrinos traveling in vacuum in the case if the initial flavor is $\nu_e$ (left), $\nu_{\mu}$ (middle) and $\nu_{\tau}$ (right) for different values of the $CP$-violating phase $\delta$ depicted by different colors. 
	}
	\label{Cost_L_Exp}
\end{figure*}

\begin{figure*}[t] 
	\centering
	\begin{tabular}{cc}
		\includegraphics[width=.325\textwidth]{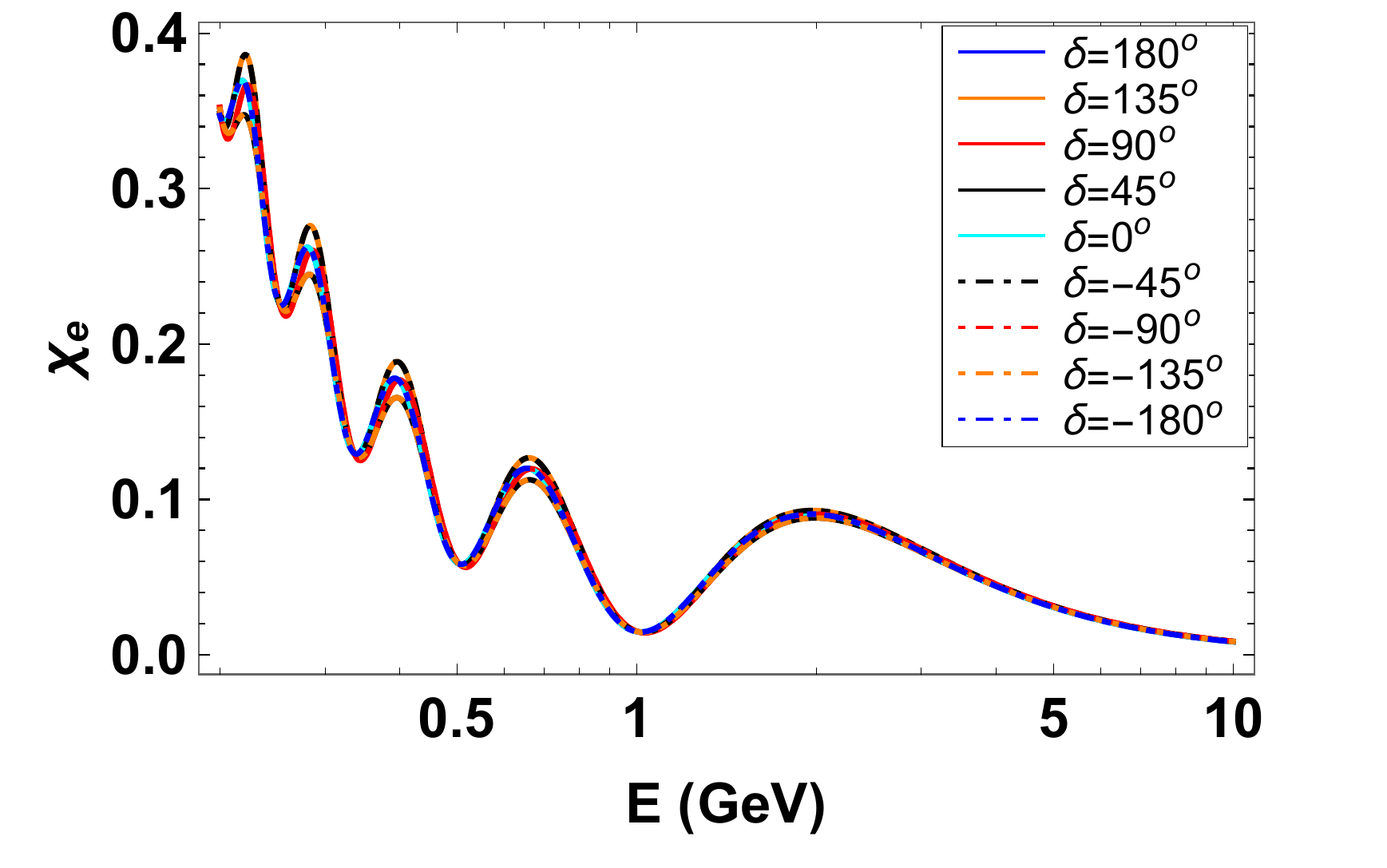}
		\includegraphics[width=.32\textwidth]{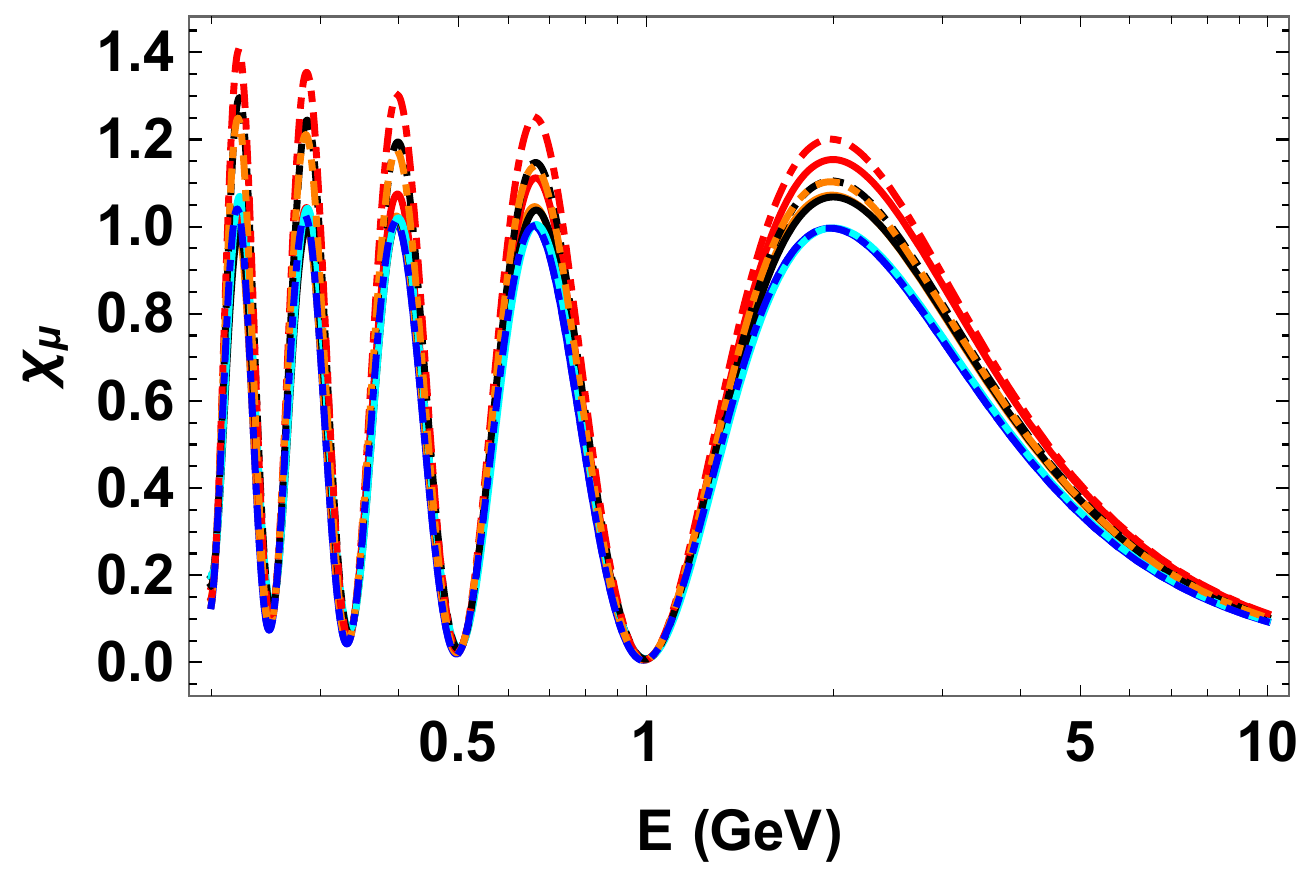}
		\includegraphics[width=.32\textwidth]{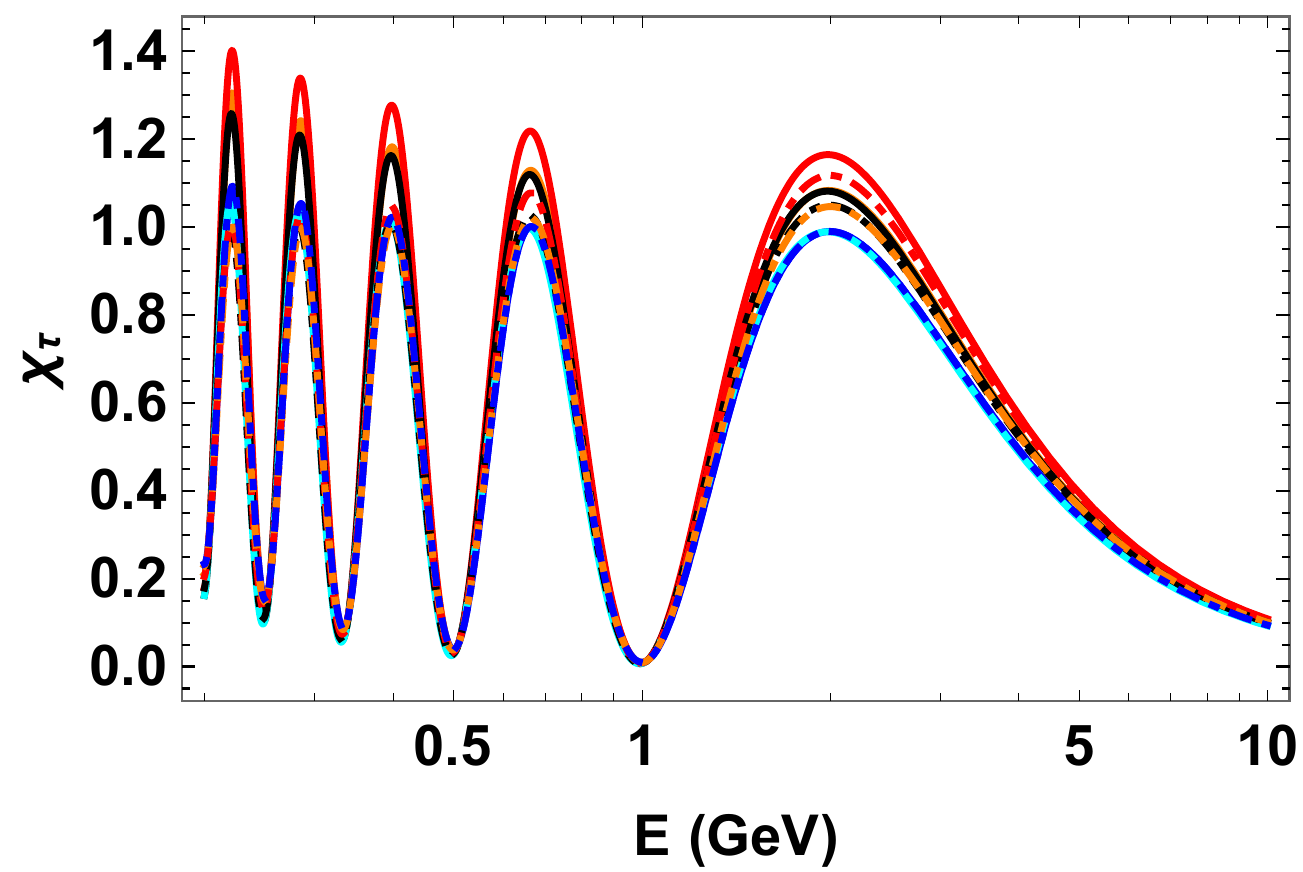}\\
		\includegraphics[width=.32\textwidth]{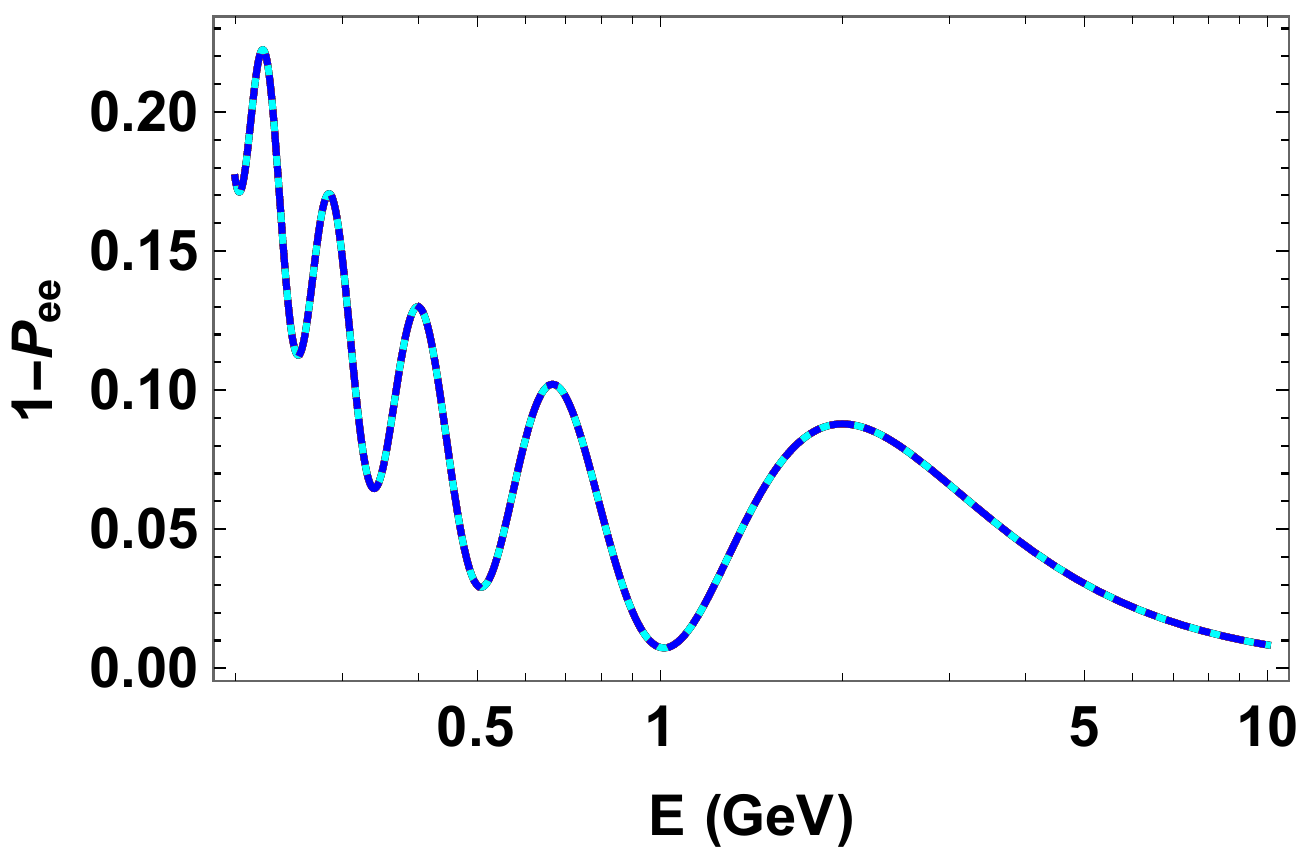}
		\includegraphics[width=.32\textwidth]{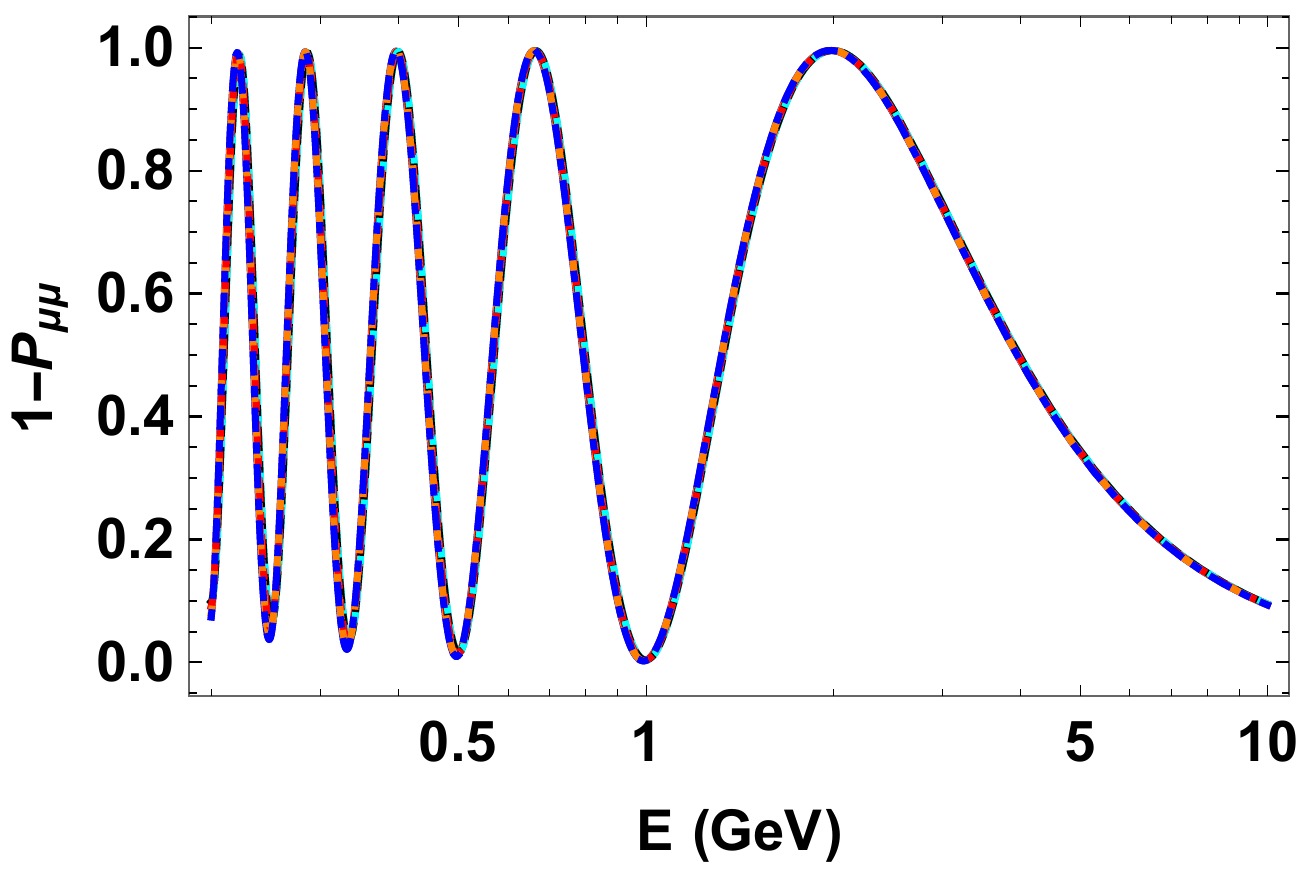}
		\includegraphics[width=.32\textwidth]{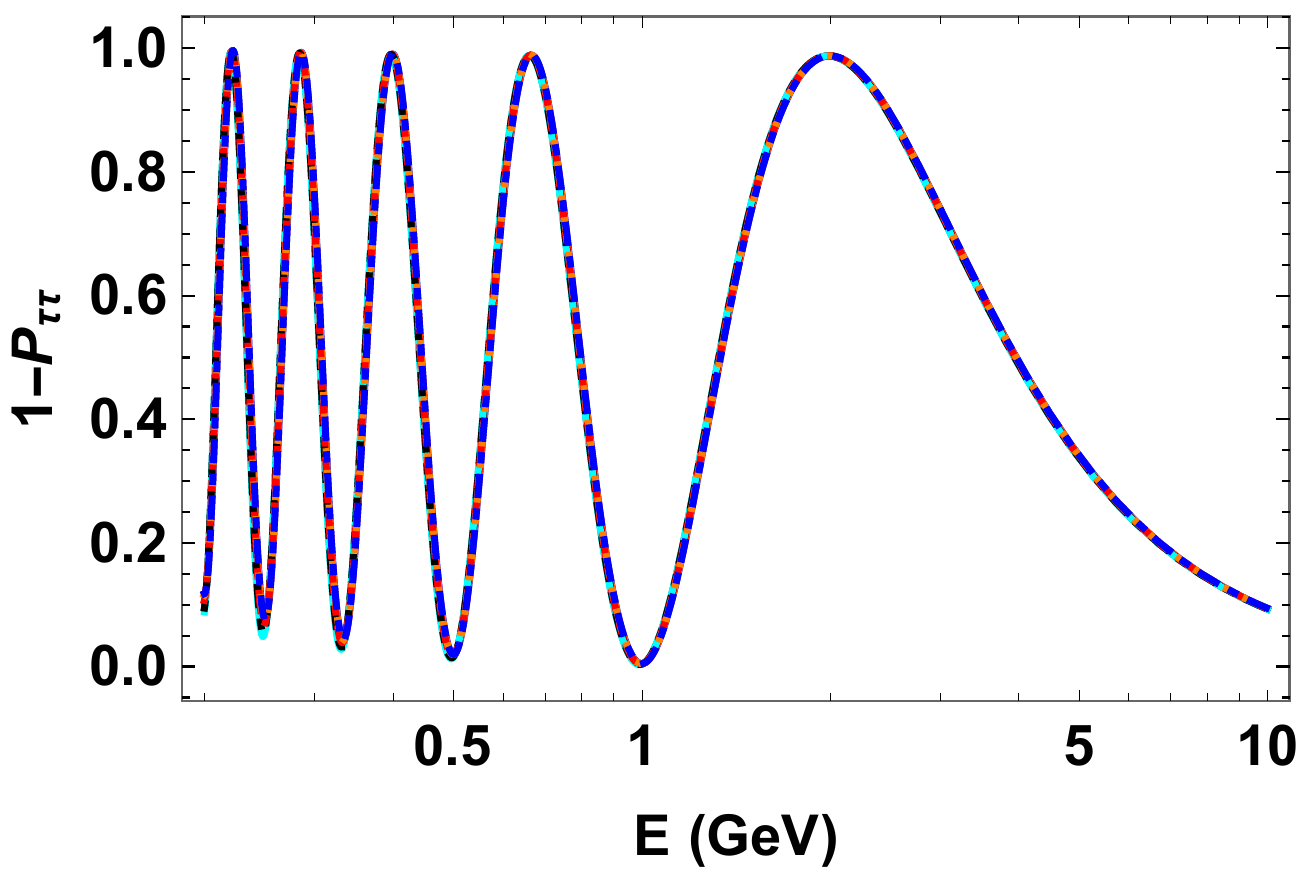}\\
		\includegraphics[width=.32\textwidth]{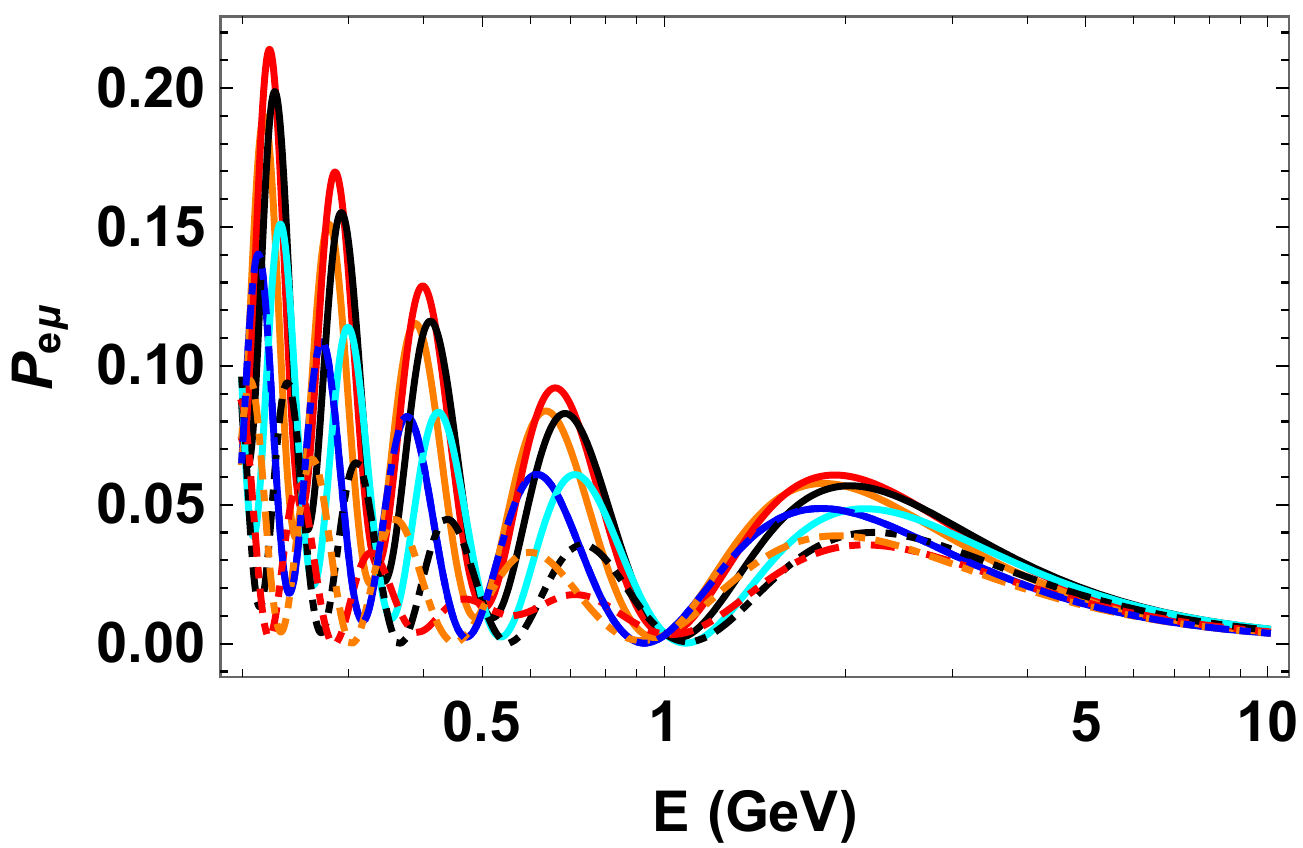}
		\includegraphics[width=.32\textwidth]{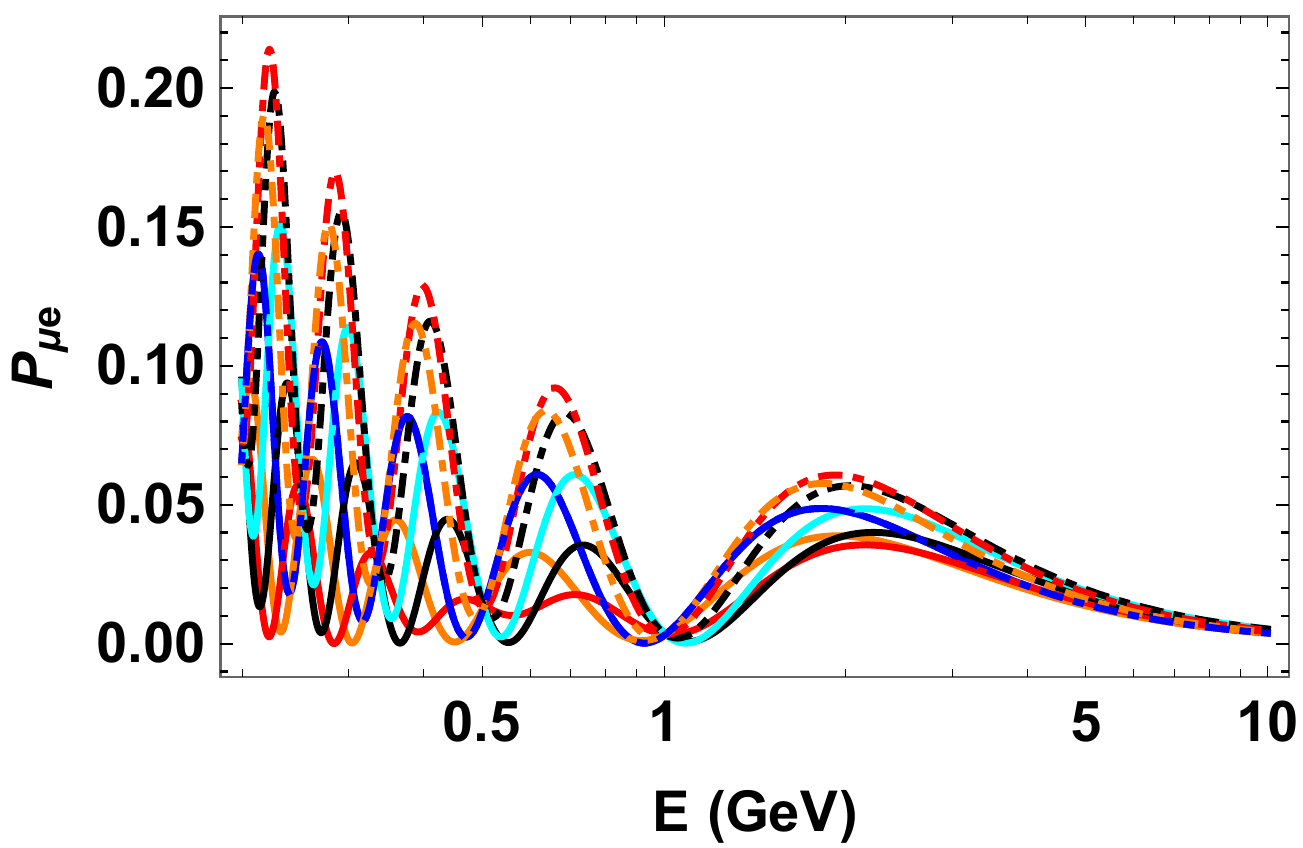}
		\includegraphics[width=.32\textwidth]{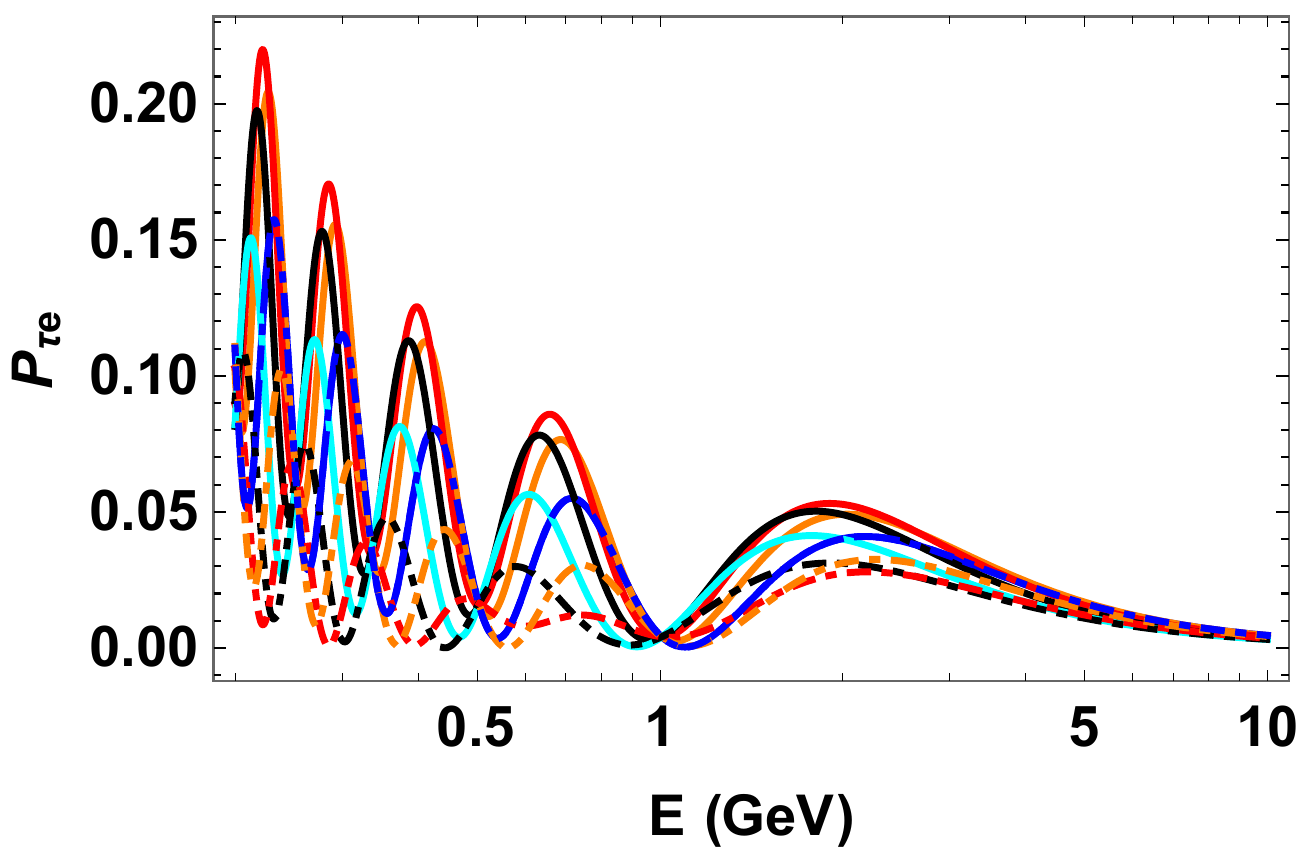}\\
		\includegraphics[width=.32\textwidth]{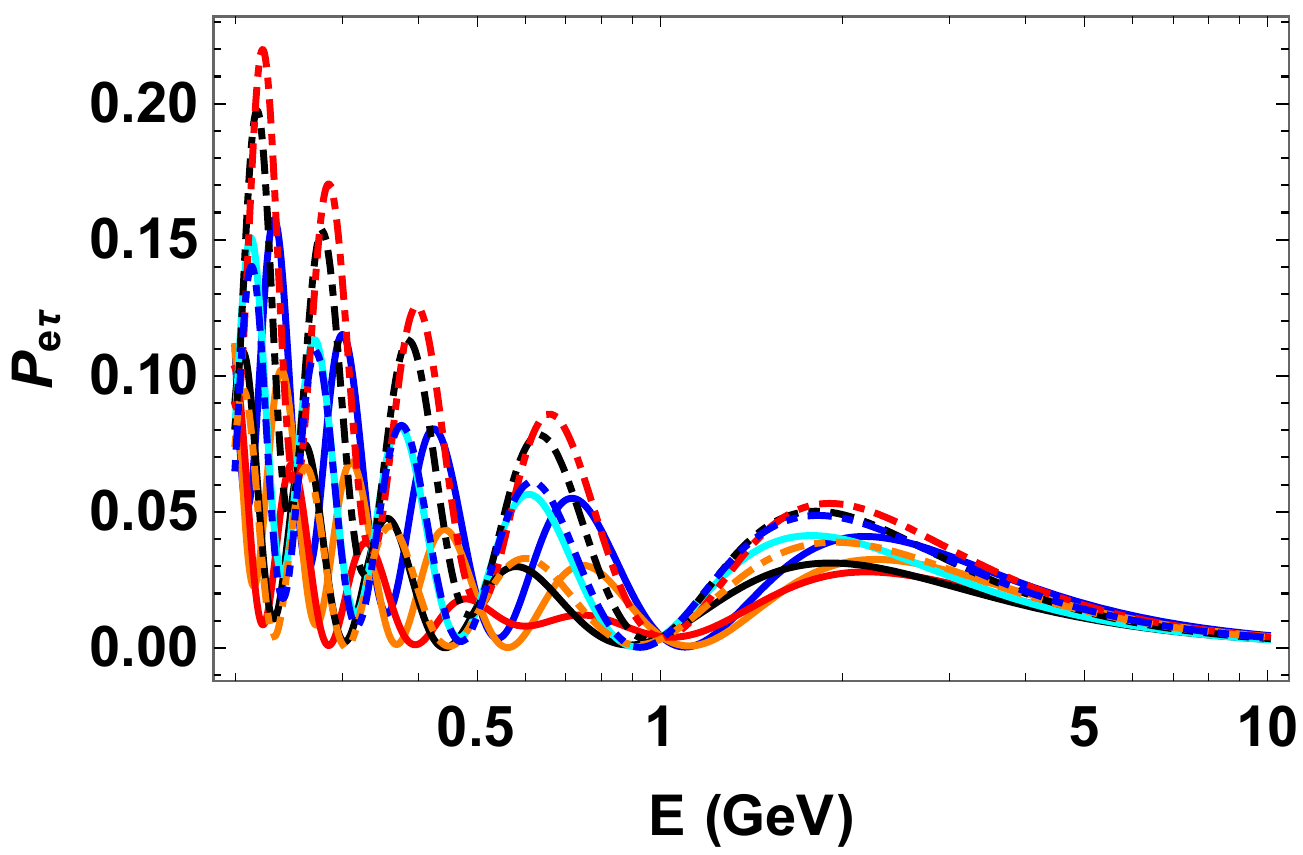}
		\includegraphics[width=.32\textwidth]{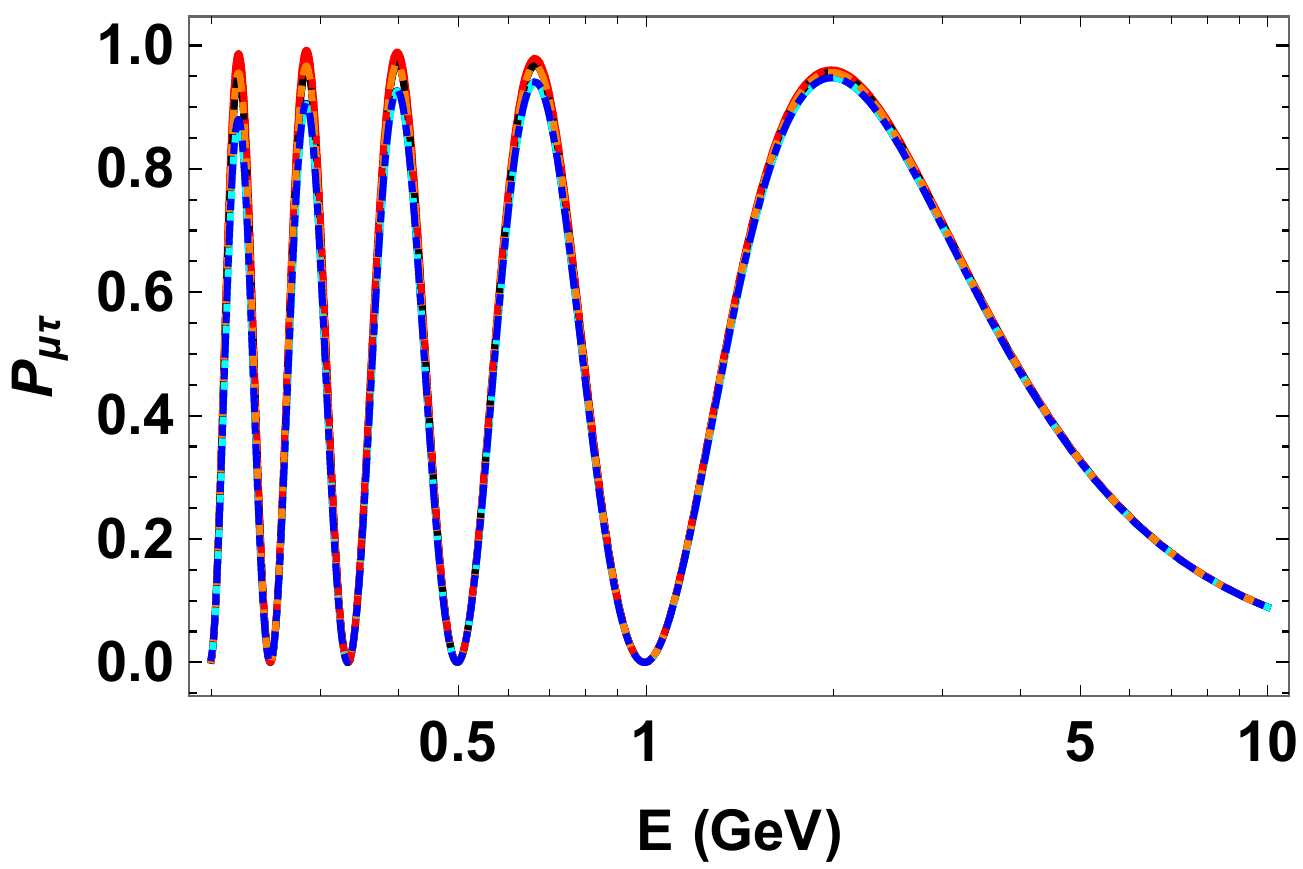}
		\includegraphics[width=.32\textwidth]{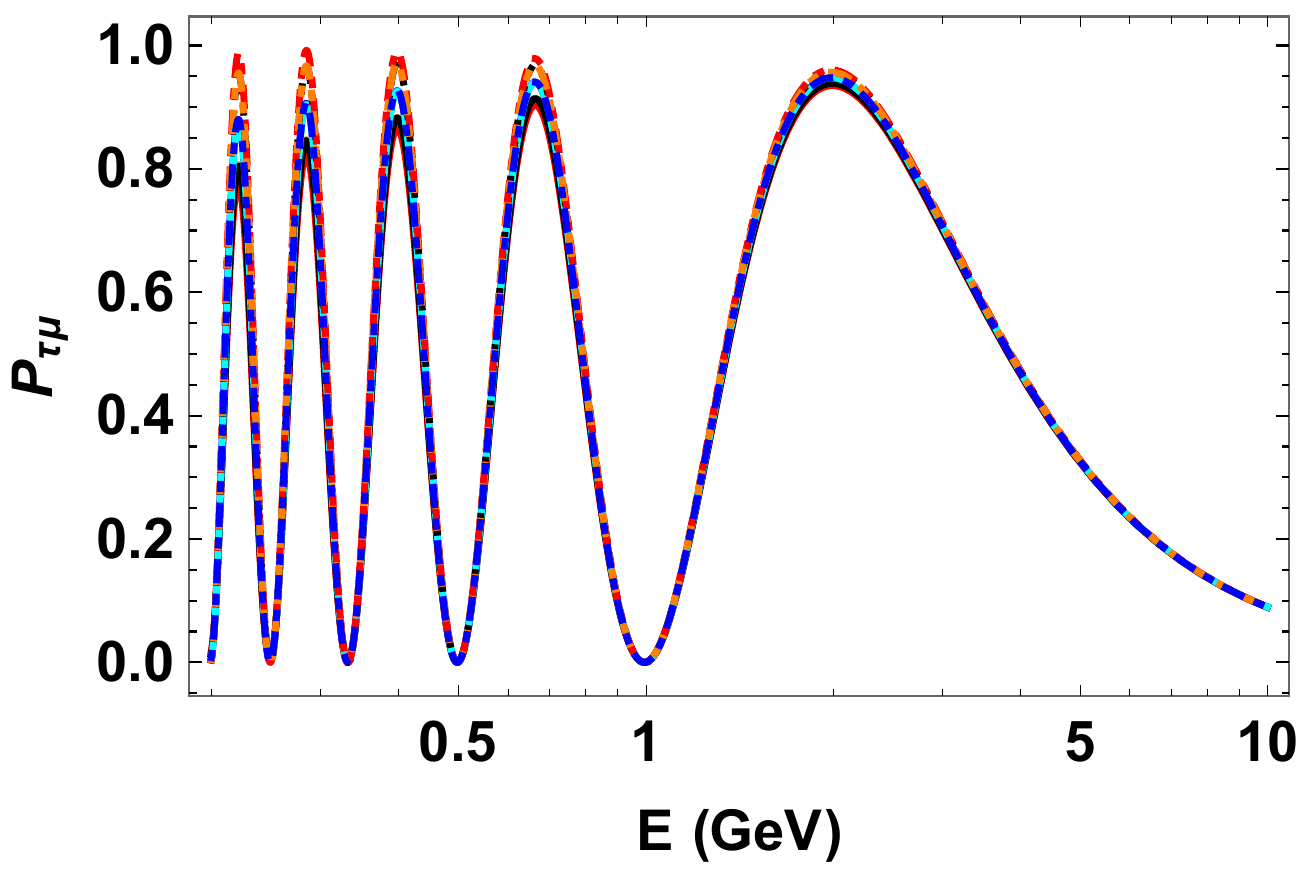}
	\end{tabular}
	\caption{(color online) Complexity (first row), 1-$P_{\alpha \alpha}$ (second row) and various transition probabilities (third and fourth rows) with respect to the neutrino-energy $E$ in case of initial flavor $\nu_e$ (left), $\nu_{\mu}$ (middle) and $\nu_{\tau}$ (right) for different values of the $CP$-violating phase $\delta$ depicted by different colors. Here, we have considered $L = 1000$ km. All other parameters are set at their best-fit values.}
	\label{Cost_E_Exp}
\end{figure*}

In Fig.~\ref{Chi_L} we have plotted the complexity $\chi_\alpha$ with respect to $L/E$ ratio, where $L$ and $E$ are the distance traveled by neutrinos in vacuum and energy of neutrino, respectively, keeping $\delta = 0^o$ in case of initial flavor $\nu_e$ (blue solid line), $\nu_{\mu}$ (red dashed line) and $\nu_{\tau}$ (green dot-dashed line). The left panel shows the general case of neutrino evolution whereas the right panel represents the scenario that is experimentally reliable, as the $L/E$ ratio corresponds to the current and planned long baseline experimental facilities. The rapid oscillation pattern seen in the left panel (zoomed-in in the right panel) is due to $\Delta m_{31}^2$ mass-squared difference in the oscillation phase, while the longer oscillation pattern is due to $\Delta m_{21}^2$ in the oscillation phase. The oscillation length is $\sim 10^3$~km at $E=1$~GeV for $\Delta m_{31}^2$ and $\sim 3\times 10^4$~km at $E=1$~GeV for $\Delta m_{21}^2$. In the general case (left panel), we can see that the complexity is maximum if the neutrino is produced initially as $\nu_e$, however, this happens only at a very large $L/E$ value of $\sim 1.6\times 10^4$~km/GeV. While, in current experimental setups (right panel), which covers roughly one oscillation length for $\Delta m_{31}^2$, the initial $\nu_e$ flavor provides the least complexity among all neutrino flavors. 

Next, in Fig.~\ref{Cost_L_Exp}, we have plotted the complexity $\chi_\alpha$ (upper two panels) and the total oscillation probability for a given flavor $\nu_\alpha$ to other flavors {\it i.e.,} $1-P_{\alpha \alpha}$ (bottom panels) with respect to the $L/E$ ratio for different values of $\delta$. We can see here that the complexity mimics the features of the total oscillation probability 
$1-P_{\alpha \alpha}$.
However, it is visible that $\chi_\alpha$ for all three flavors provide more information regarding the $CP$-violating phase $\delta$. For the large $L/E$ range (top panels) the complexities are maximized and the corresponding $\delta = +90^o$ or $-90^o$ for $\chi_\mu$ and $\chi_\tau$, and at $\delta = \pm 90^o$ for $\chi_e$. Note that the CP is maximally violated at approximately these $\delta$ values.\footnote{For $\chi_\mu$ and $\chi_\tau$ the CP-violation is maximum for $\delta \approx \pm 95^o$ because of the cross-terms in the Krylov states.}
In the limited $L/E$ range (middle panels) $\chi_\mu$ and $\chi_\tau$ are maximized at $\delta = -90^o$ (red-dashed line) and at $\delta = +90^o$ (red-solid line), respectively, where CP is maximally violated. However, $\chi_e$ is maximized at $\delta = +135^o$ and at $-45^o$. The reason is that the complexity is rather low for $\chi_e$ in the low $L/E$ range, as discussed before, and cannot probe the $\delta = \pm90^o$ value for which $\chi_e$ is maximized (upper left panel).  

\begin{figure*}[t] 
	\centering
	\begin{tabular}{cc}
		\includegraphics[width=.32\textwidth]{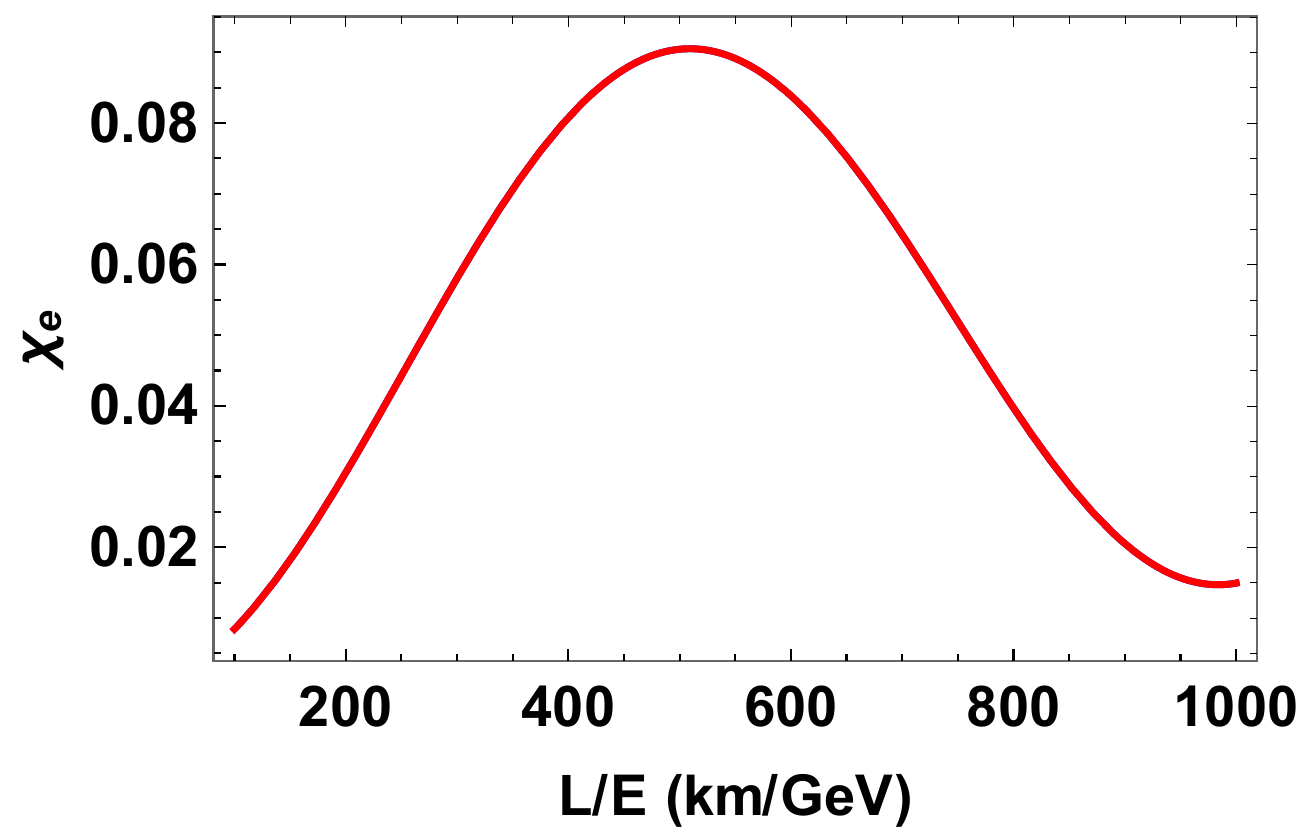}
		\includegraphics[width=.32\textwidth]{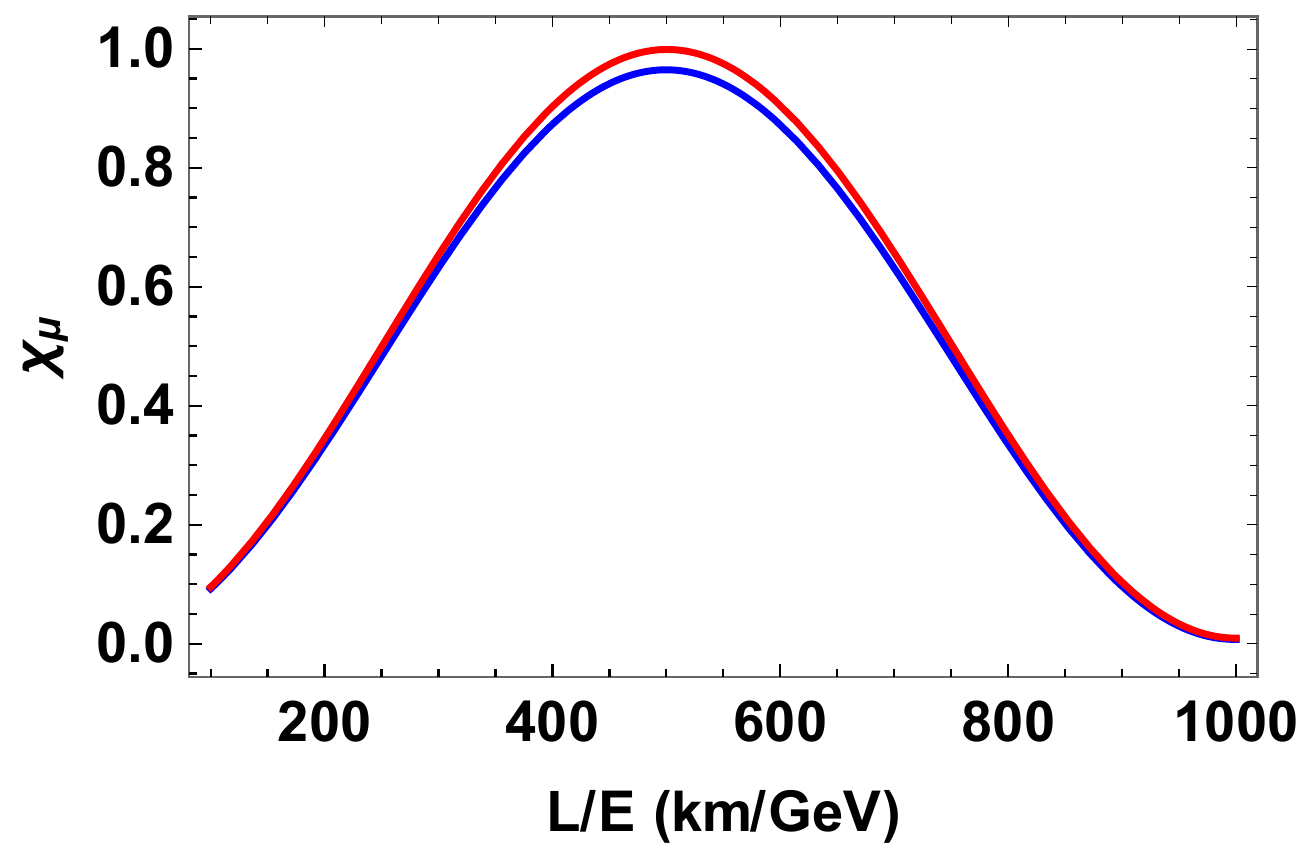}
		\includegraphics[width=.32\textwidth]{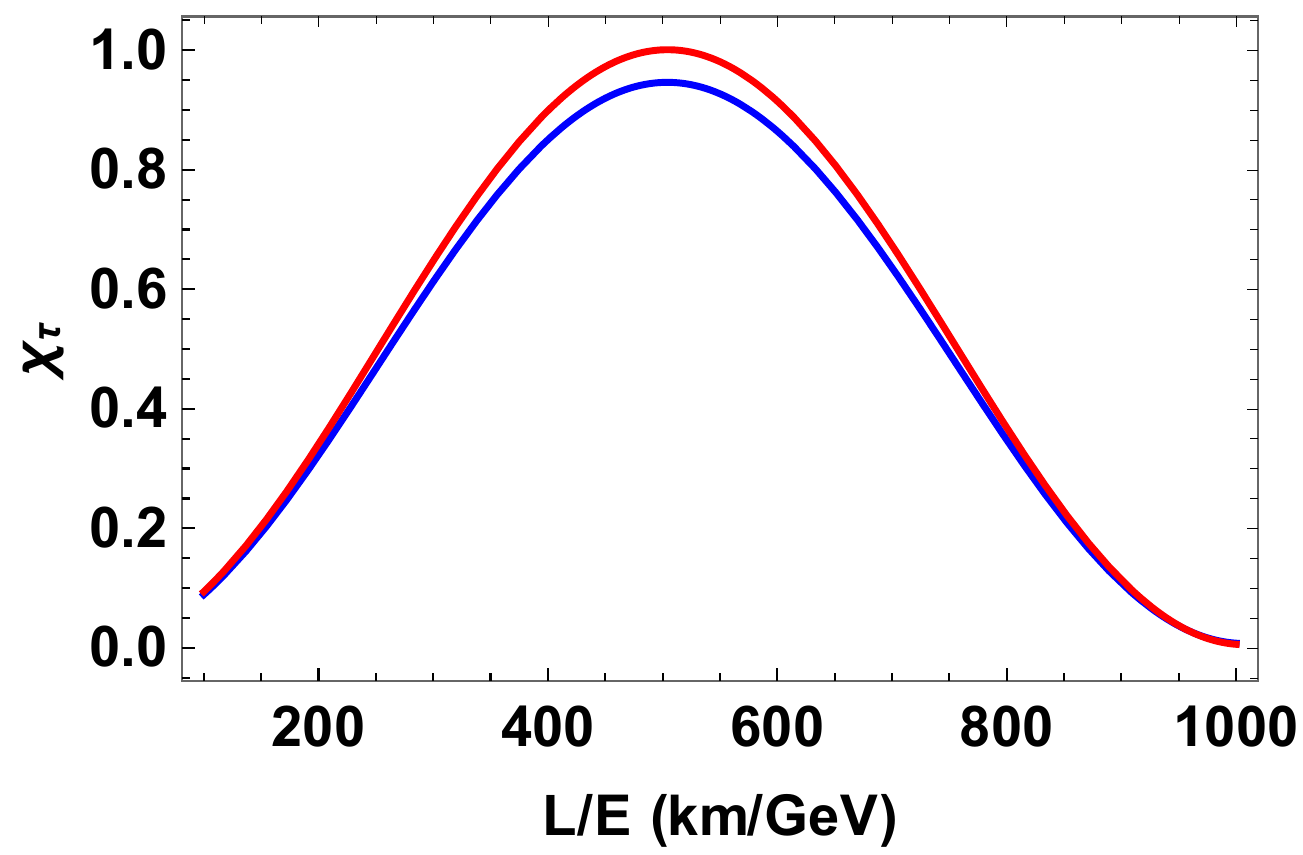}\\
		\includegraphics[width=.32\textwidth]{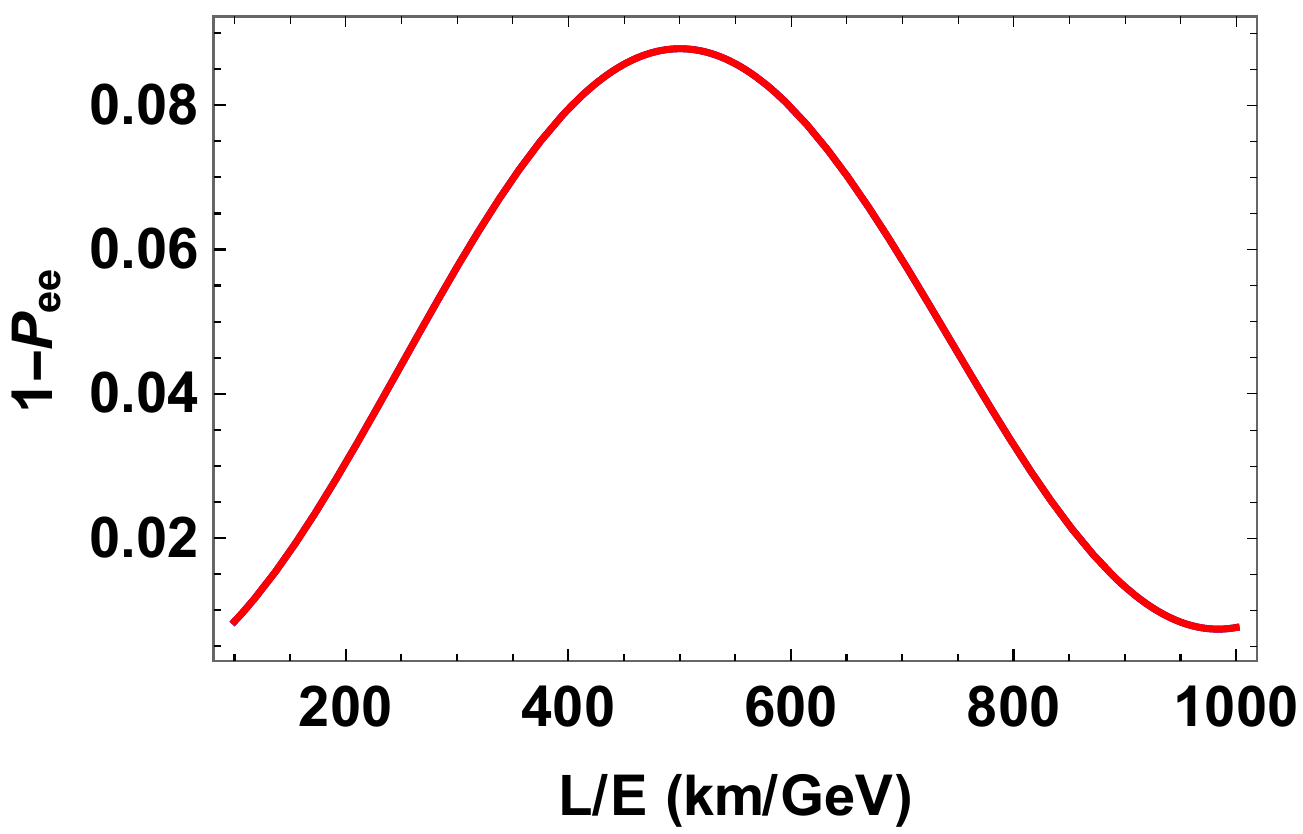}
		\includegraphics[width=.32\textwidth]{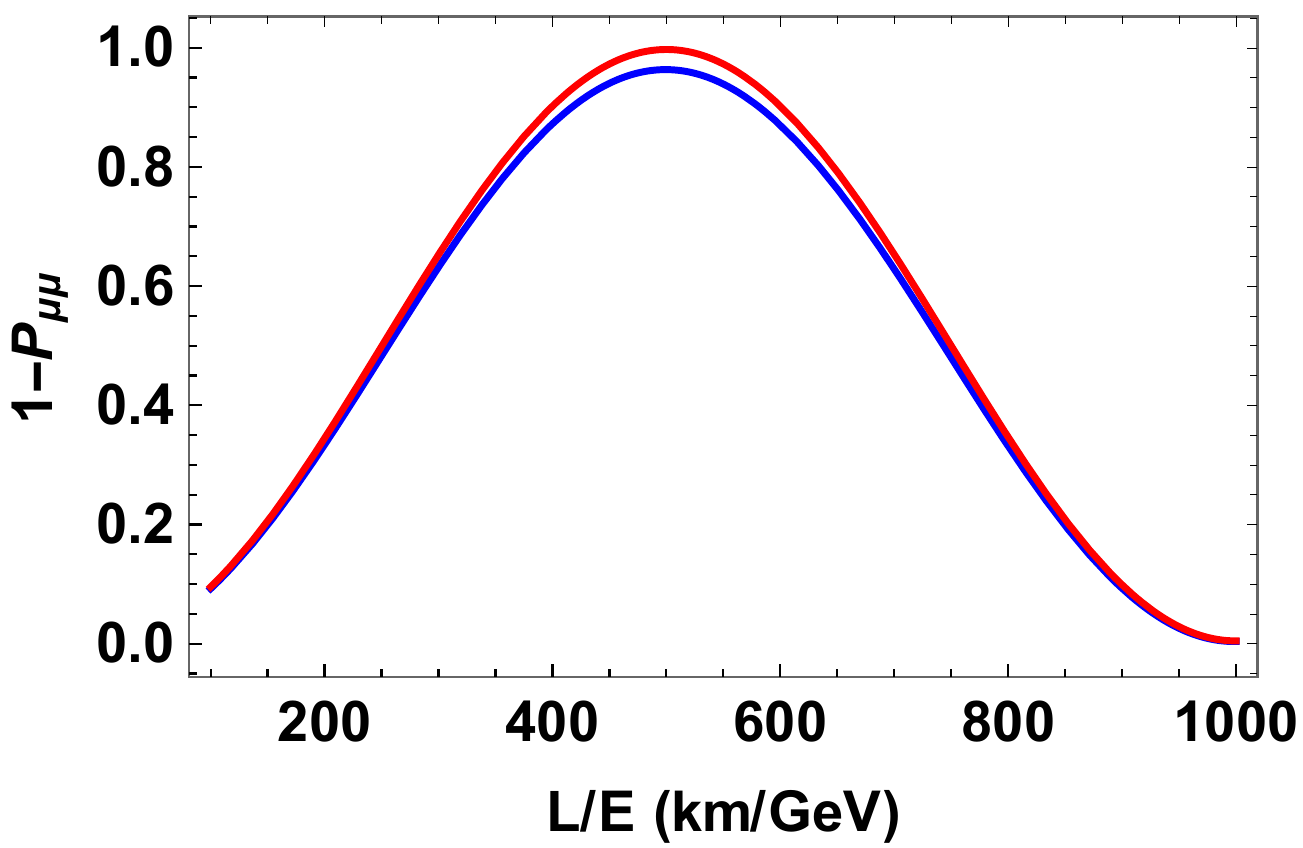}
		\includegraphics[width=.32\textwidth]{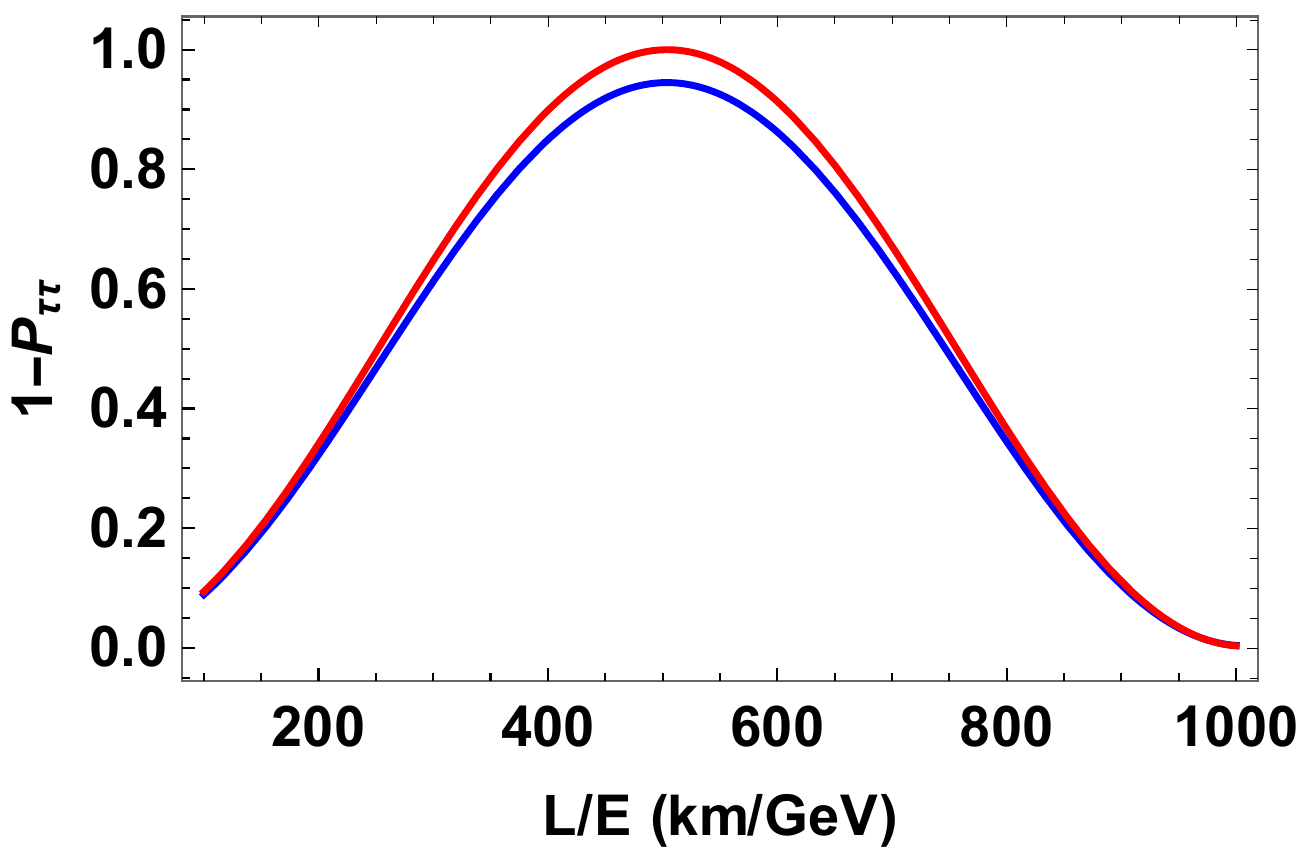}
	\end{tabular}
	\caption{(color online) Complexity (upper panel) and 1-$P_{\alpha \alpha}$ (lower panel) with respect to the $L/E$ ratio in case of initial flavor $\nu_e$ (left), $\nu_{\mu}$ (middle) and $\nu_{\tau}$ (right) where the effects of higher octant ($\theta_{23}=51.295^o$) and lower octant ($\theta_{23}=44.026^o$) of $\theta_{23}$ are represented by blue and red curves, respectively.}
	\label{Cost_theta23_Exp}
\end{figure*}

Dependence of complexity on $\delta$ can also be seen in the first and second rows of Fig.~\ref{Cost_E_Exp} where we have shown the variations of $\chi_\alpha$ and their corresponding total oscillation probabilities $1-P_{\alpha\alpha}$ with energy $E$ for a fixed baseline of $L=1000$~km. 
It is clear from these plots that the effect of $\delta$ is significantly distinguishable if the initial flavor is either $\nu_{\mu}$ or $\nu_{\tau}$. In the case of initial $\nu_e$, this effect of non-zero $\delta$ is again quite small. The non-zero $\delta$ value notably enhances the complexity of the system for $\nu_{\mu}$ and $\nu_{\tau}$ flavors and these are maximum for $\delta = -90^o$ and $\delta = 90^o$, respectively. As mentioned earlier, these are also the values for which CP is maximally violated.

In Fig.~\ref{Cost_E_Exp}, we have also compared the complexities with corresponding (individual) oscillation probabilities $P_{\alpha\beta}$. For example, $\chi_e$ can be compared with $P_{e\mu}$ and $P_{e\tau}$, $\chi_{\mu}$ can be compared with $P_{\mu e}$ and $P_{\mu\tau}$ and so on. It can be seen that the oscillation probabilities $P_{\alpha\beta}$ where $\alpha\neq \beta$, indicate specific values of $\delta$-phase to be maximum. Specifically, $P_{e\mu}$, $P_{\tau e}$ and $P_{\mu \tau}$ are maximum for $\delta=90^o$ whereas $P_{\mu e}$, $P_{e\tau}$ and $P_{\tau\mu}$ are maximum for $\delta=-90^o$. On the other hand, $\chi_{\mu}$, which is a combination of $P_{\mu e}$ and $P_{\mu \tau}$, is maximum at $\delta=-90^o$ showing more inclination towards $P_{\mu e}$. Similarly, $\chi_{\tau}$, which is a combination of $P_{\tau e}$ and $P_{\tau\mu}$ approaches its maximum value at $\delta=90^o$. The variation of $\chi_e$ with respect to $\delta$ is  
different than $P_{e\mu}$ and $P_{e\tau}$ as $\chi_e$ achieves its maximum value at both $\delta = 135^o$ and $-45^o$ for the adopted $L$ in these plots. However, this variation of $\chi_e$ with $\delta$ is very small. The oscillation maxima and minima for $P_{e\mu}$, $P_{e\tau}$, $P_{\mu e}$ and $P_{\tau e}$ also varies with $\delta$. This is because the CP phase $\delta$ gets added in the expressions for the oscillation phase. Therefore, depending on the sensitivity of an experiment to a certain energy range, measurements involving $\nu_e$ can result in higher probability for a certain value of $\delta$ other than $\pm 90^o$ where $\chi_e$ has the global maximum (see Fig.~\ref{Cost_L_Exp}). 

We have also analyzed the effects of the octant of $\theta_{23}$ on complexity. In Fig.~\ref{Cost_theta23_Exp} we plot $\chi_\alpha$ (upper panels) and their corresponding $1-P_{\alpha\alpha}$ (lower panels) with respect to the $L/E$ ratio. In this figure, blue and red curves represent the case of upper ($\theta_{23} = 51.295^o$) and lower ($\theta_{23} = 44.026^o$) octants of $\theta_{23}$, respectively. The $\theta_{23}$-values we considered here are the extreme points associated with 3$\sigma$ allowed range. It can be seen that for $\chi_e$ there is no sensitivity for the $\theta_{23}$ octant, however, the complexities associated to $\nu_{\mu}$ and $\nu_{\tau}$ flavors can distinguish between blue and red curves, $i.e.,$ $\chi_{\mu}$ and $\chi_{\tau}$ show some sensitivity to the octant of $\theta_{23}$. However, this feature of complexities is almost similar to that of $1-P_{\alpha \alpha}$.
Therefore, complexity does not provide additional information for the parameter $\theta_{23}$.
\begin{figure*}[t] 
	\centering
	\begin{tabular}{cc}
		\includegraphics[width=.32\textwidth]{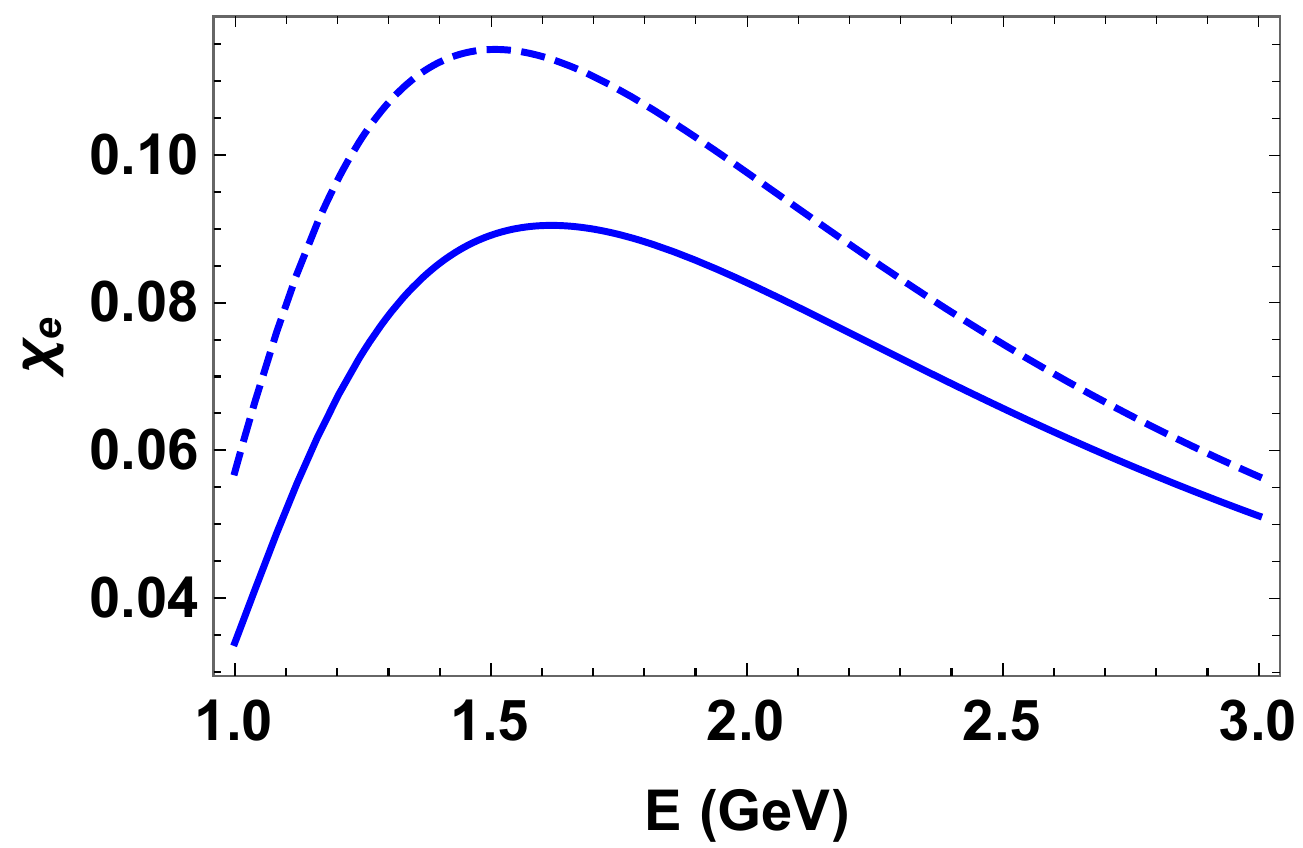}
		\includegraphics[width=.32\textwidth]{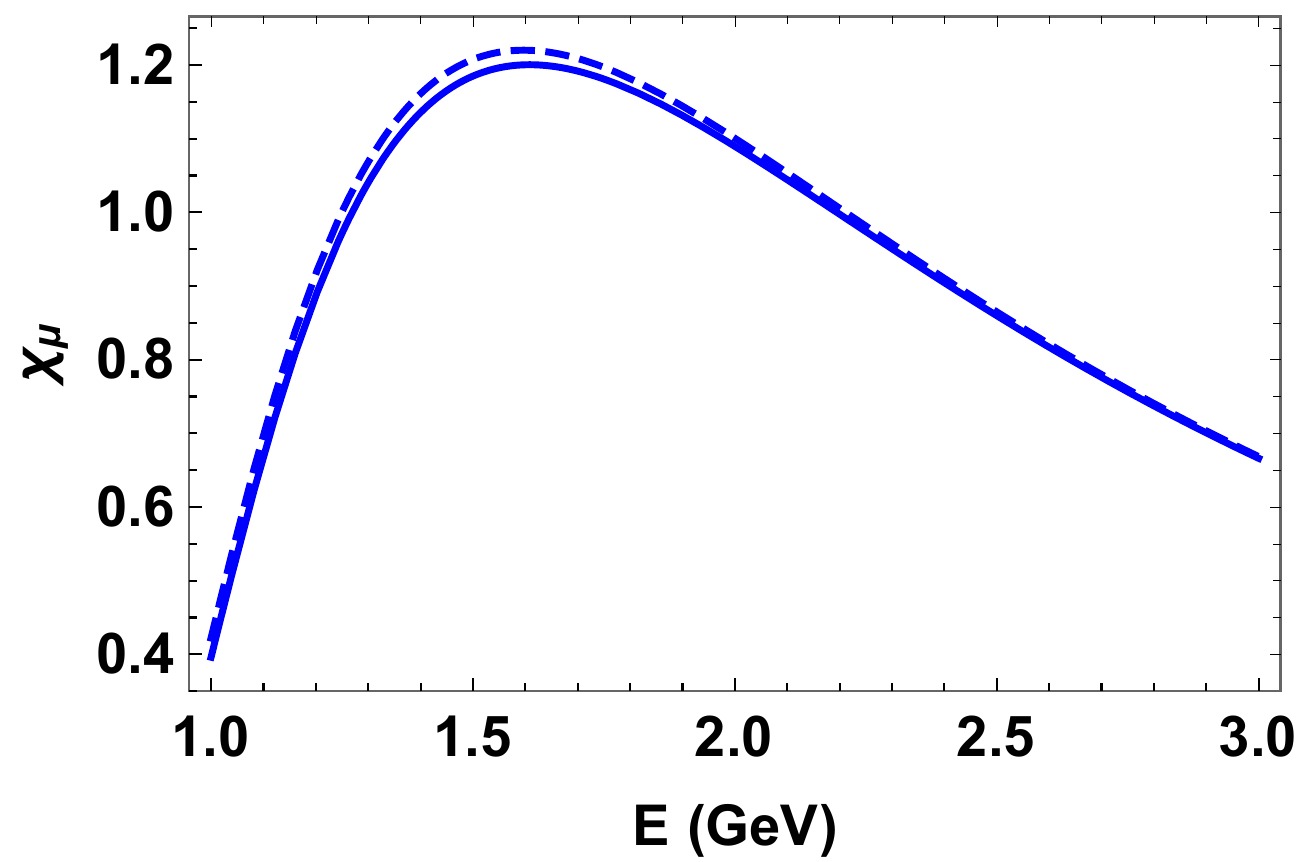}
		\includegraphics[width=.32\textwidth]{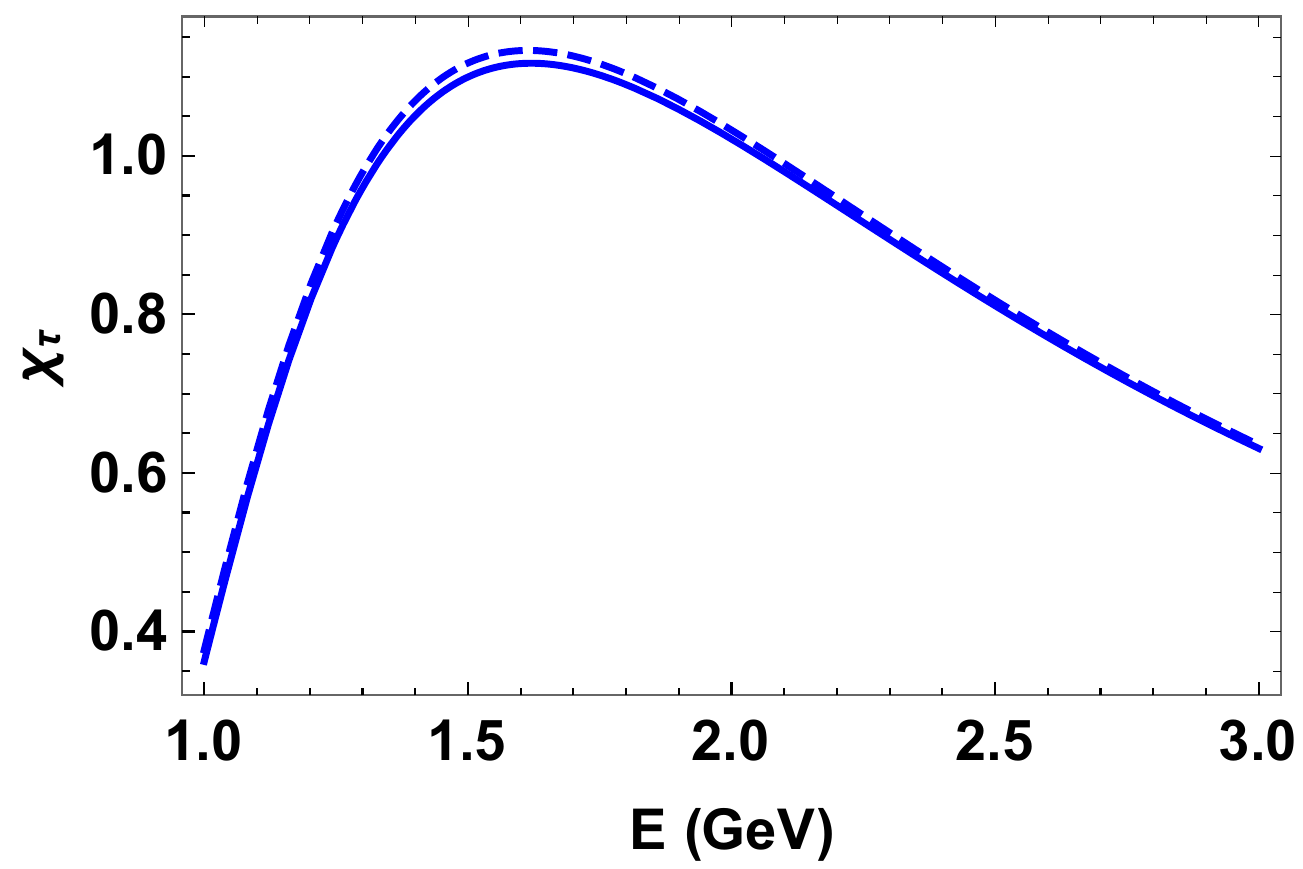}
	\end{tabular}
	\caption{(color online) Complexity $\chi_e$ (left), $\chi_{\mu}$ (middle) and $\chi_{\tau}$ (right) w. r. t. neutrino-energy $E$ is shown. Here, $L=810$ km, $\delta=-90^o$ and matter potential $V=1.01 \times 10^{-13}$ eV have been considered. Solid and dashed curves represent the case of vacuum and matter oscillations, respectively.}
	\label{Cost_matter}
\end{figure*}

\subsection{Complexity estimates for specific experiments}
The currently operating two long baseline neutrino oscillation experiments, T2K in Japan \cite{T2K} and NOvA in the USA \cite{article}, are poised to measure oscillation parameters such as $\delta$, $\theta_{23}$ and the mass hierarchy, {\it i.e.,} the sign of the mass-squared difference $|\Delta m_{31}^2|$. T2K has a baseline of $L=295$ km while that of NOvA is $L=810$ km. Muon neutrinos are produced in these experiments through charged pion decays. The flux of these neutrinos peaks at approximately 0.6~GeV and 1.8~GeV, respectively, for T2K and NOvA. Latest results from T2K hint a measurement of the CP-violating phase $\delta = -2.14^{+0.90}_{-0.69}$ radians and a preference for normal hierarchy~\cite{T2K:2021xwb}. The NOvA experiment in its latest analysis~\cite{NOvA:2021nfi}, however, rejects the T2K best-fit value of $\delta$ by more than $2\sigma$ confidence and prefers instead $\delta = 0.82^{+0.27}_{-0.87}~\pi$, again with a preference for normal hierarchy. See, e.g., reference~\cite{Rahaman:2022rfp} for a review of this tension between the T2K and NOvA results and plausible solutions.

In this subsection, we explore complexity in the context of the T2K and NOvA experiments, and sensitivity of complexity on the oscillation parameters, especially the CP phase $\delta$. Note that the matter effect discussed in Sec.~4.3 is important for the NOvA experiment, where neutrinos propagate through the crust of the Earth over a distance of 810~km  
from their production point to the detector. Matter effects can be considered negligible for T2K due to its shorter baseline and lower energy range of neutrinos. In Fig.~\ref{Cost_matter} we plot the complexities $\chi_e$ (left panel), $\chi_{\mu}$ (middle panel) and $\chi_{\tau}$ (right panel) calculated without (solid lines) and with (dashed lines) matter effect with respect to the neutrino-energy $E$ for the NOvA baseline. 
The matter potential, in this case, is $V=1.01\times 10^{-13}$ eV for an average density of 2.8~g/cm$^3$.
It is clear that the matter effect increases complexity of the system in all cases of initial flavors of the neutrino, but most significantly for $\nu_e$ as expected.

\begin{figure*}[t] 
	\centering
	\begin{tabular}{cc}
		\includegraphics[width=.32\textwidth]{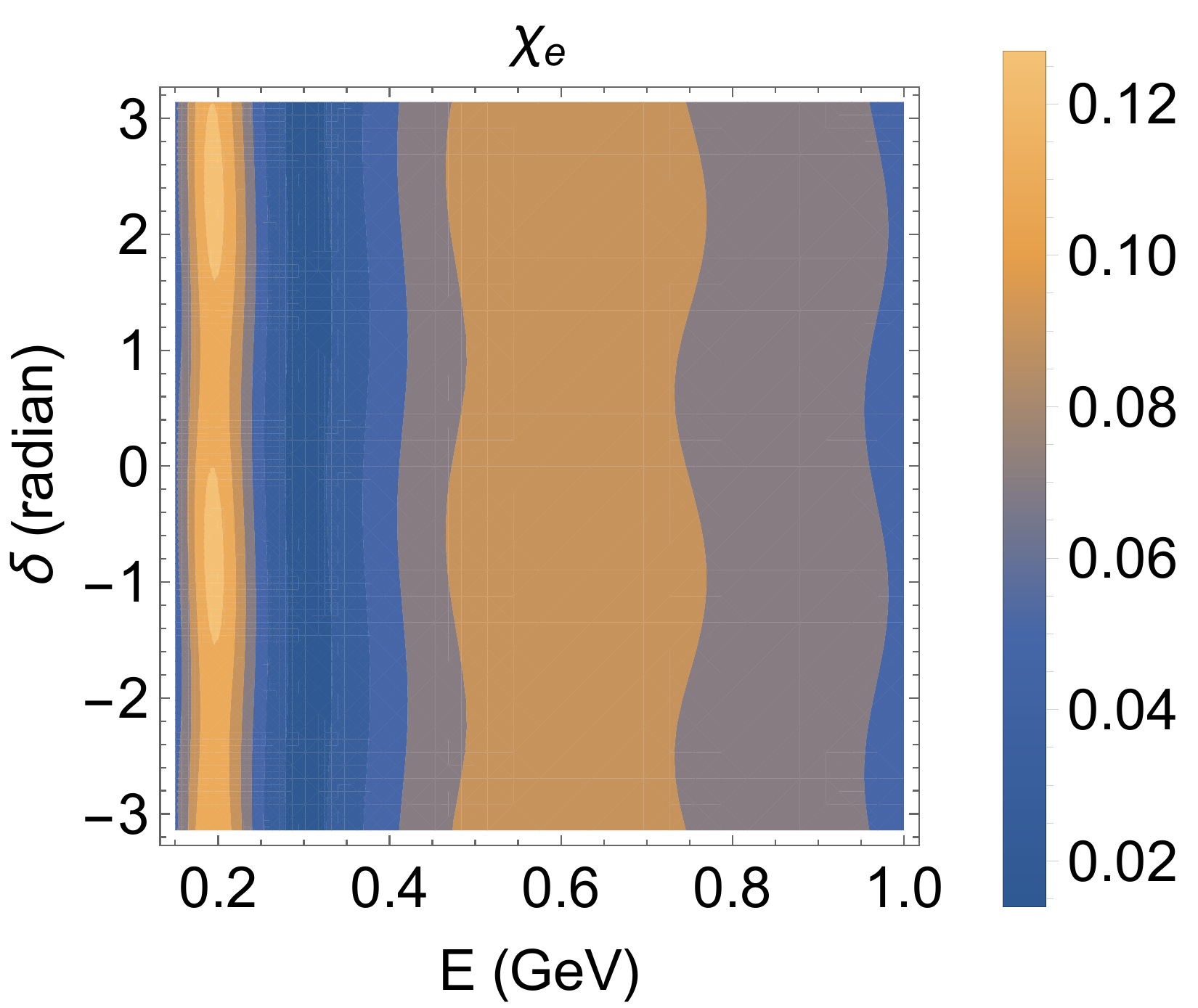}
		
		\includegraphics[width=.32\textwidth]{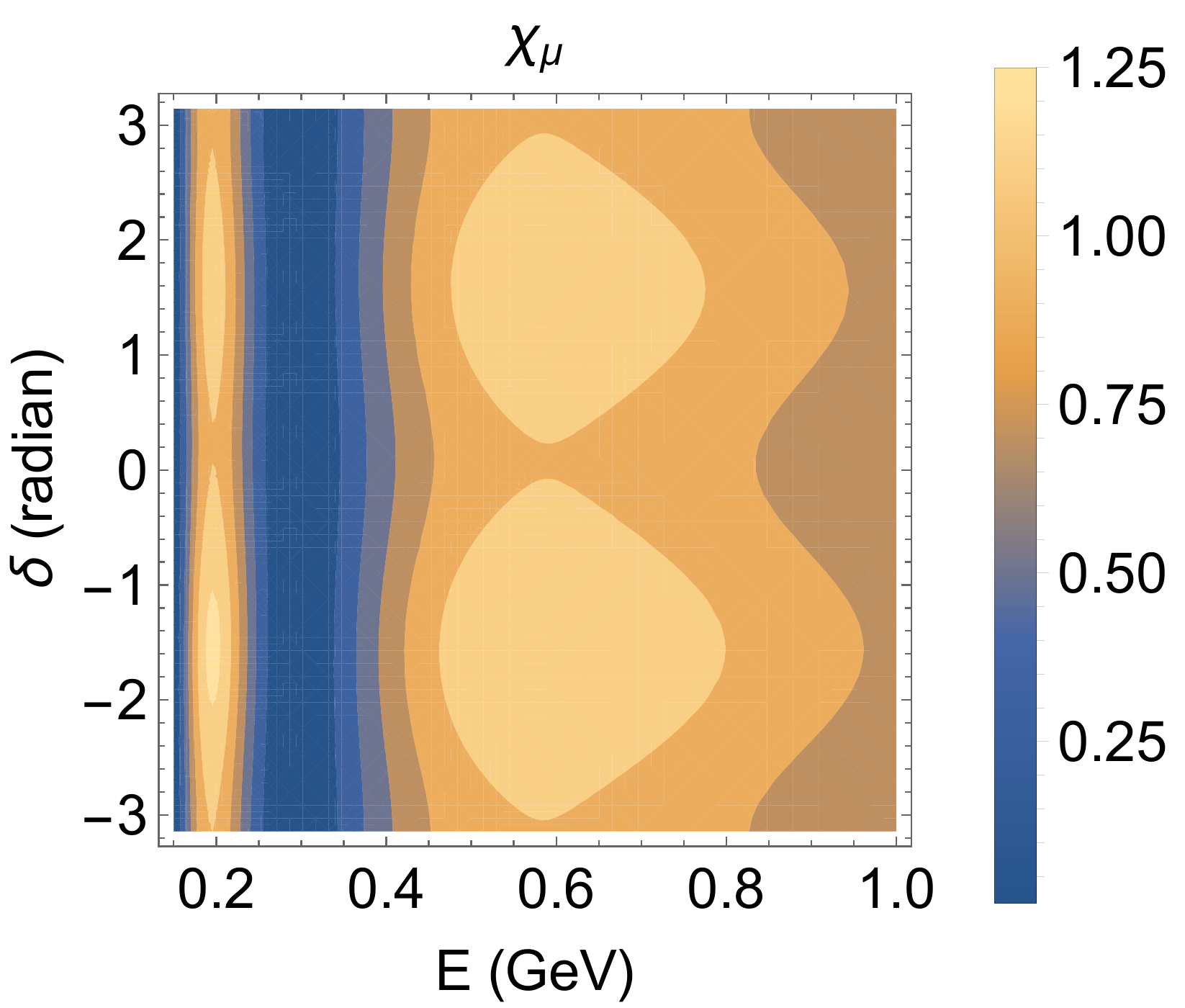}
		
		\includegraphics[width=.32\textwidth]{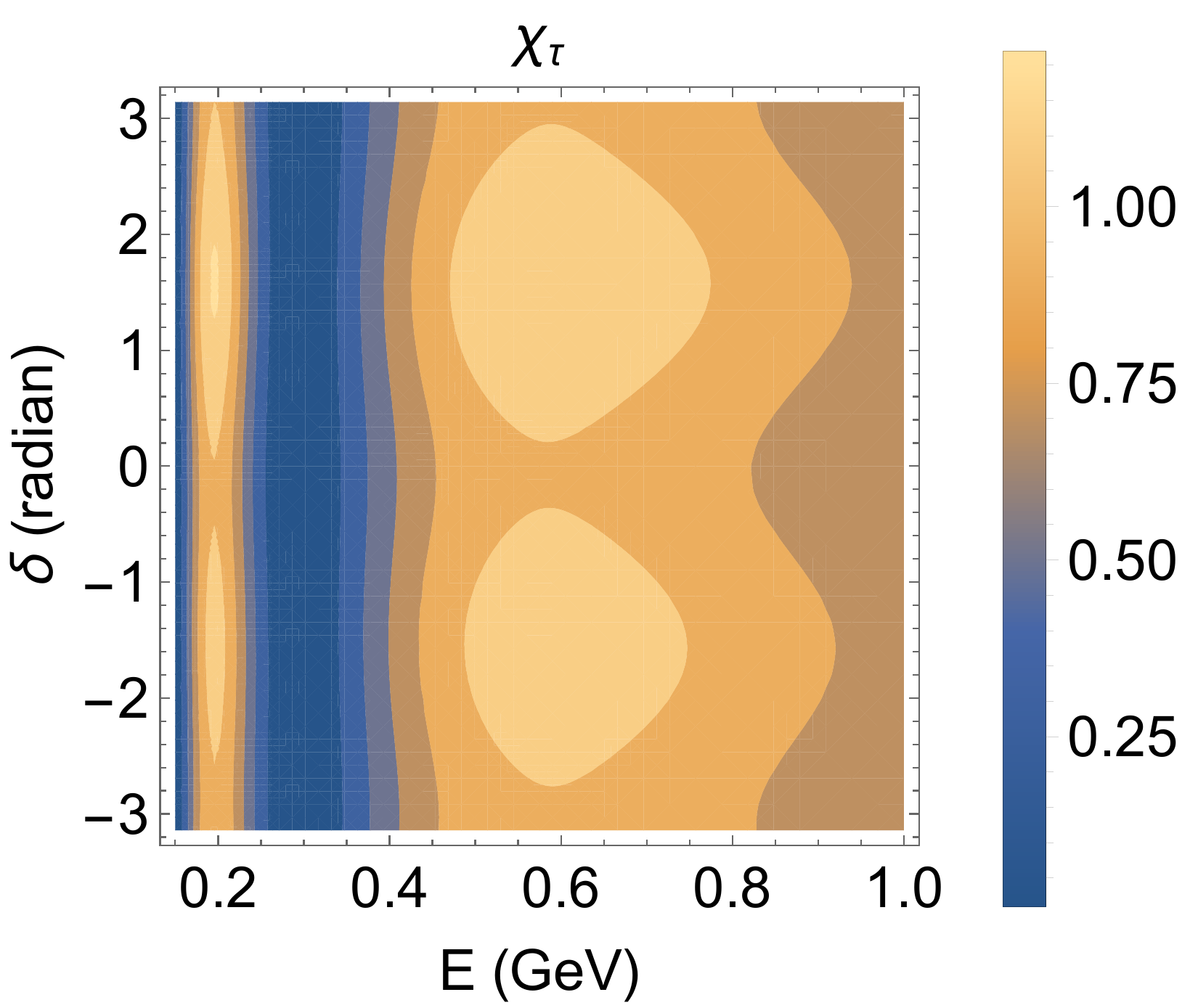}\\
		\includegraphics[width=.32\textwidth]{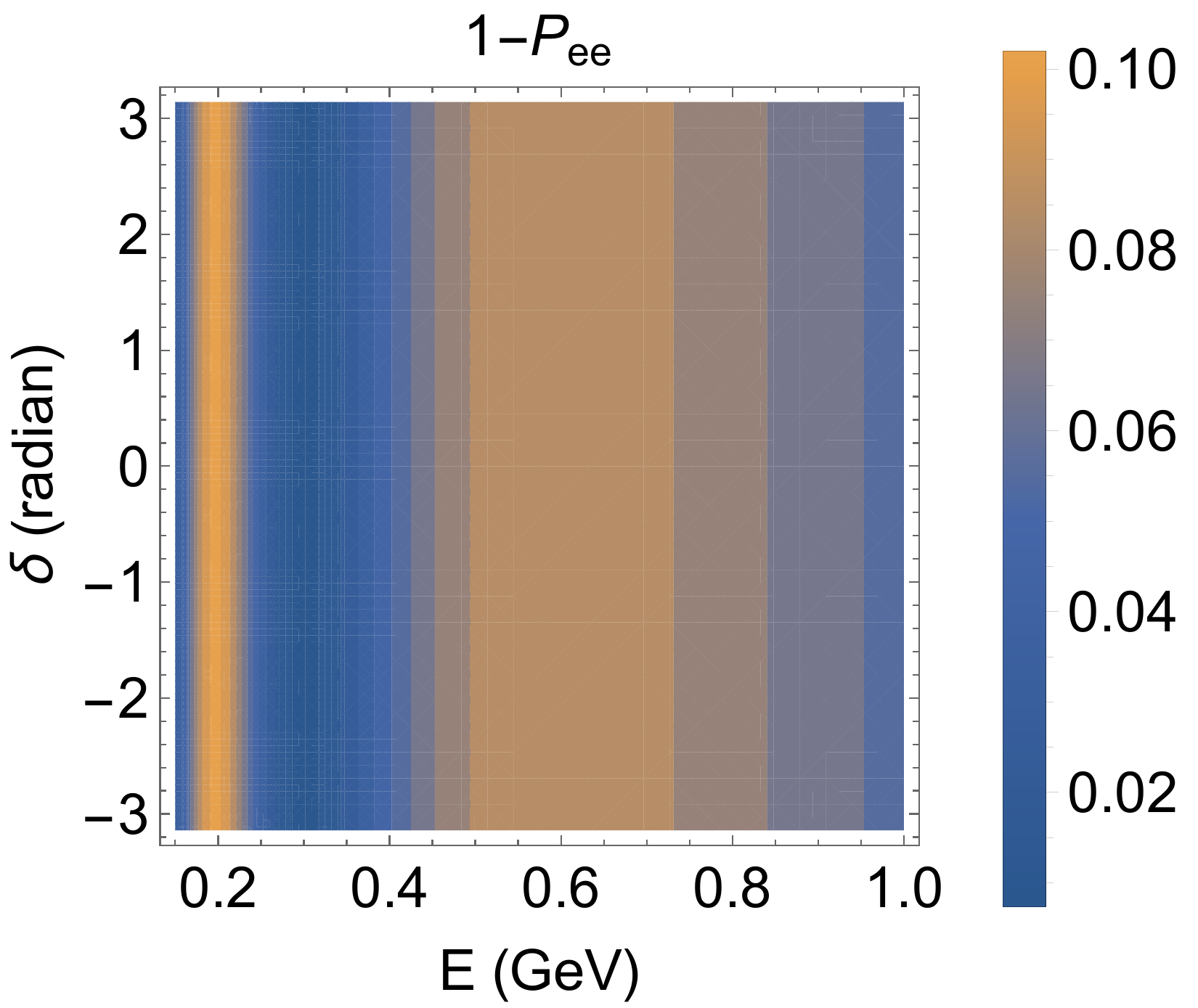}
		
		\includegraphics[width=.32\textwidth]{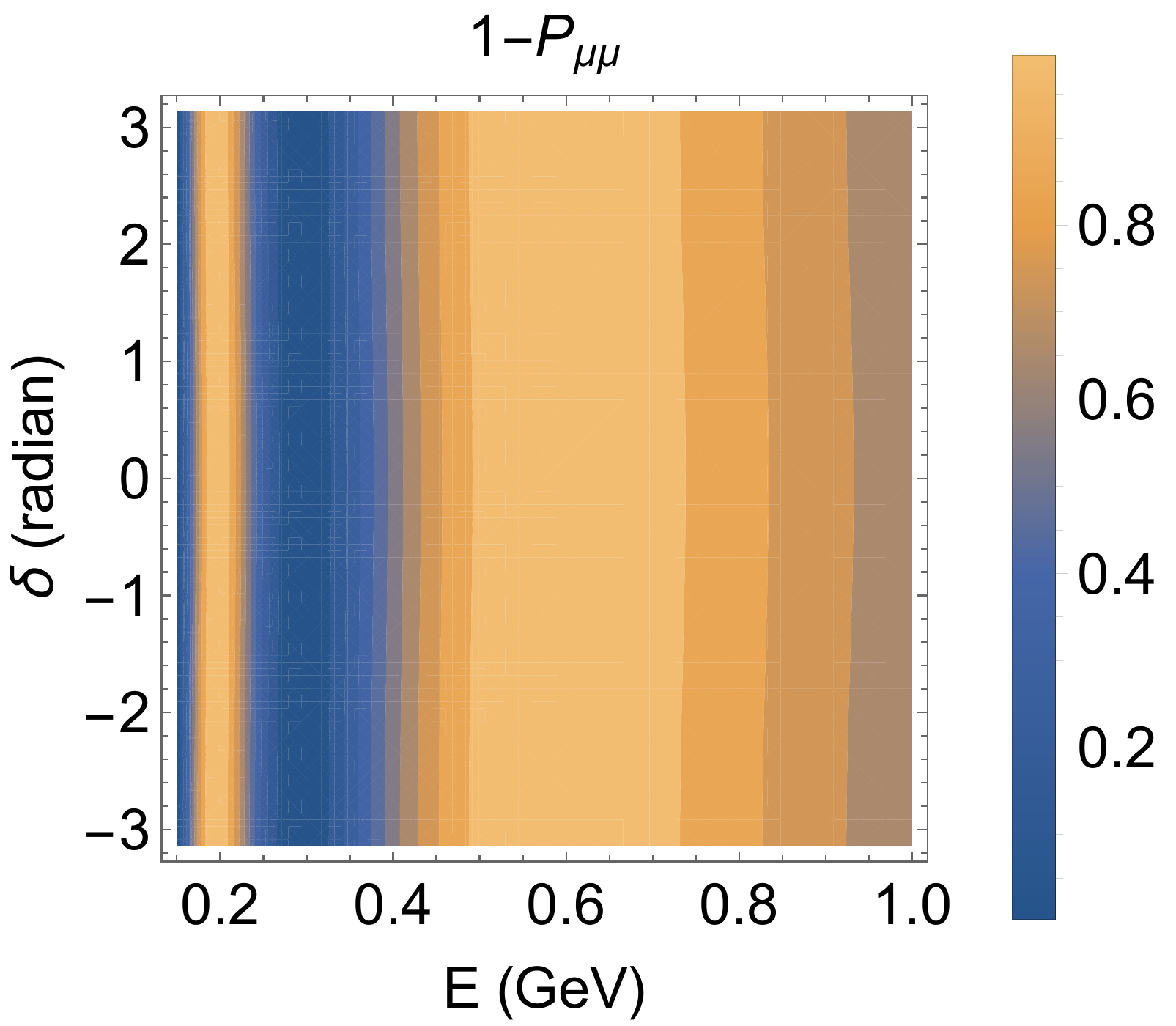}
		
		\includegraphics[width=.32\textwidth]{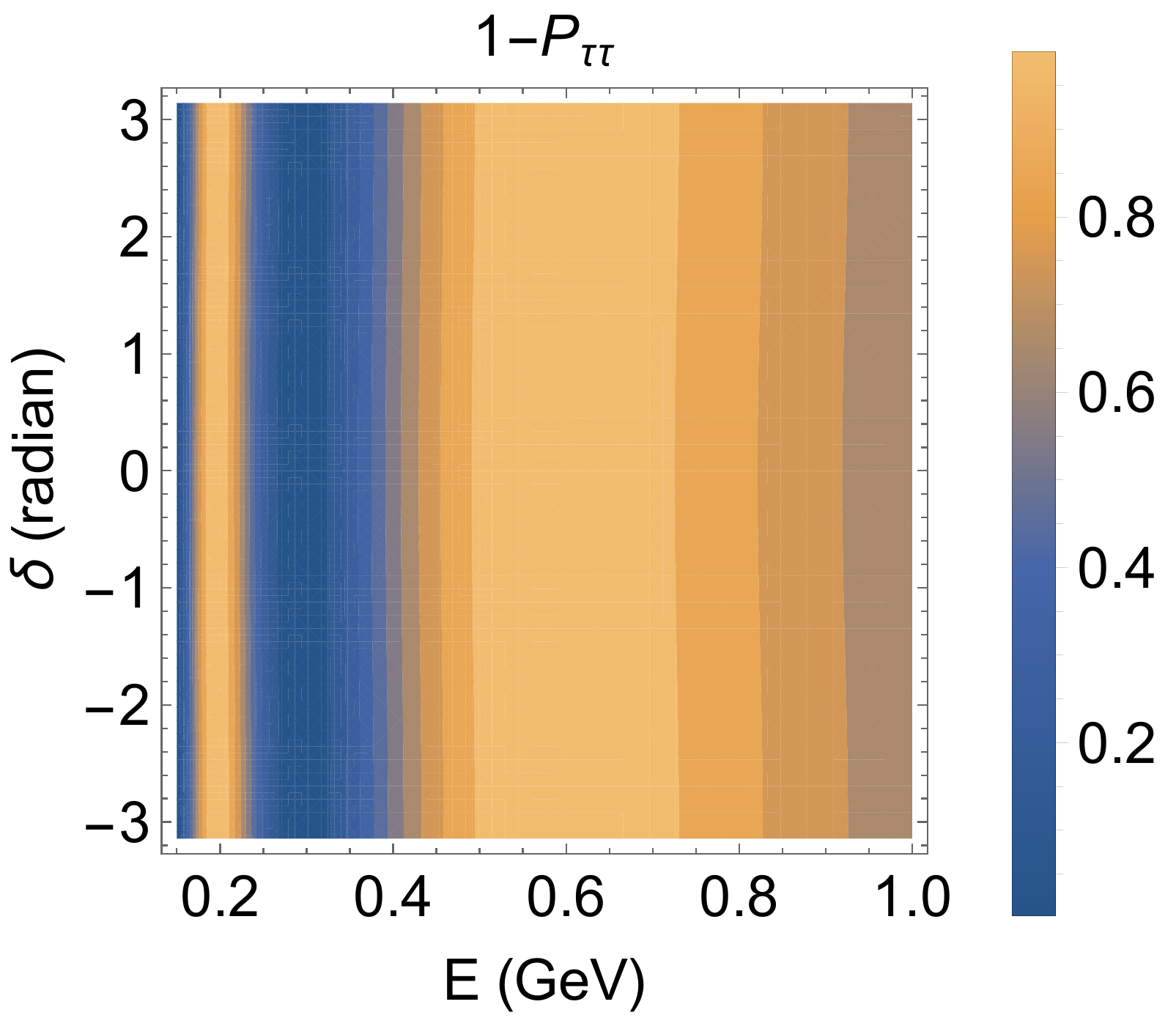}\\
		\includegraphics[width=.32\textwidth]{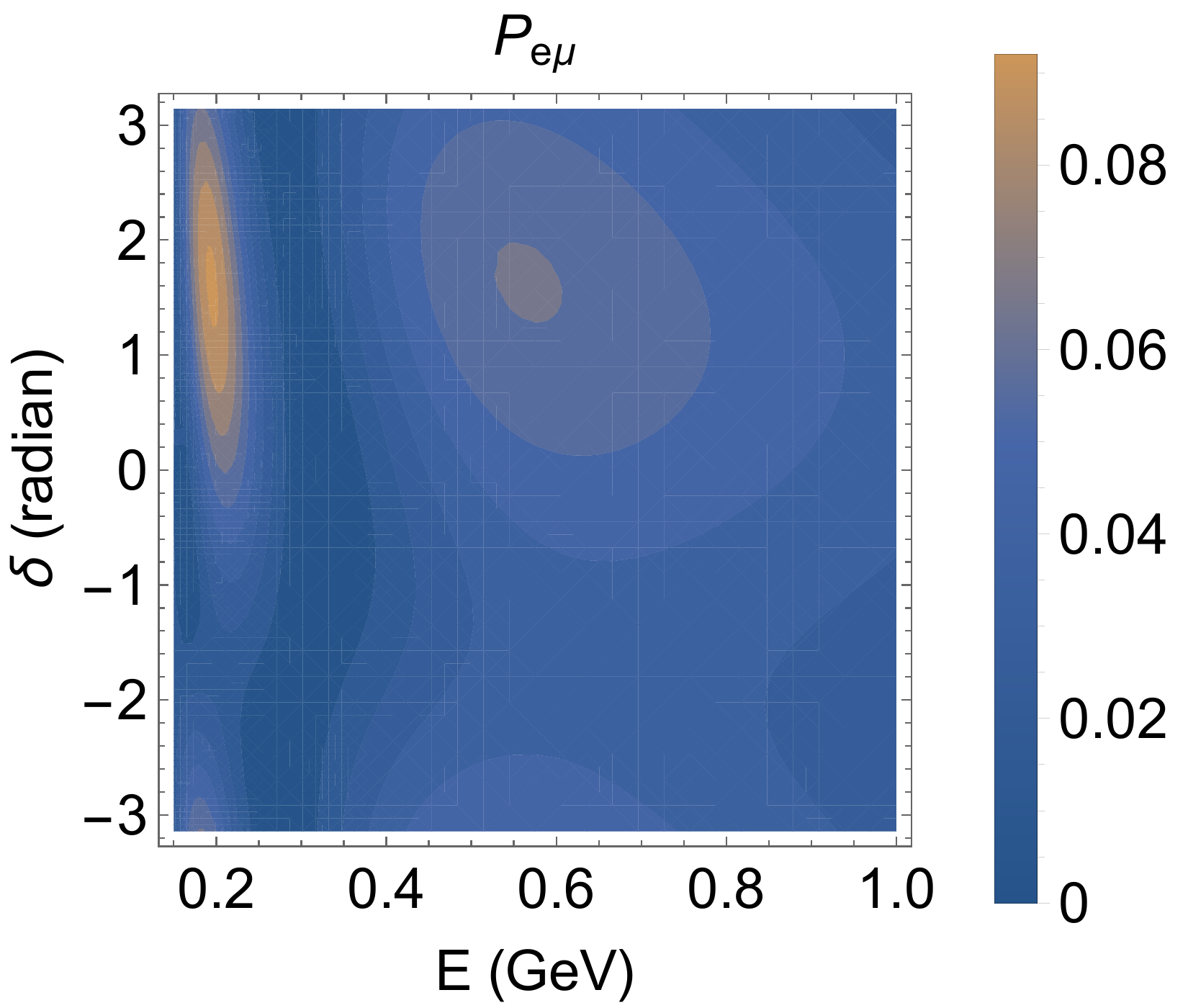}
		
		\includegraphics[width=.32\textwidth]{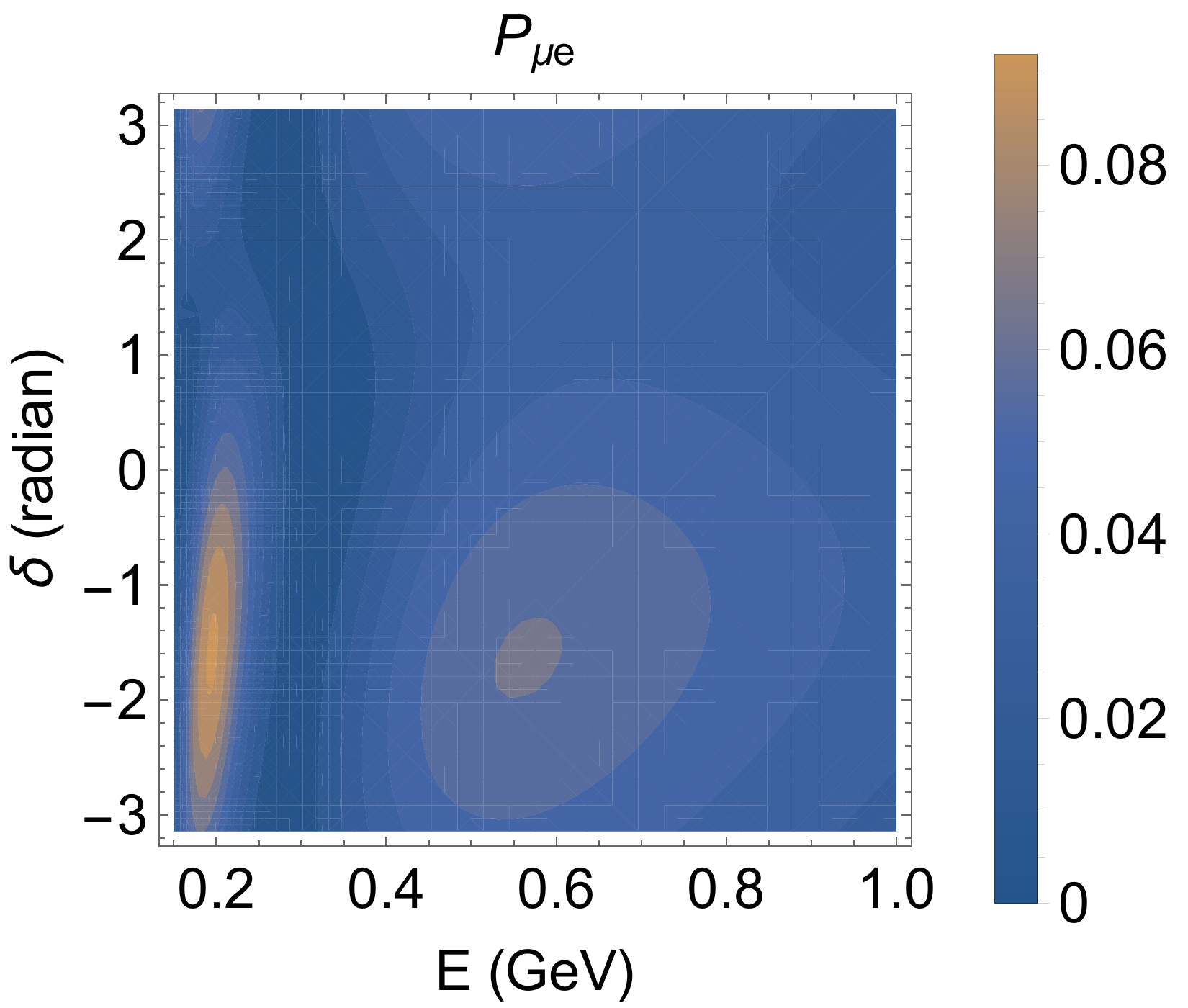}
		
		\includegraphics[width=.32\textwidth]{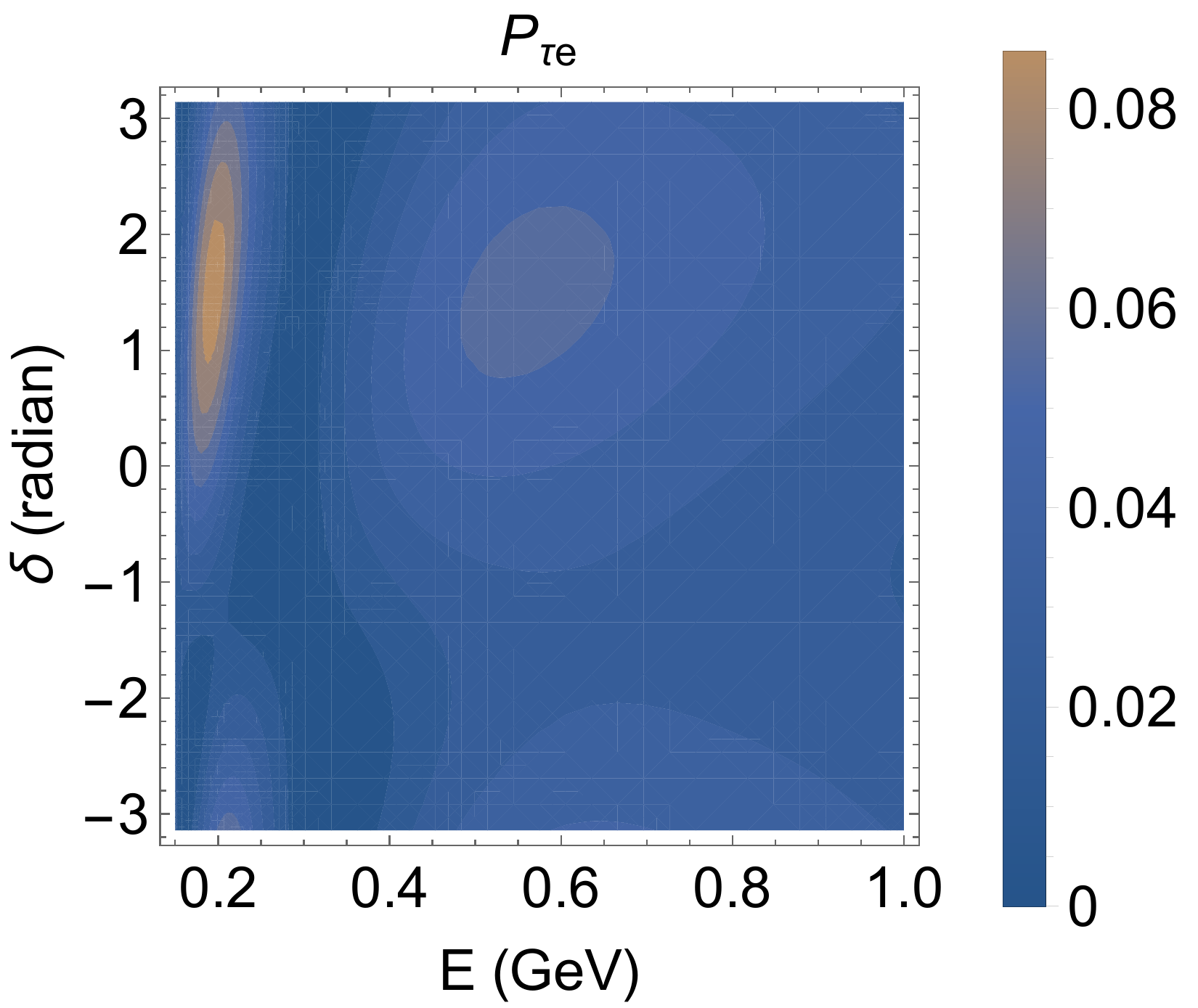}\\
		
		\includegraphics[width=.32\textwidth]{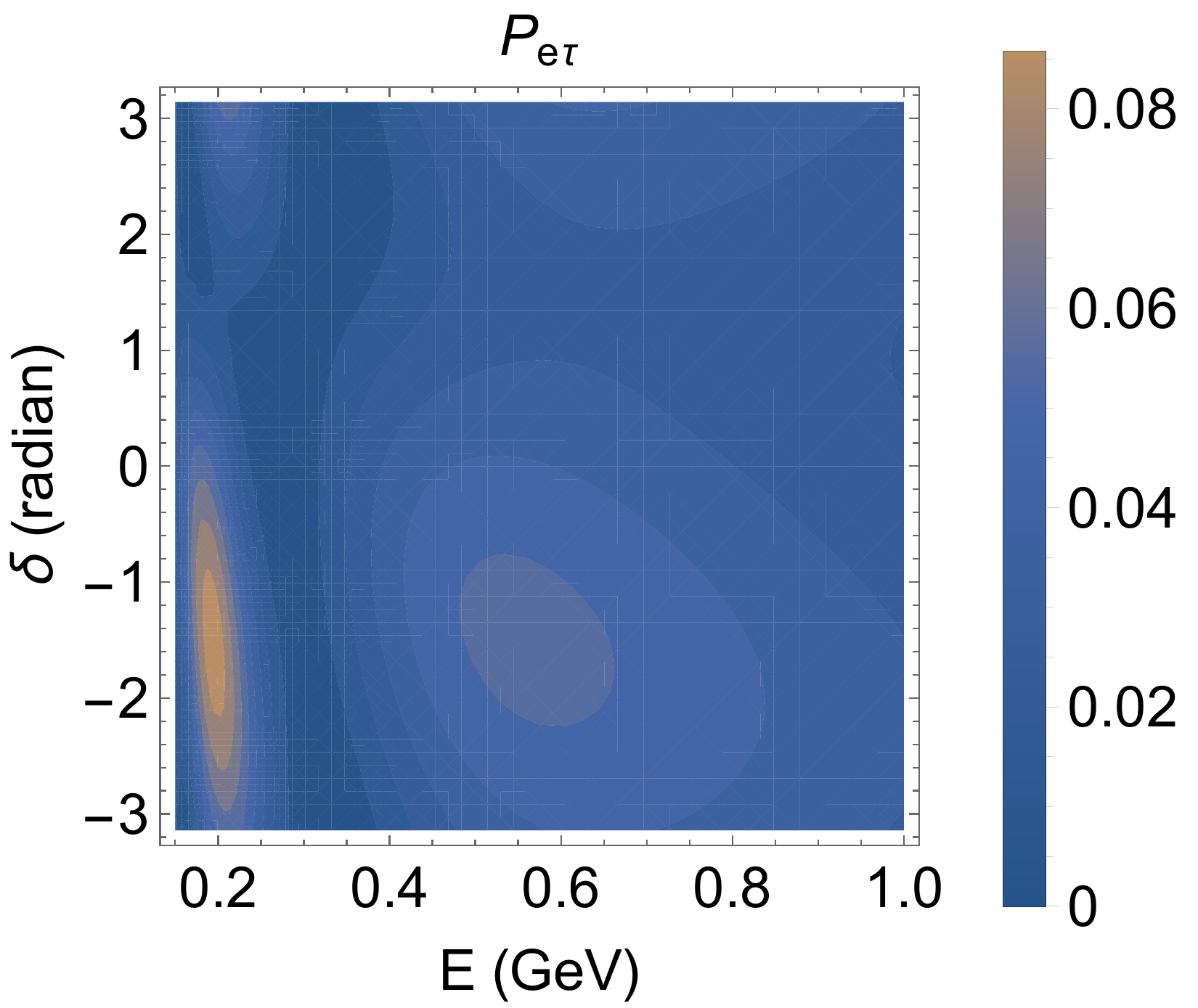}
		
		\includegraphics[width=.32\textwidth]{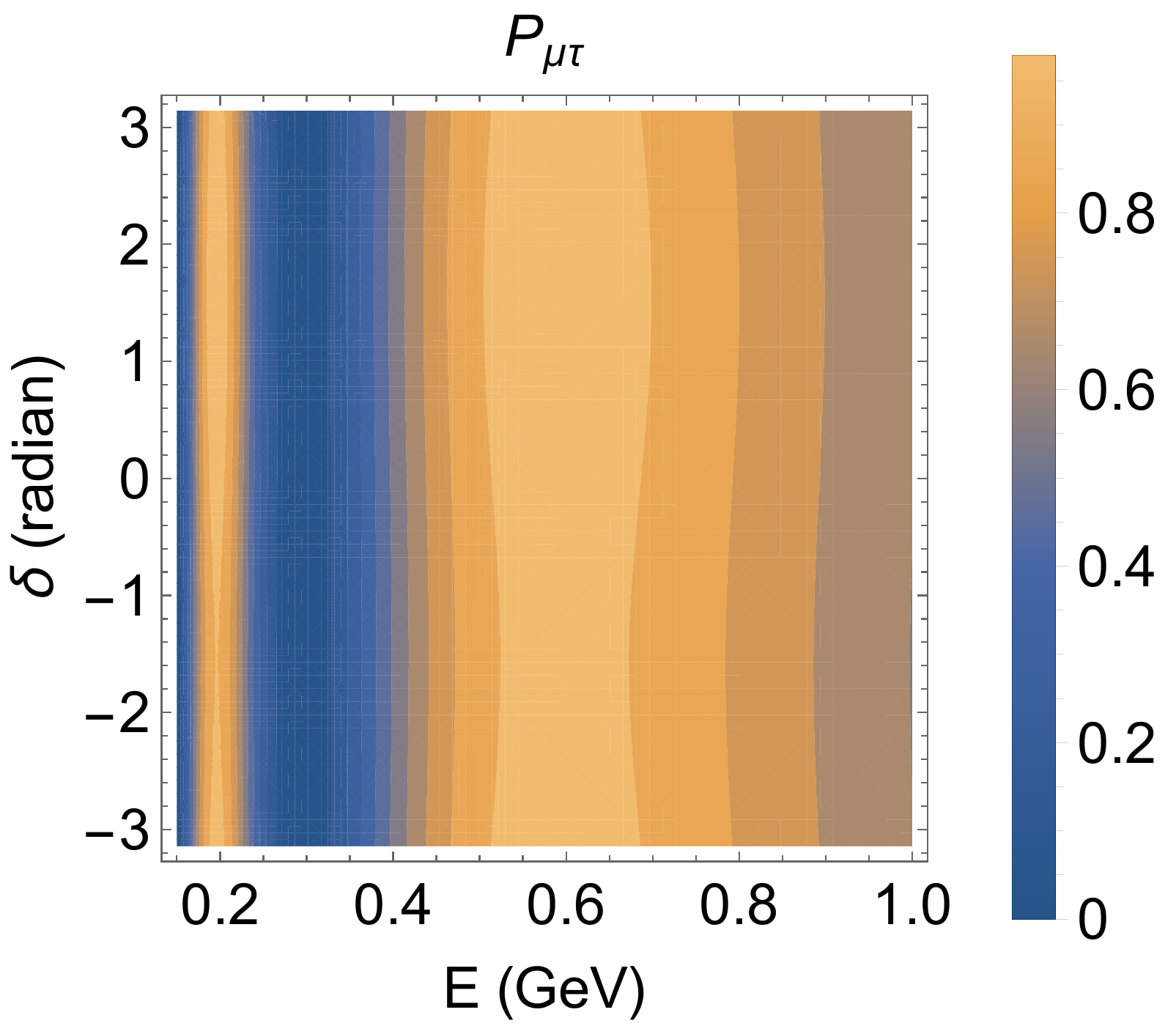}
		
		\includegraphics[width=.32\textwidth]{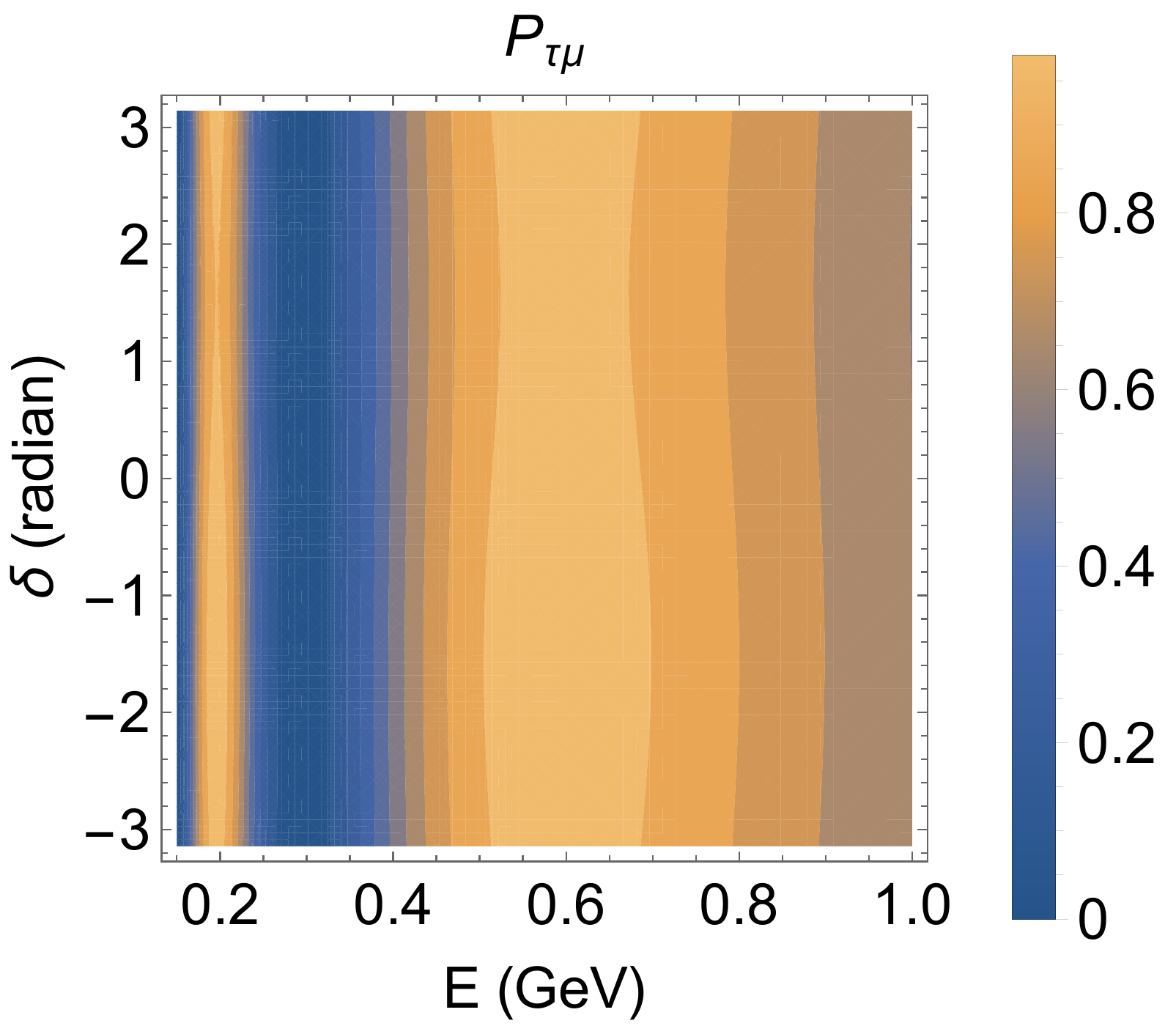}
		
	\end{tabular}
	\caption{T2K: Complexity (first row), 1-$P_{\alpha \alpha}$ (second row) and oscillation probabilities $P_{\alpha\beta}$ ($\alpha\neq \beta$) (third and fourth row) are manifested in the plane of $E-\delta$ in case of initial flavor $\nu_e$ (left), $\nu_{\mu}$ (middle) and $\nu_{\tau}$ (right). Here, we have considered $L=295$ km corresponding to the T2K experimental setup.}
	\label{Cost_Edelta_t2k}
\end{figure*}

\begin{figure*}[t] 
	\centering
	\begin{tabular}{cc}
		\includegraphics[width=.32\textwidth]{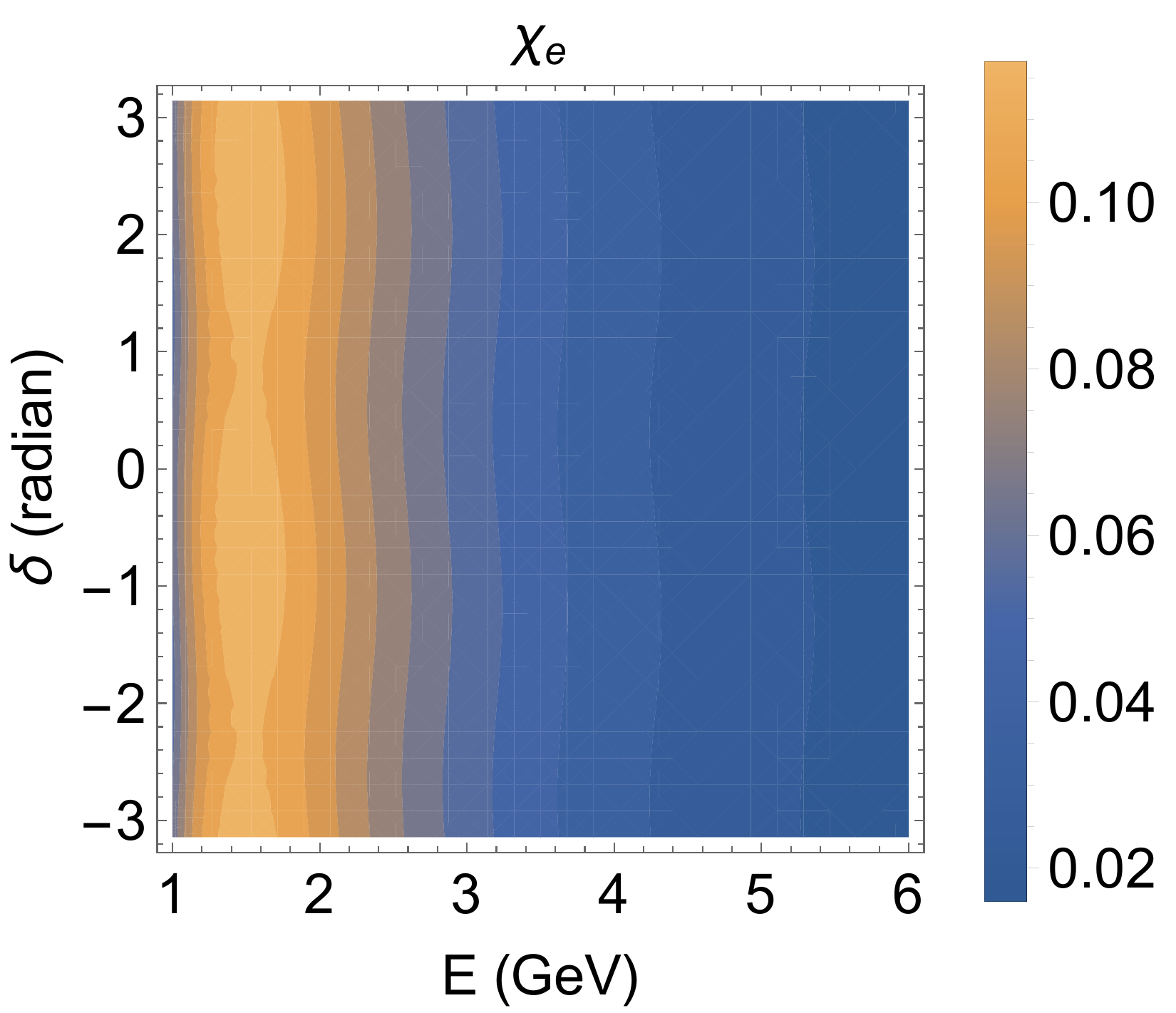}
		\includegraphics[width=.32\textwidth]{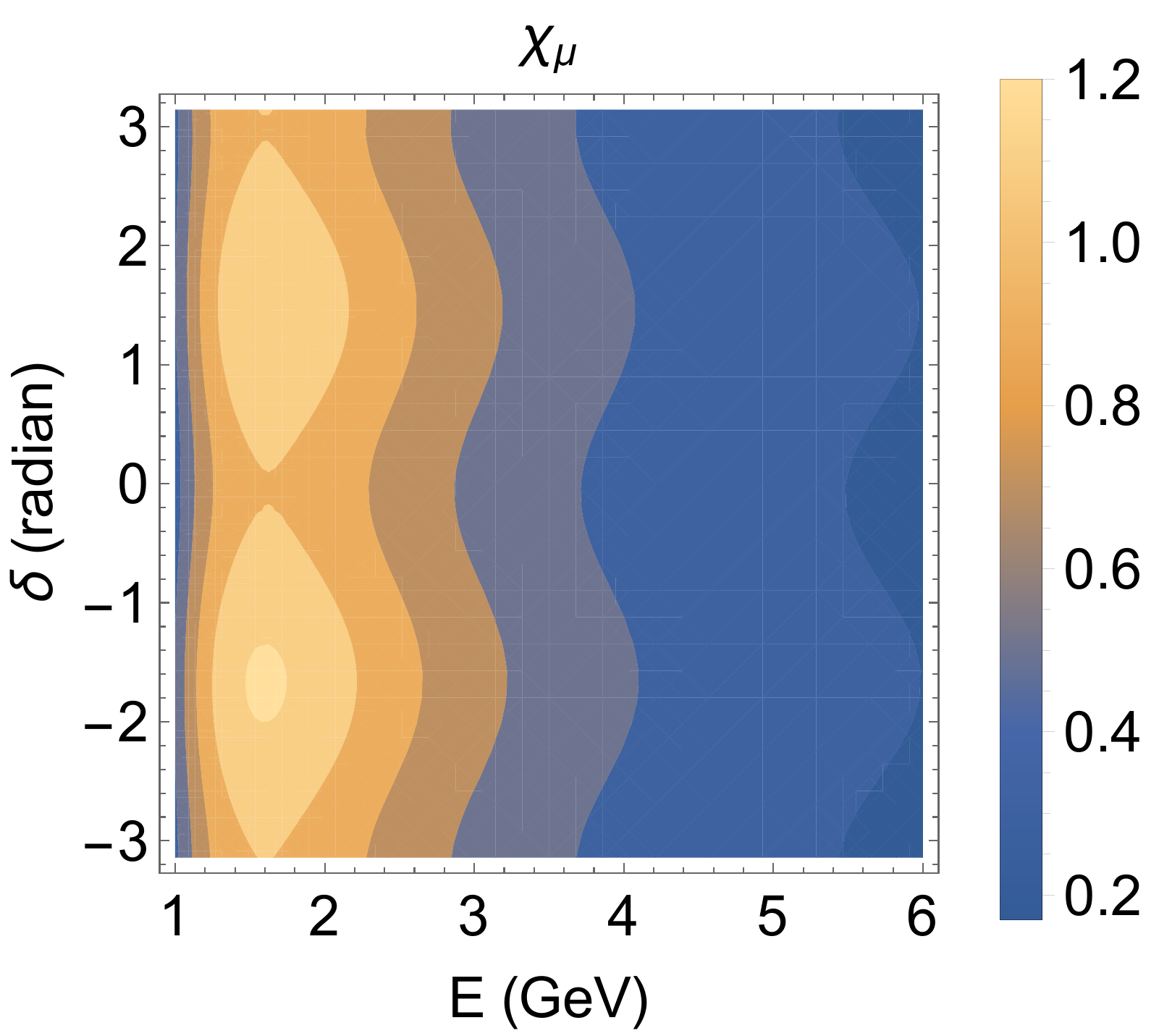}
		\includegraphics[width=.32\textwidth]{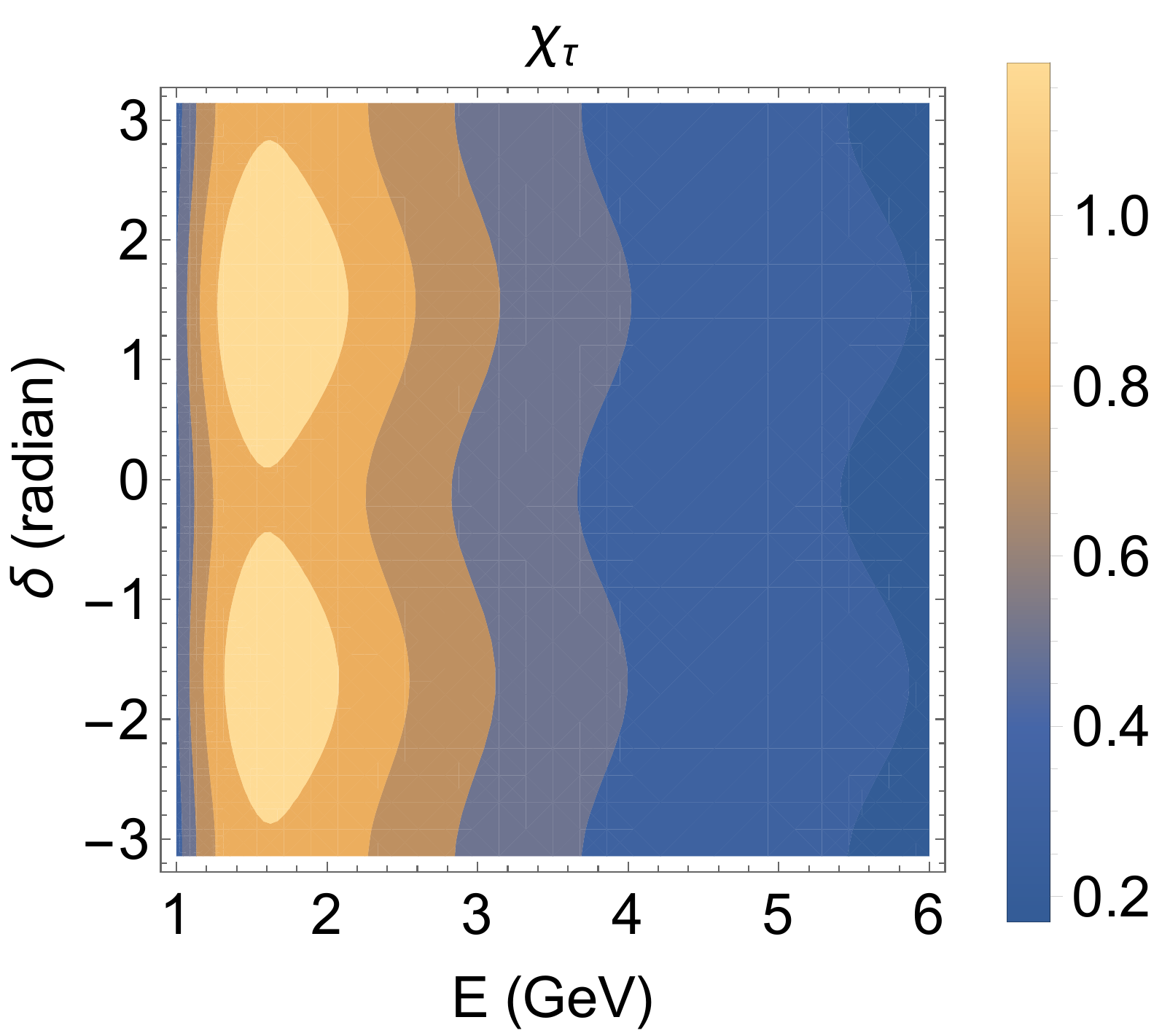}\\
		\includegraphics[width=.32\textwidth]{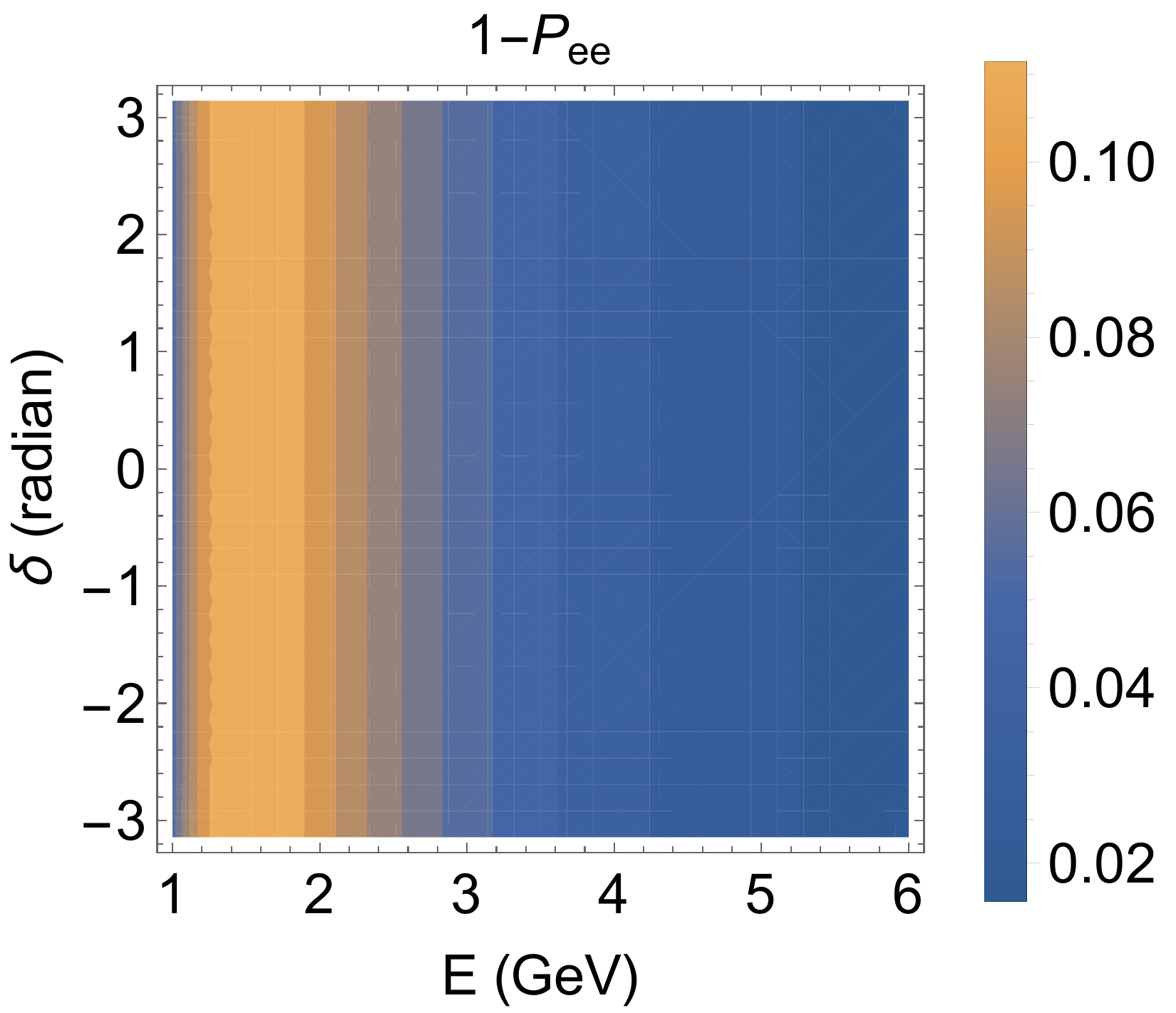}
		\includegraphics[width=.32\textwidth]{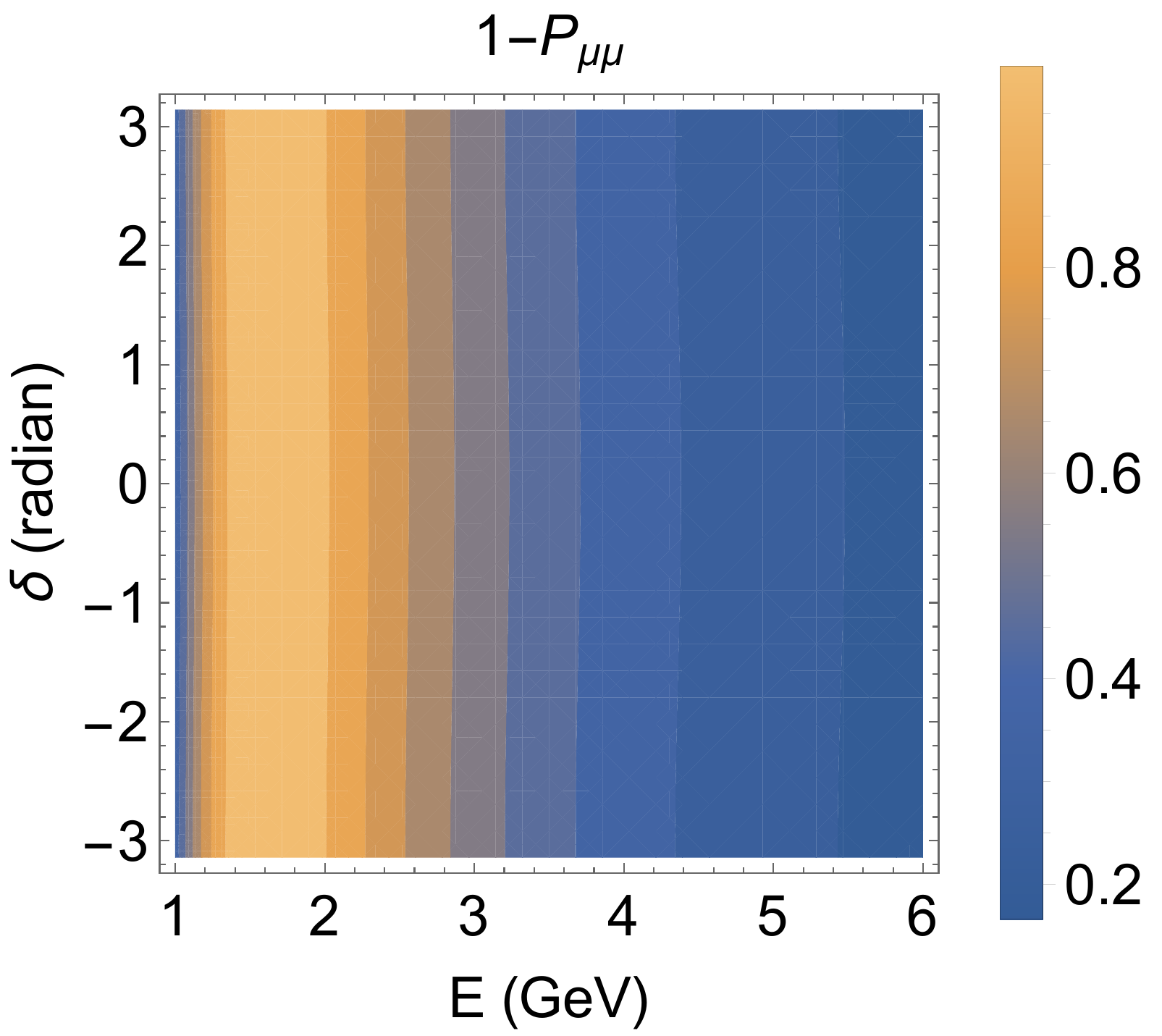}
		\includegraphics[width=.32\textwidth]{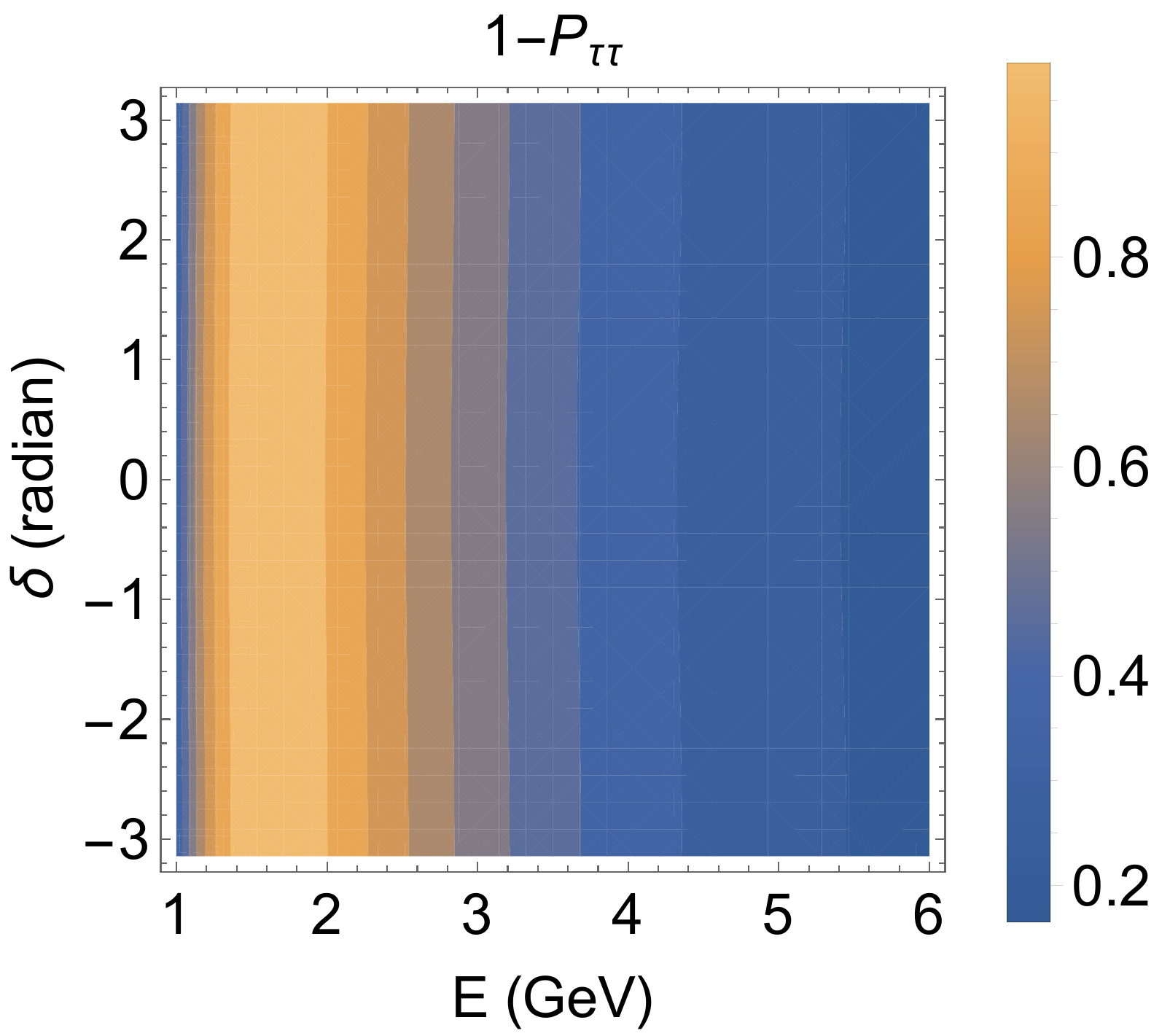}\\
		\includegraphics[width=.32\textwidth]{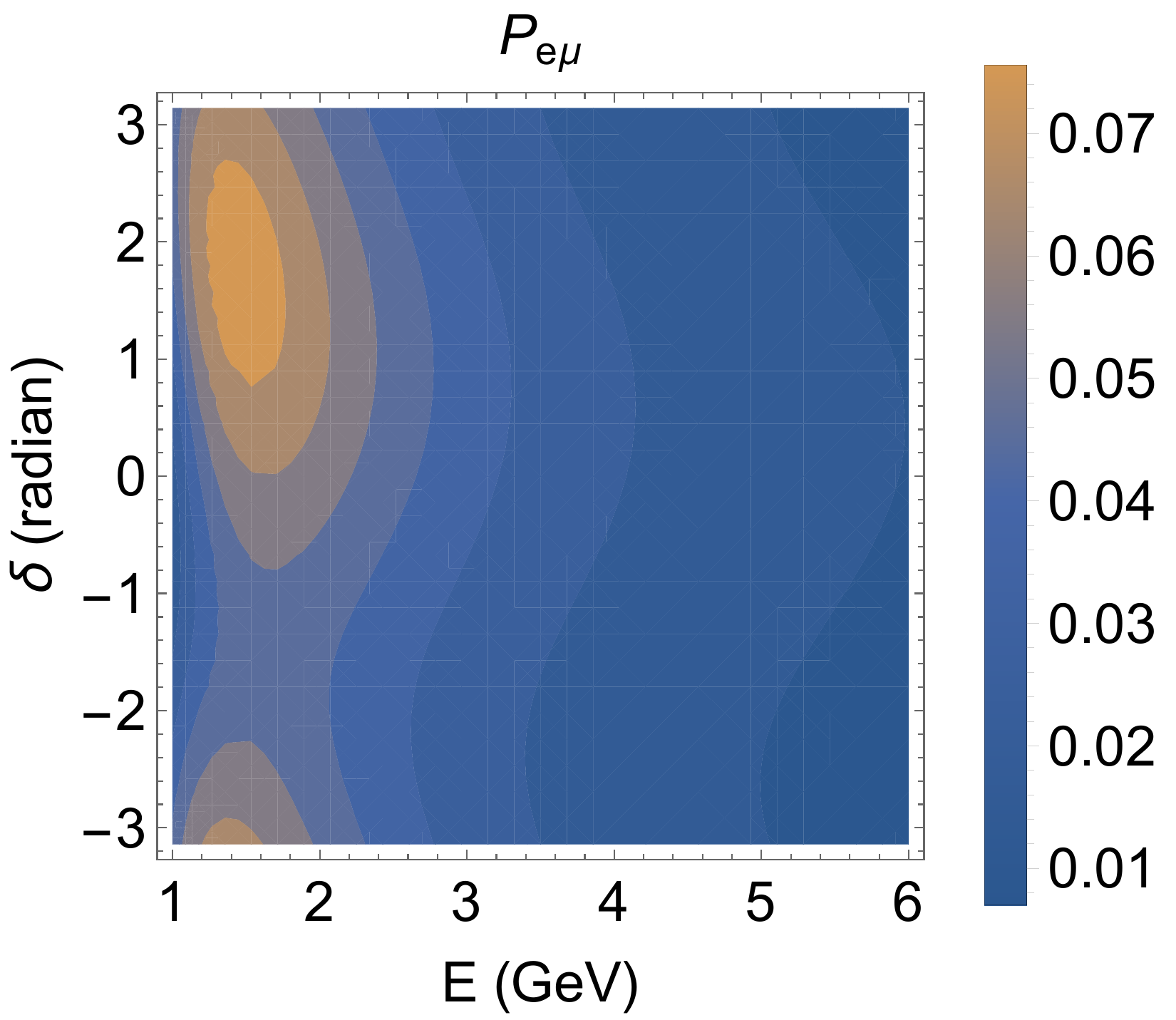}		
		\includegraphics[width=.32\textwidth]{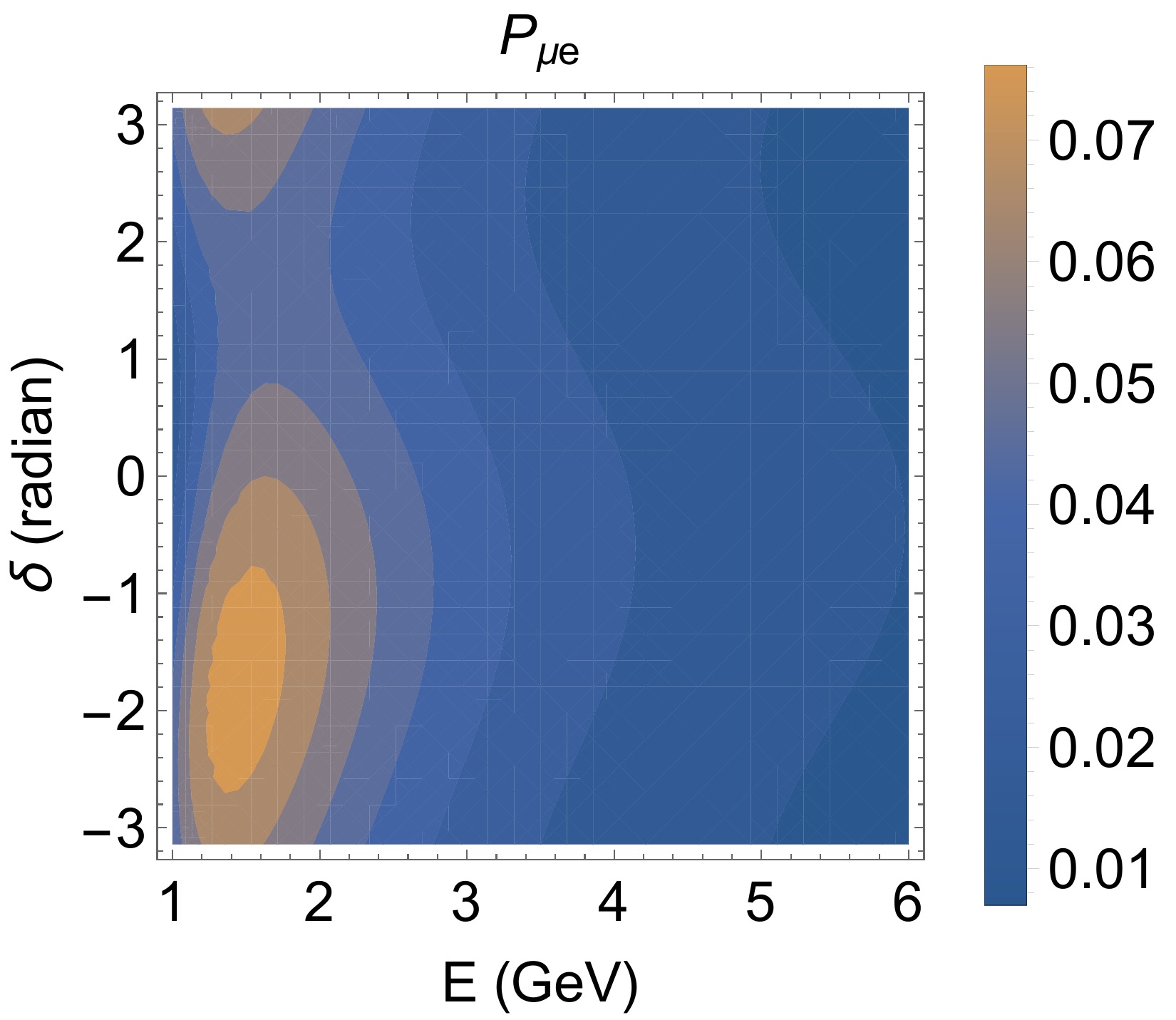}
		\includegraphics[width=.32\textwidth]{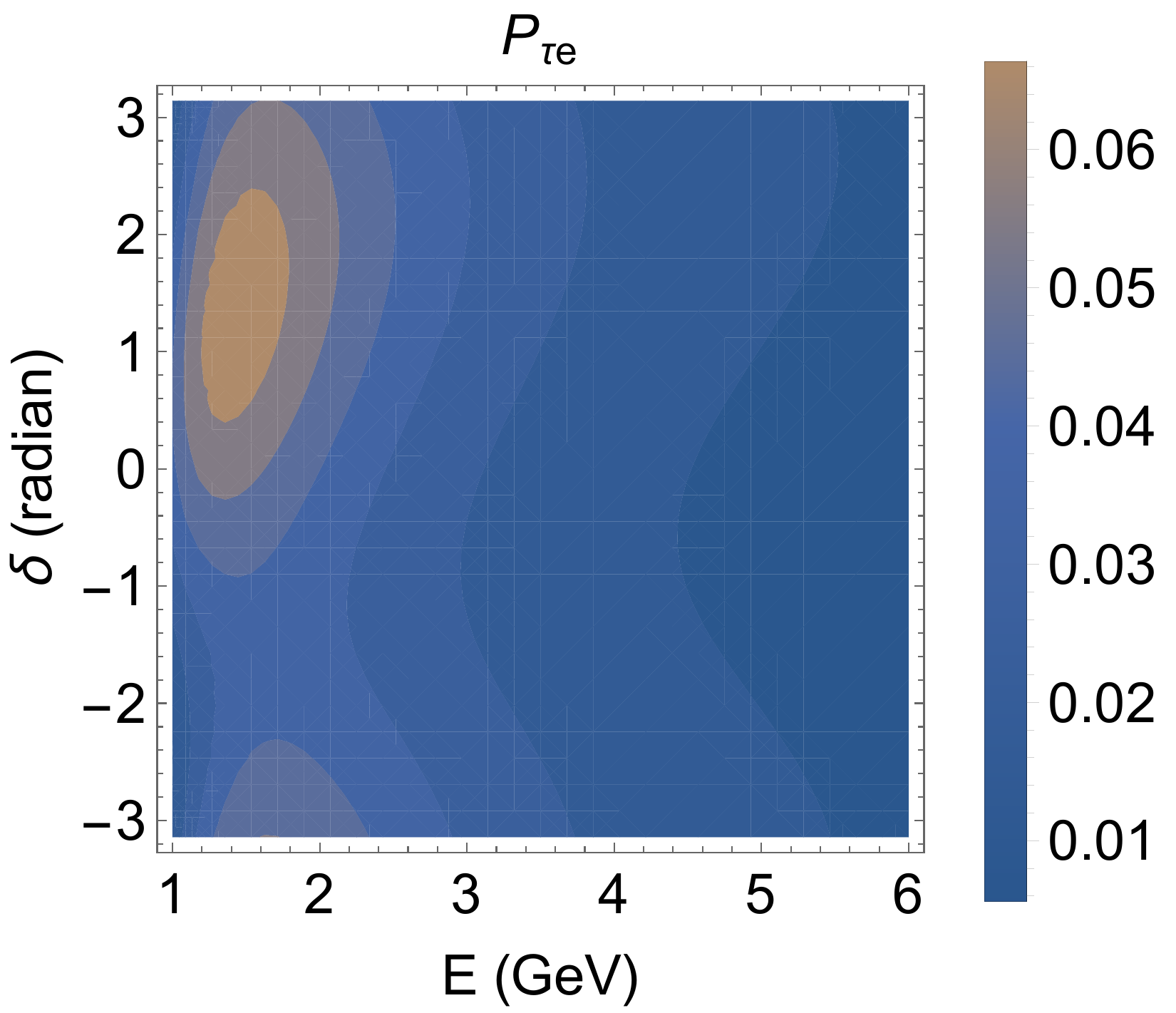}\\
		\includegraphics[width=.32\textwidth]{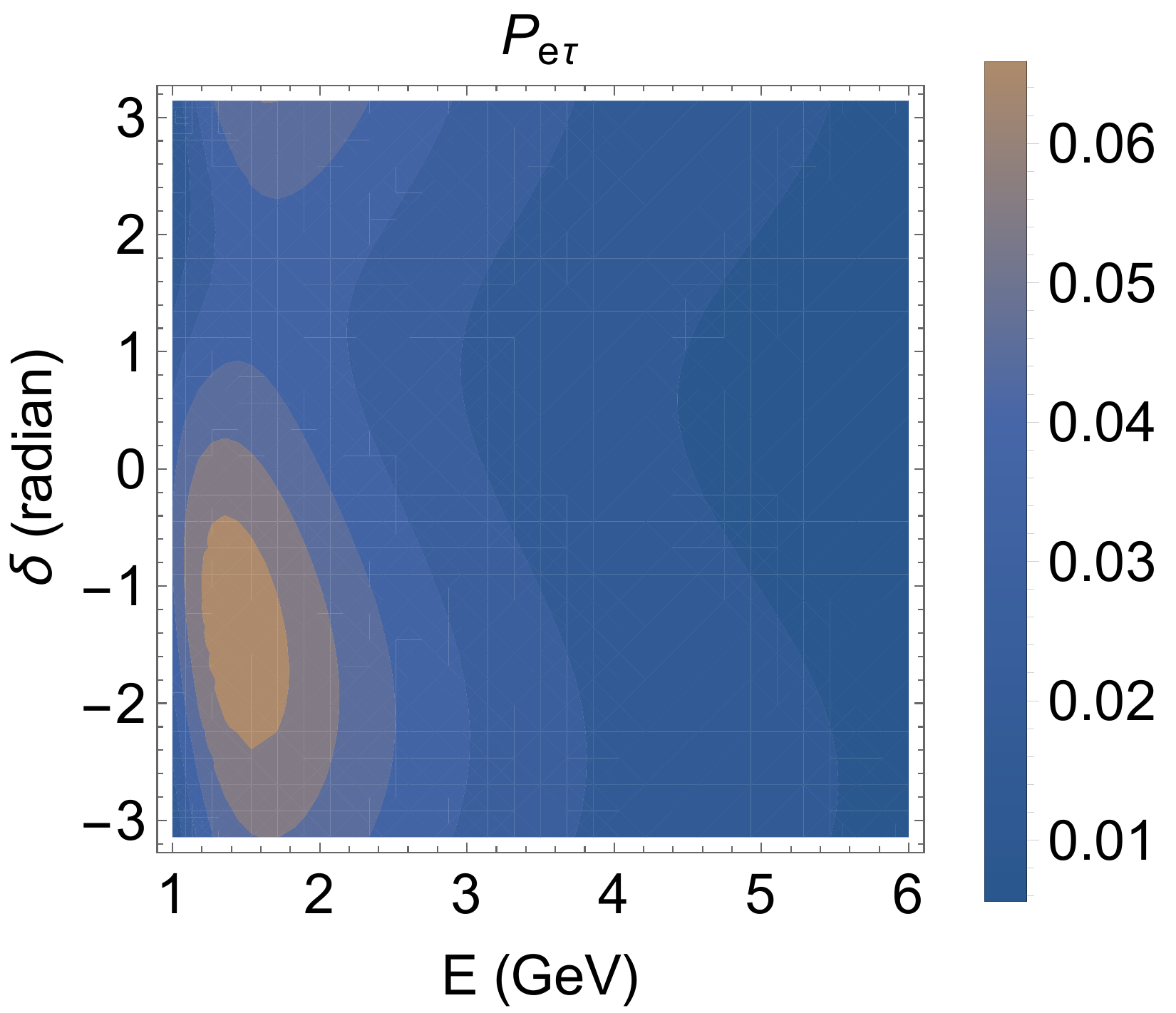}
		\includegraphics[width=.32\textwidth]{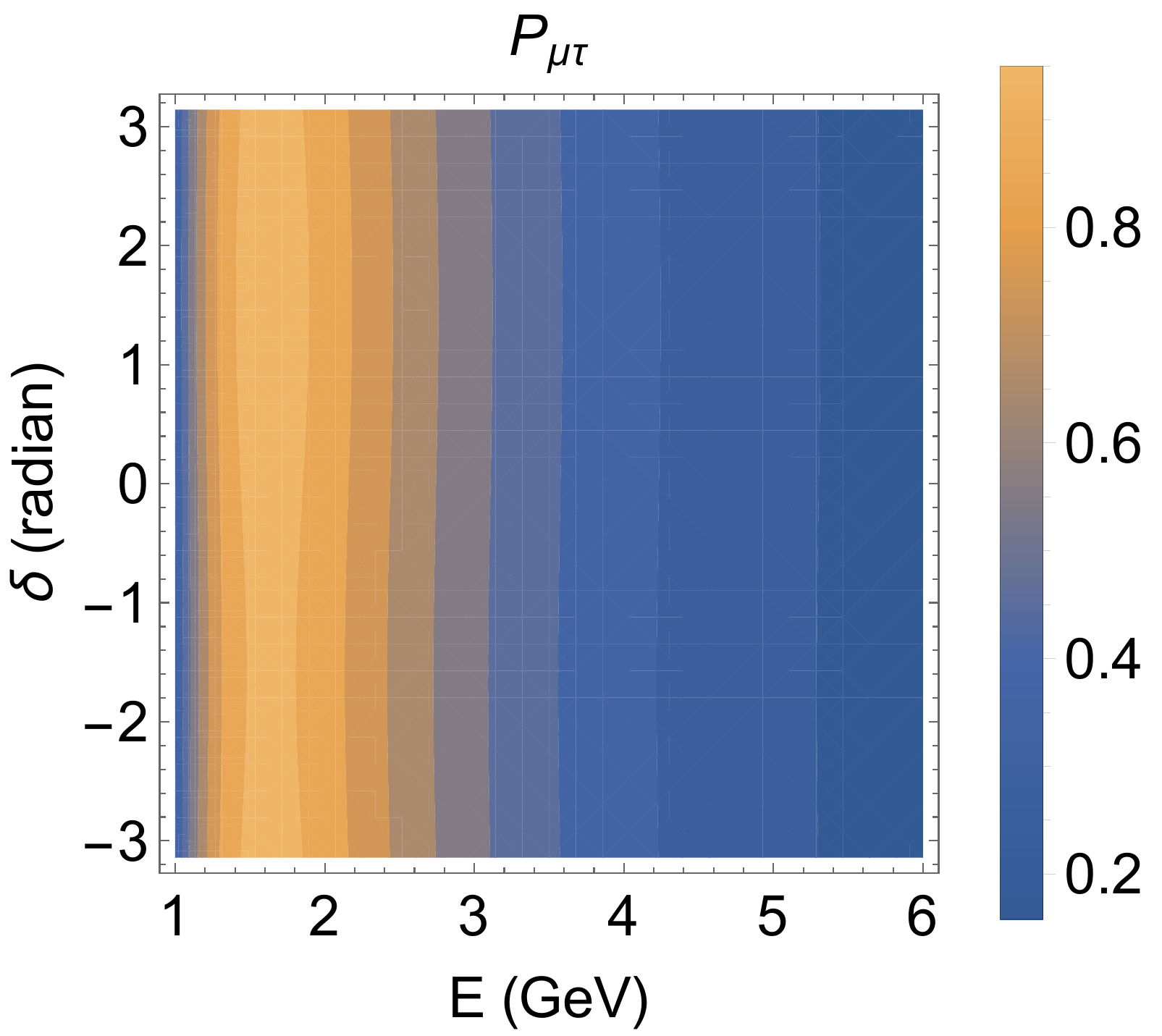}
		\includegraphics[width=.32\textwidth]{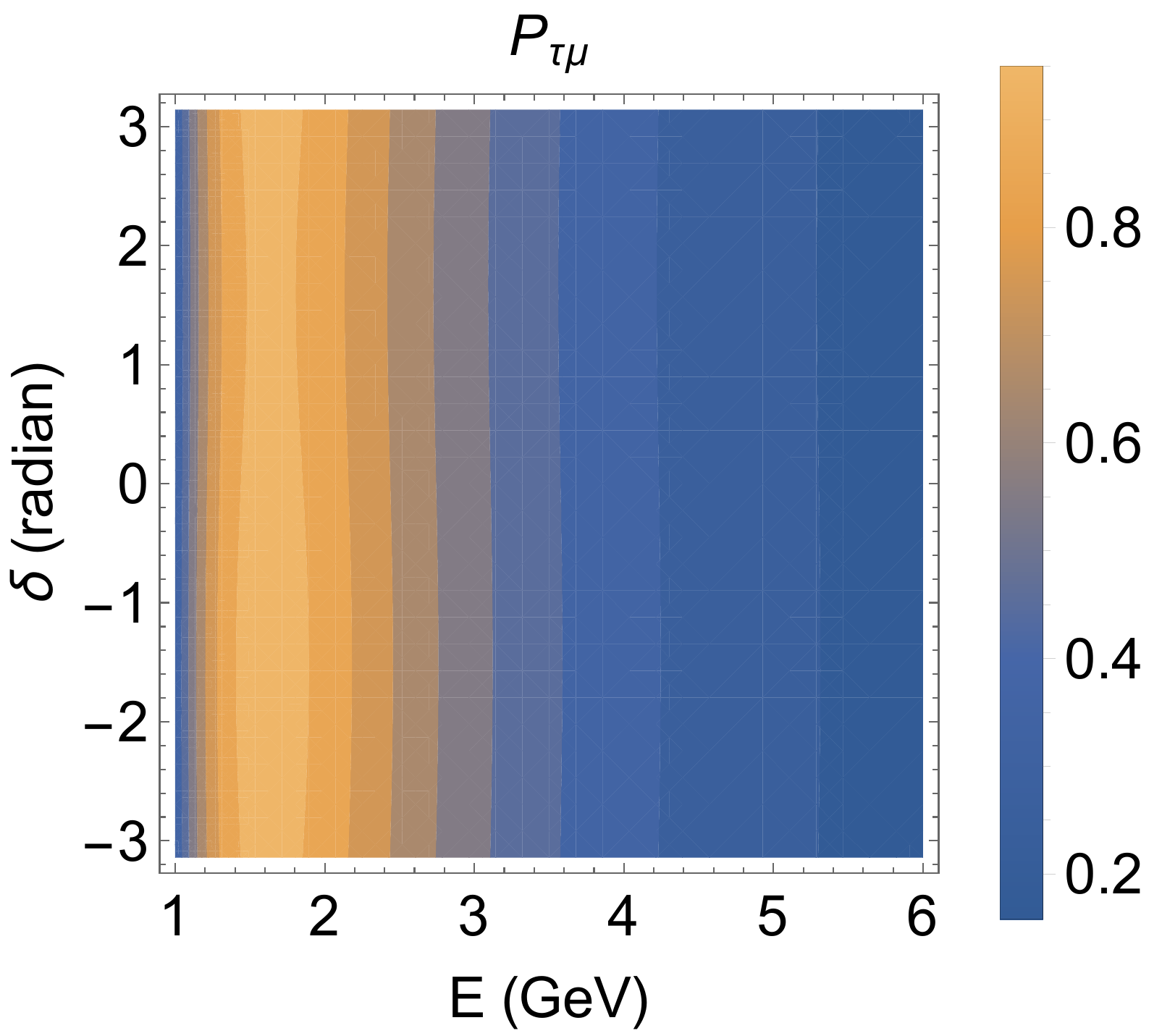}
	\end{tabular}
	\caption{NOvA: Complexity (first row), 1-$P_{\alpha \alpha}$ (second row) and oscillation probabilities $P_{\alpha\beta}$ ($\alpha\neq \beta$) (third and fourth row) are manifested in the plane of $E-\delta$ in case of initial flavor $\nu_e$ (left), $\nu_{\mu}$ (middle) and $\nu_{\tau}$ (right). Here, we have considered $L=810$ km corresponding to the NOvA experimental setup.}
	\label{Cost_Edelta_NOvA}
\end{figure*}

In Figs.~\ref{Cost_Edelta_t2k} and \ref{Cost_Edelta_NOvA} we show contour plots of $\chi_\alpha$ as functions of the CP-phase $\delta$ and neutrino energy $E$, respectively for the T2K and NOvA experiments. We have also compared complexities with the total oscillation probability $1-P_{\alpha\alpha}$ and individual oscillation probabilities $P_{\alpha\beta}$.  
One can see that $\chi_e$ shows less variations with respect to $\delta$ while this sensitivity is largely enhanced in the case of $\chi_{\mu}$ and $\chi_{\tau}$ at the relevant flux energies of $E\approx 0.6$~GeV and $E\approx 1.8$~GeV, respectively, for T2K and NOvA. For both the experiments, the maxima of $\chi_{\mu}$ and $\chi_{\tau}$ are found at $\delta \approx -1.5$ radian and $\delta = 1.5$ radian, respectively. This means that the matter effect just enhances the magnitude of complexities (as shown in Fig.~\ref{Cost_matter}), however, the characteristics of $\chi_\alpha$ with respect to $\delta$ are almost similar for both T2K and NOvA experiments.
We have also compared the complexities with corresponding flavor transition probabilities to specific flavors, for example, $\chi_e$ is compared with $P_{e\mu}$ and $P_{e\tau}$. Note that $1-P_{\alpha\alpha}$ are essentially featureless and do not provide much information on $\delta$, the reason being a cancellation of features in individual probabilities $P_{\alpha\beta}$ during the summation.

\FloatBarrier

\FloatBarrier

Let us compare results from the complexities with experimental results and probabilities. In the T2K and NOvA experimental setups, where only $\nu_\mu$ beams are produced, the only relevant complexity is $\chi_\mu$. For both the T2K and NOvA $\chi_\mu$ is maximized at $\delta \approx -1.5$ radian at the relevant experimental energies. The T2K best-fit value of $\delta = -2.14^{+0.90}_{-0.69}$ radian is consistent with this expectation. The NOvA best-fit, however, is at $\delta \approx 2.58$ radian which is far away from the maximum $\chi_\mu$ in the lower-half plane of $\delta$ but is still within a region of high $\chi_\mu$ value in the upper-half plane of $\delta$. Now, if we look at $P_{\mu e}$, which is the only oscillation probability accessible to the T2K and NOvA setups, it becomes maximum at $\delta \approx -1.5$ radian. This is compatible with T2K best-fit but is in odd with the NOvA best-fit. In fact, $P_{\mu e}$ is significantly lower at the NOvA best-fit point. It is interesting to see that complexity, which is an information-theoretic measure, provides correct prediction for the $\delta$ in experimental setups. We would also like to mention here that Fig. (\ref{Cost_Edelta_NOvA}) is obtained for the case of normal mass hierarchy, however, we have also noticed that $\chi_{\mu}$ exhibits the same characteristic in case of inverted mass hierarchy. 

\begin{figure*}[t] 
	\centering
	\begin{tabular}{cc}
		\includegraphics[width=.32\textwidth]{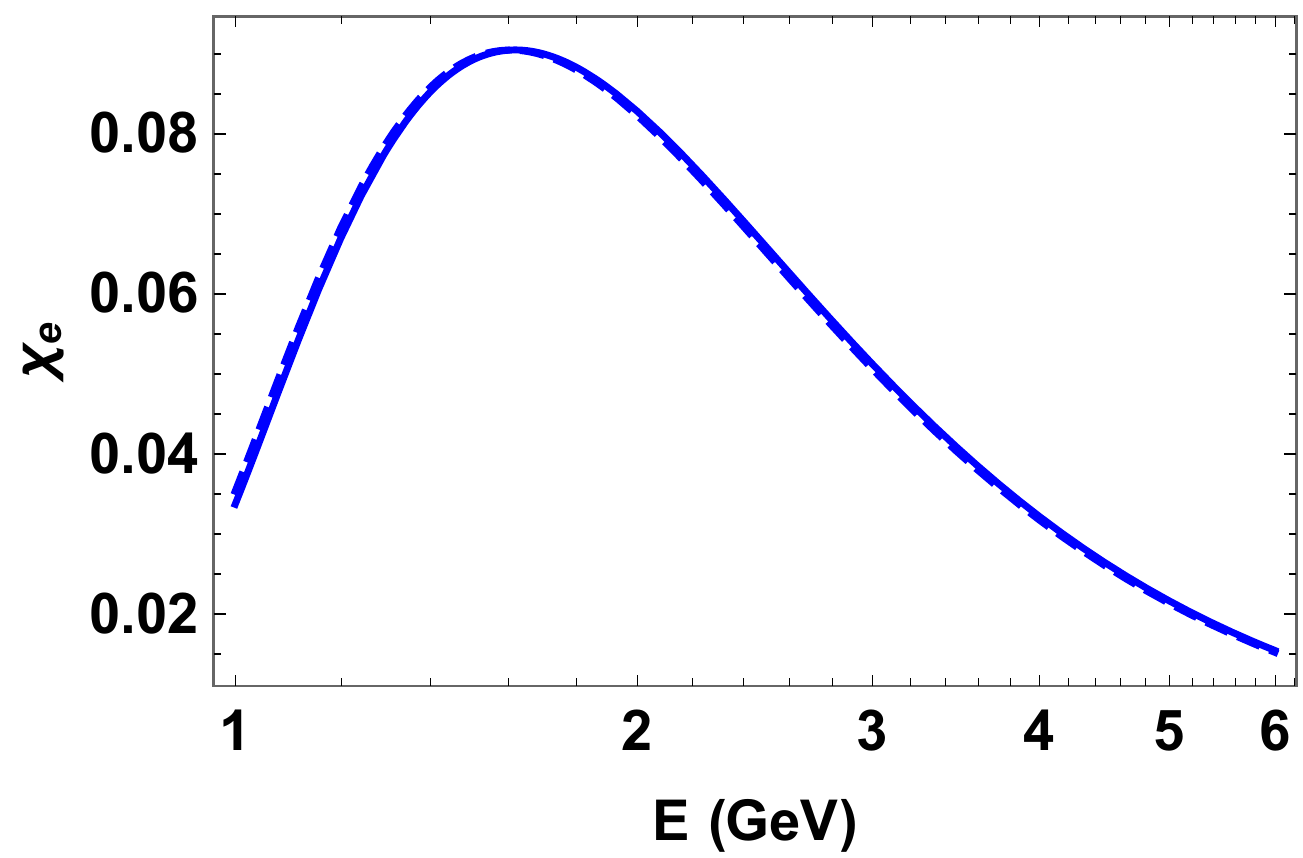}
		\includegraphics[width=.32\textwidth]{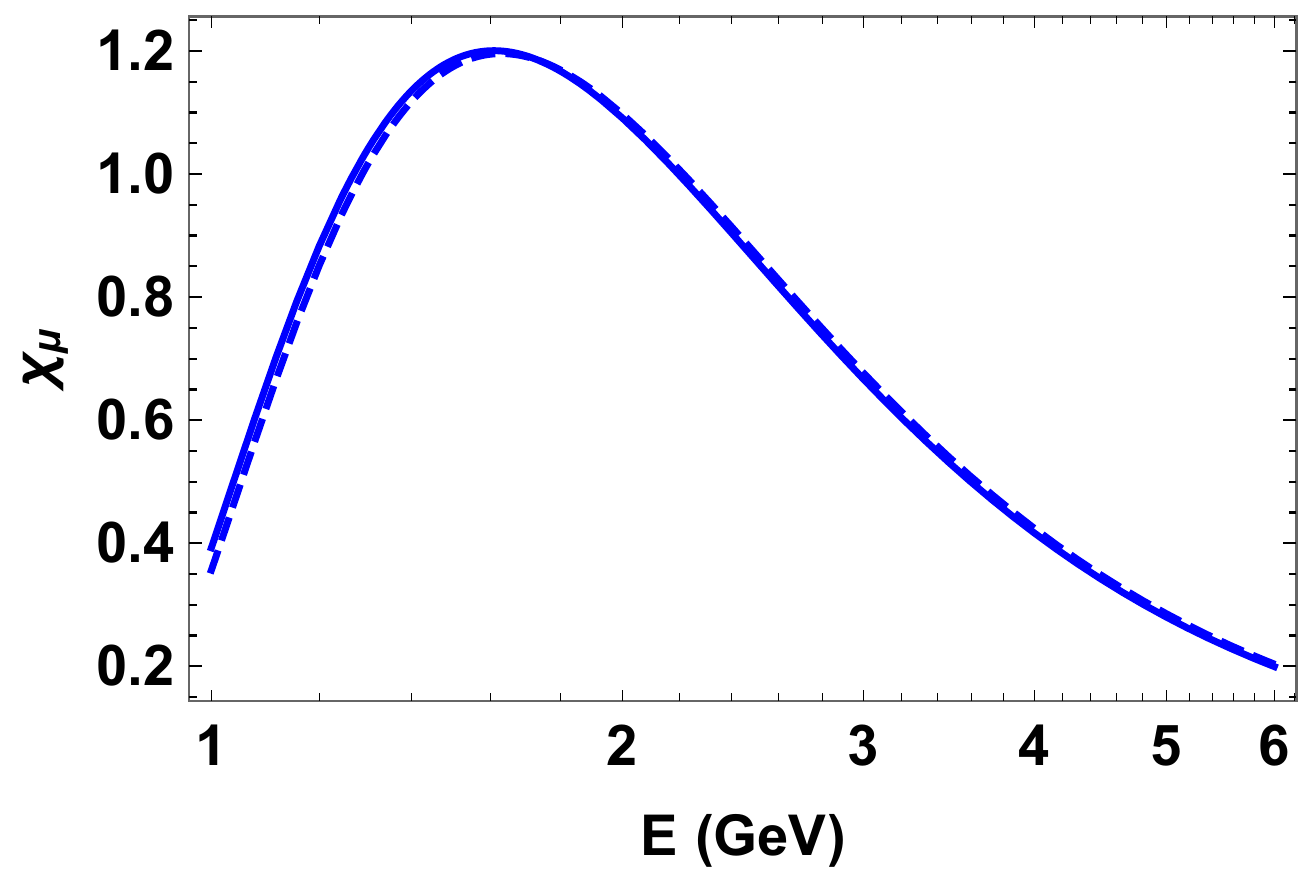}
		\includegraphics[width=.32\textwidth]{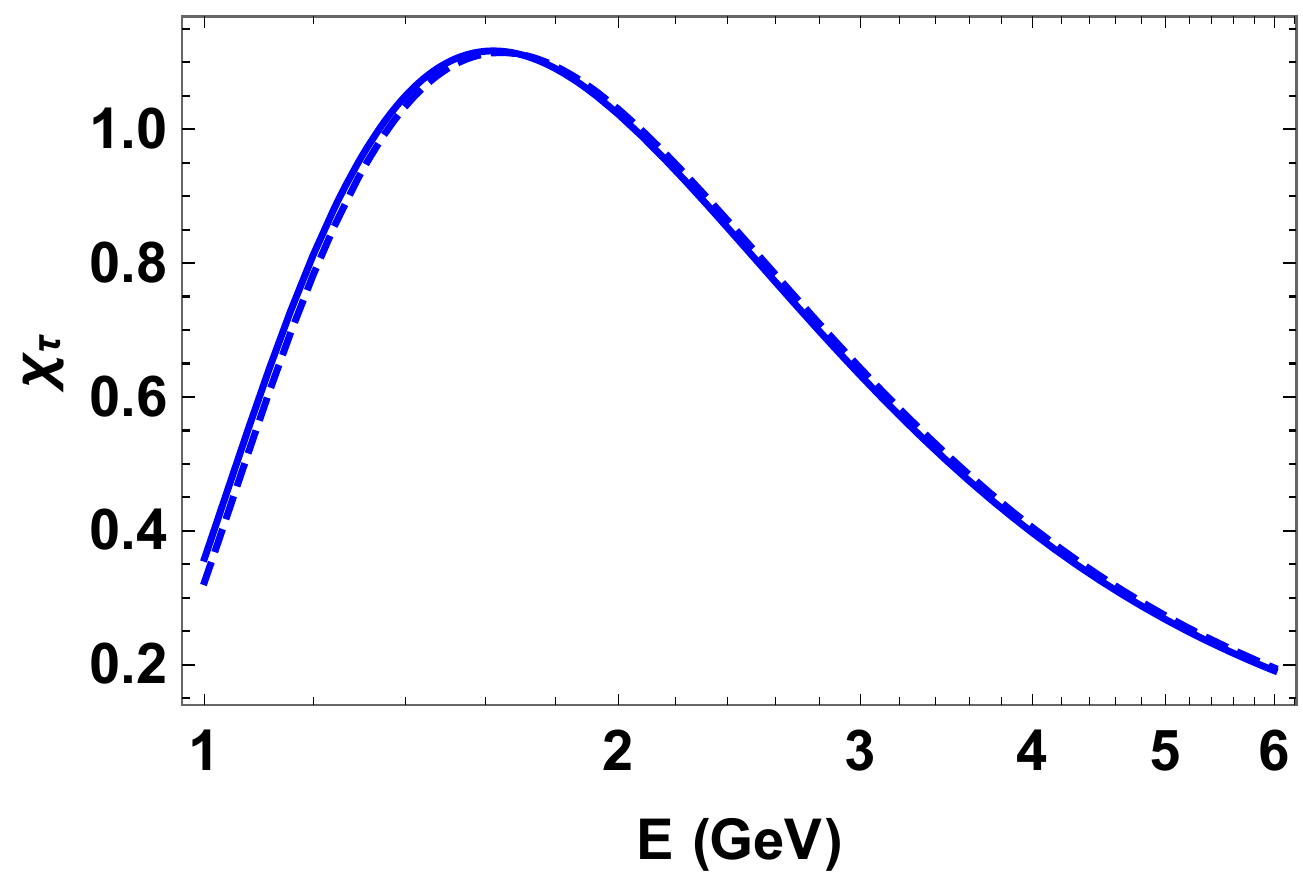}\\
		\includegraphics[width=.32\textwidth]{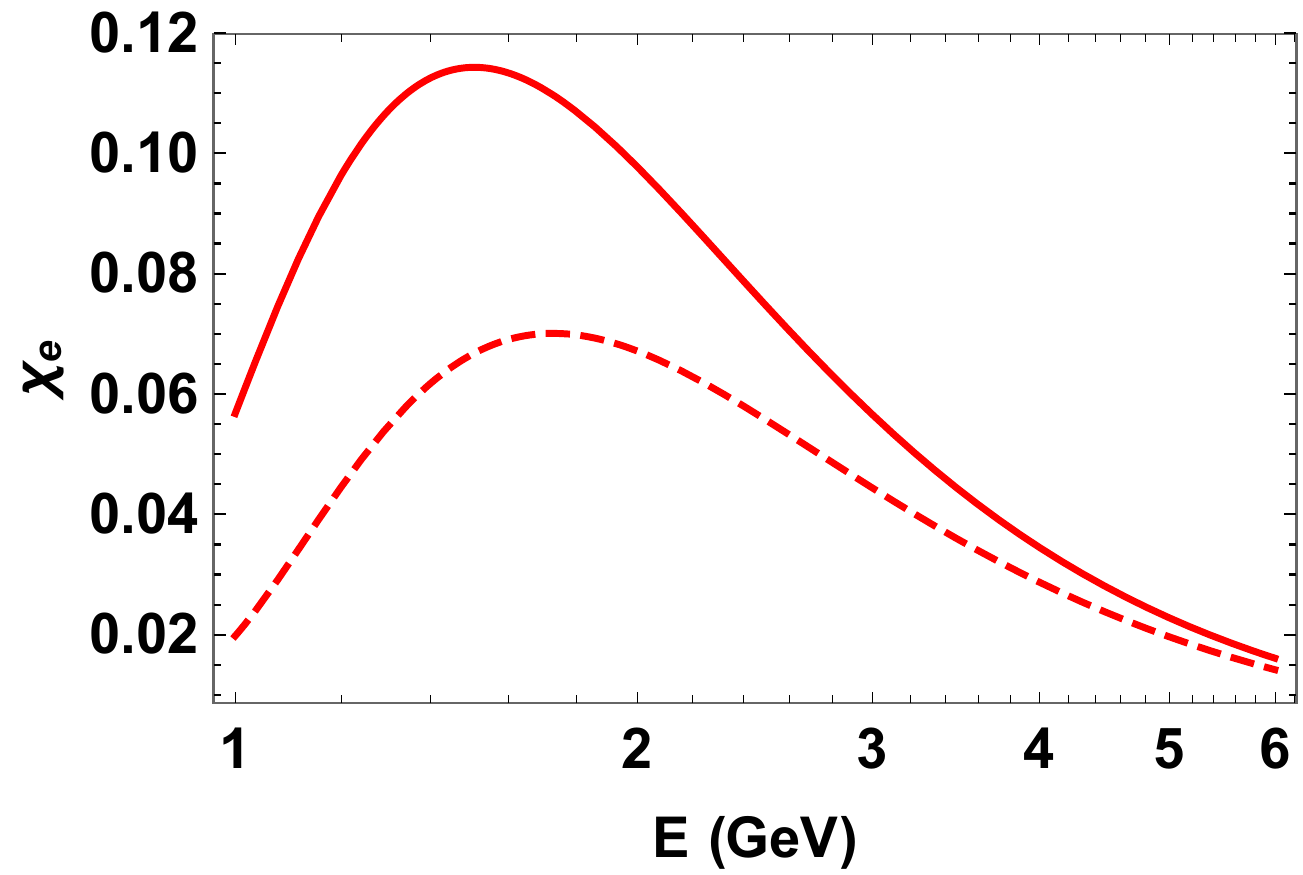}
		\includegraphics[width=.32\textwidth]{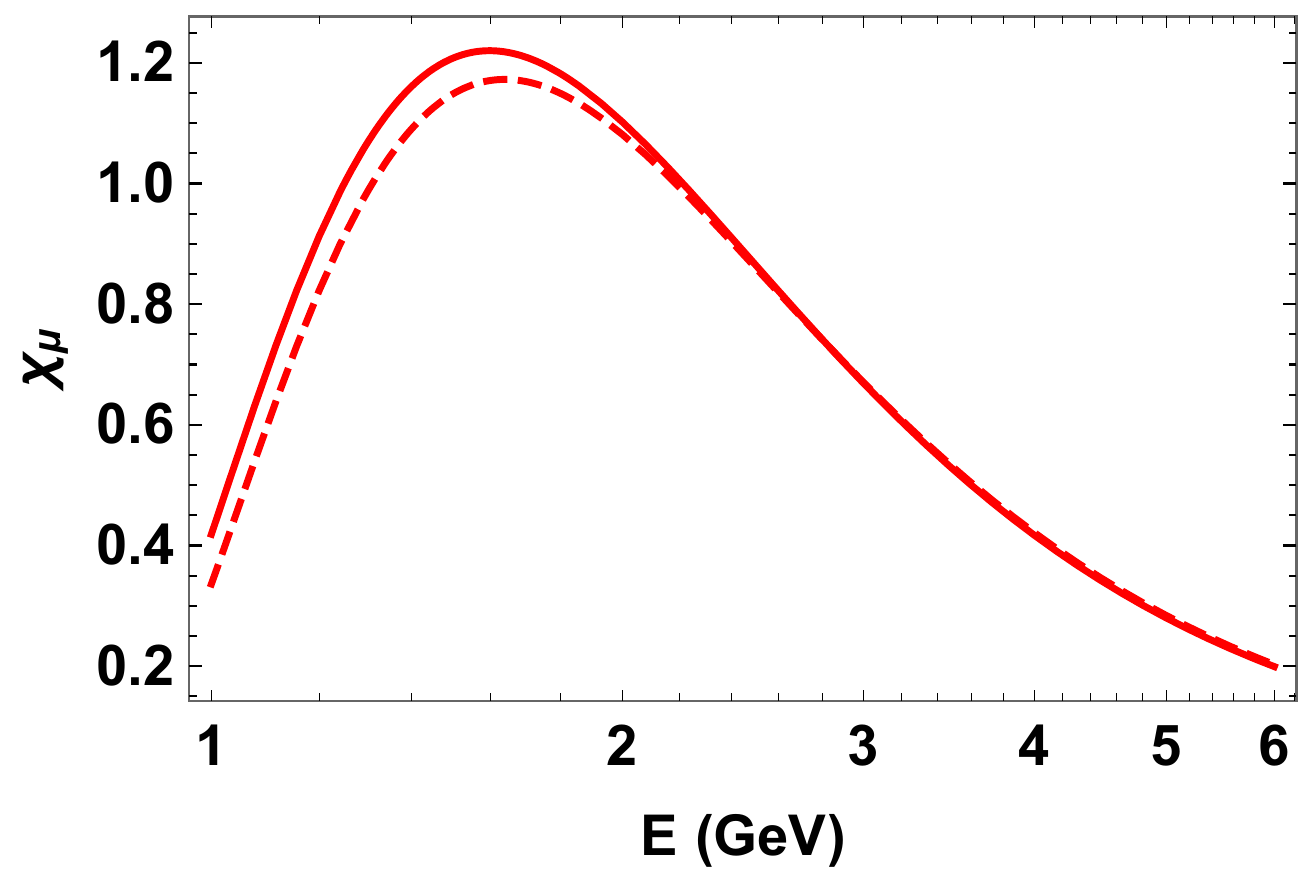}
		\includegraphics[width=.32\textwidth]{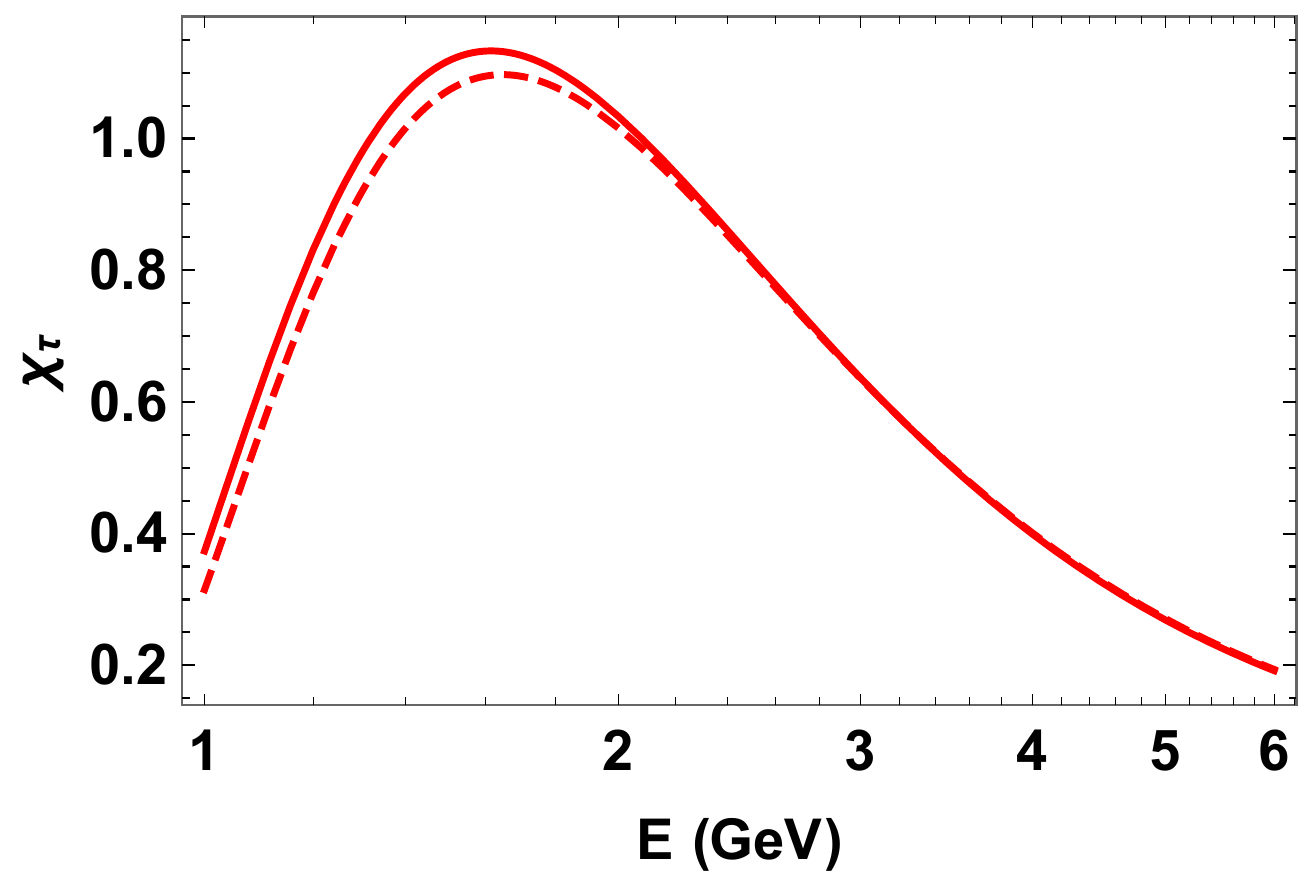}
	\end{tabular}
	\caption{NOvA: Complexity with respect to neutrino-energy $E$ in case of initial flavor $\nu_e$ (left), $\nu_{\mu}$ (middle) and $\nu_{\tau}$ (right) with $L=810$ km and $\delta=-90^o$. The upper and lower panel represent the case of vacuum and matter oscillations, respectively. Solid curves are associated with normal mass ordering (NO) and dashed curves depict the inverted ordering (IO).}
	\label{Chi_NOvA_massorder}
\end{figure*}

Further, we have also analyzed the effects of the neutrino mass hierarchy. In Fig.~\ref{Chi_NOvA_massorder} we plot $\chi_e$, $\chi_{\mu}$ and $\chi_{\tau}$ with respect to neutrino energy $E$ in the context of NOvA. Solid and dashed curves are representing normal hierarchy (NH) and inverted hierarchy (IH) of the neutrino mass eigenstates. In the upper panel, we considered the vacuum oscillation framework whereas the lower panel is depicting the case of matter oscillations. Here we can see that the complexity can distinguish between the effects due to NH and IH in the presence of non-zero matter potential.

Finally, we also compare the effects of mass hierarchy in neutrino and antineutrino oscillations scenarios. In Fig.~\ref{Chi_massoreder_antinu}, $\chi_{e}/\chi_{\bar{e}}$ (left panel), $\chi_{\mu}/\chi_{\bar{\mu}}$ (middle panel) and $\chi_{\tau}/\chi_{\bar{\tau}}$ (right panel) are plotted with respect to $E$. The red and blue curves represent the cases of neutrino and antineutrino, respectively with NH (solid line) and IH (dashed line). It can be seen that in either case of neutrino or antineutrino, the effects of NH and IH are significantly distinguishable for all three flavors. Apart from this, in the case of $\chi_e$, red-solid line (neutrinos for NH) and blue-dashed line (antineutrinos for IH) exhibit more complexity. In fact, we can see a complete swap between the NH (IH) hierarchy and $\nu$ ($\bar{\nu}$). This is a unique character of $\chi_e$ and is different from the probability $P_{\mu e}$, also shown in Fig.~\ref{Chi_massoreder_antinu}. On the other hand, for $\chi_{\mu}$ and $\chi_{\tau}$ the maximum is achieved in case of neutrinos with NH and Antineutrinos with IH, respectively. Note that complexity for antineutrinos can be achieved by replacing the matter potential $V\rightarrow -V$, and the CP phase $\delta \rightarrow -\delta$. Therefore, $\chi_e$ for neutrino in NH coincide with antineutrino in IH. There is an (almost) overlap between neutrino and antineutrino curves for IH in the case of $\chi_{\mu}$ and with NH in the case of $\chi_{\tau}$.

\section{Summary}
In this section, we summarize the results of our analysis of spread complexity in the context of neutrino oscillations.

\begin{figure*}[t] 
	\centering
	\begin{tabular}{cc}
		
		\includegraphics[width=.35\textwidth]{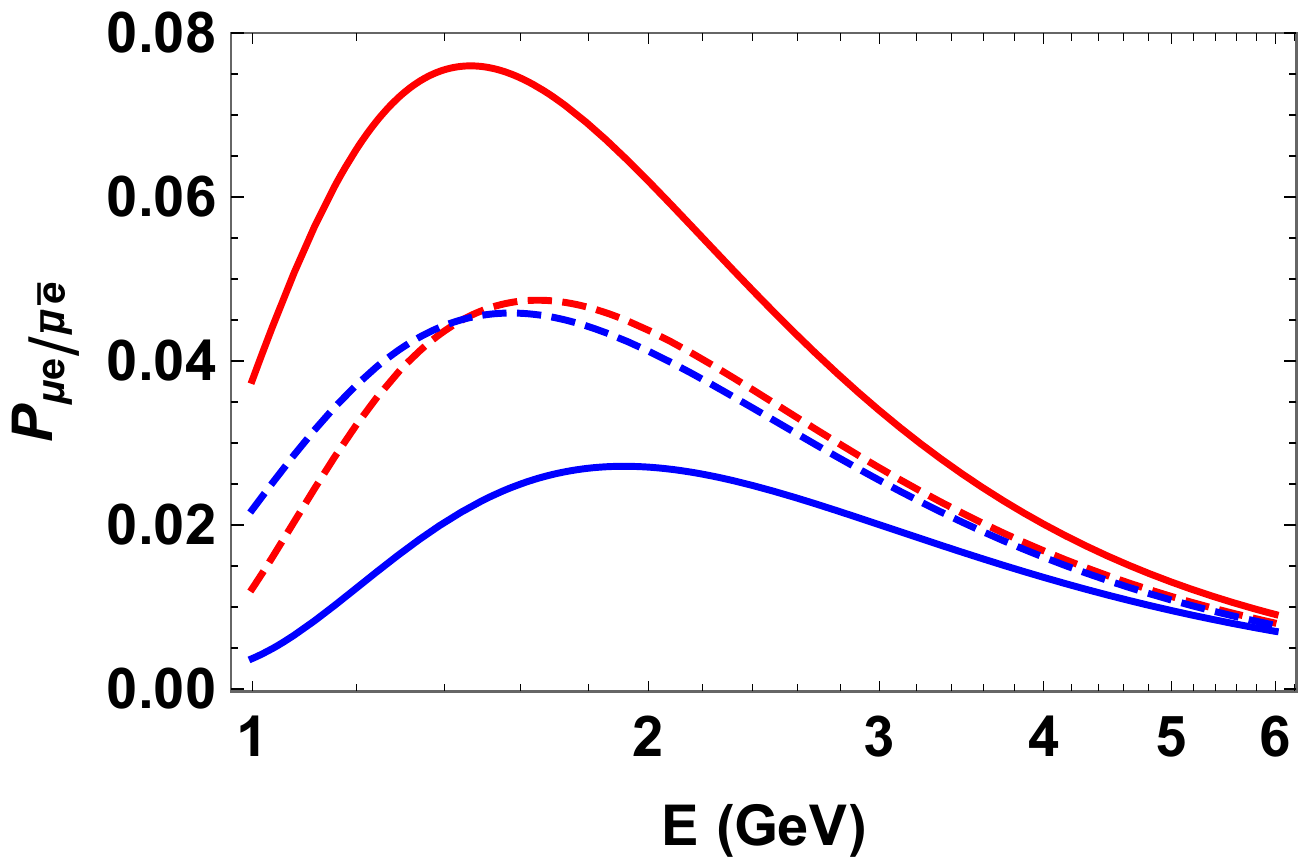}
		\includegraphics[width=.35\textwidth]{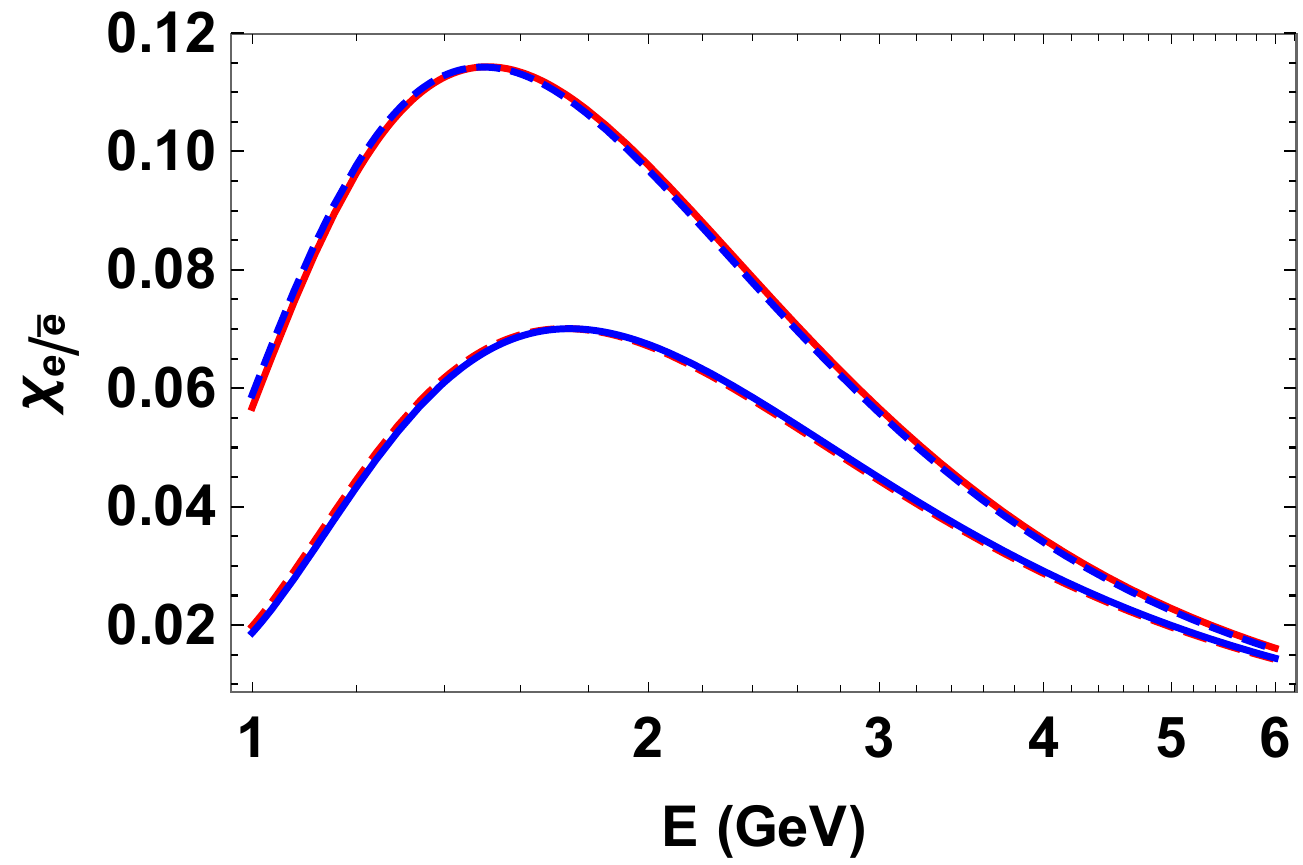}\\
		\includegraphics[width=.35\textwidth]{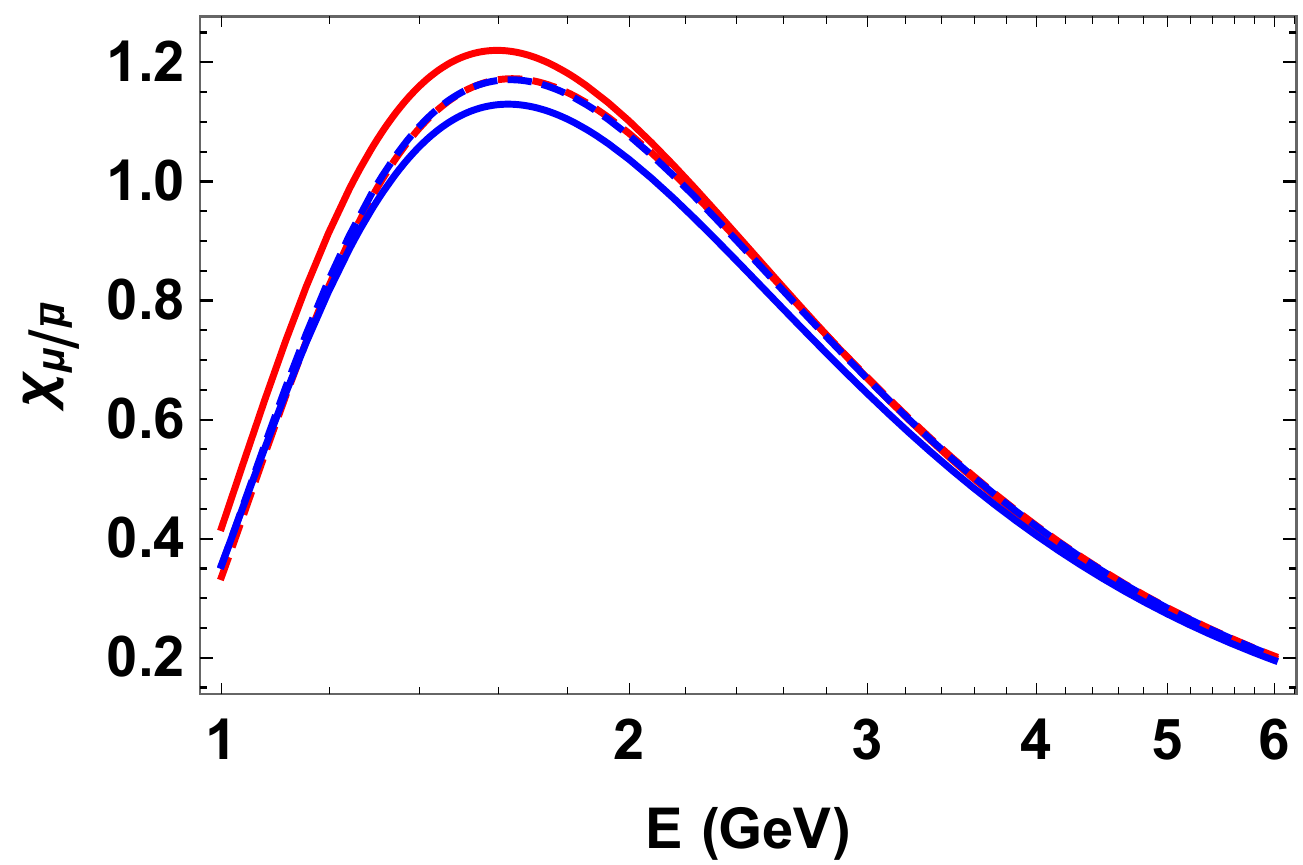}
		\includegraphics[width=.35\textwidth]{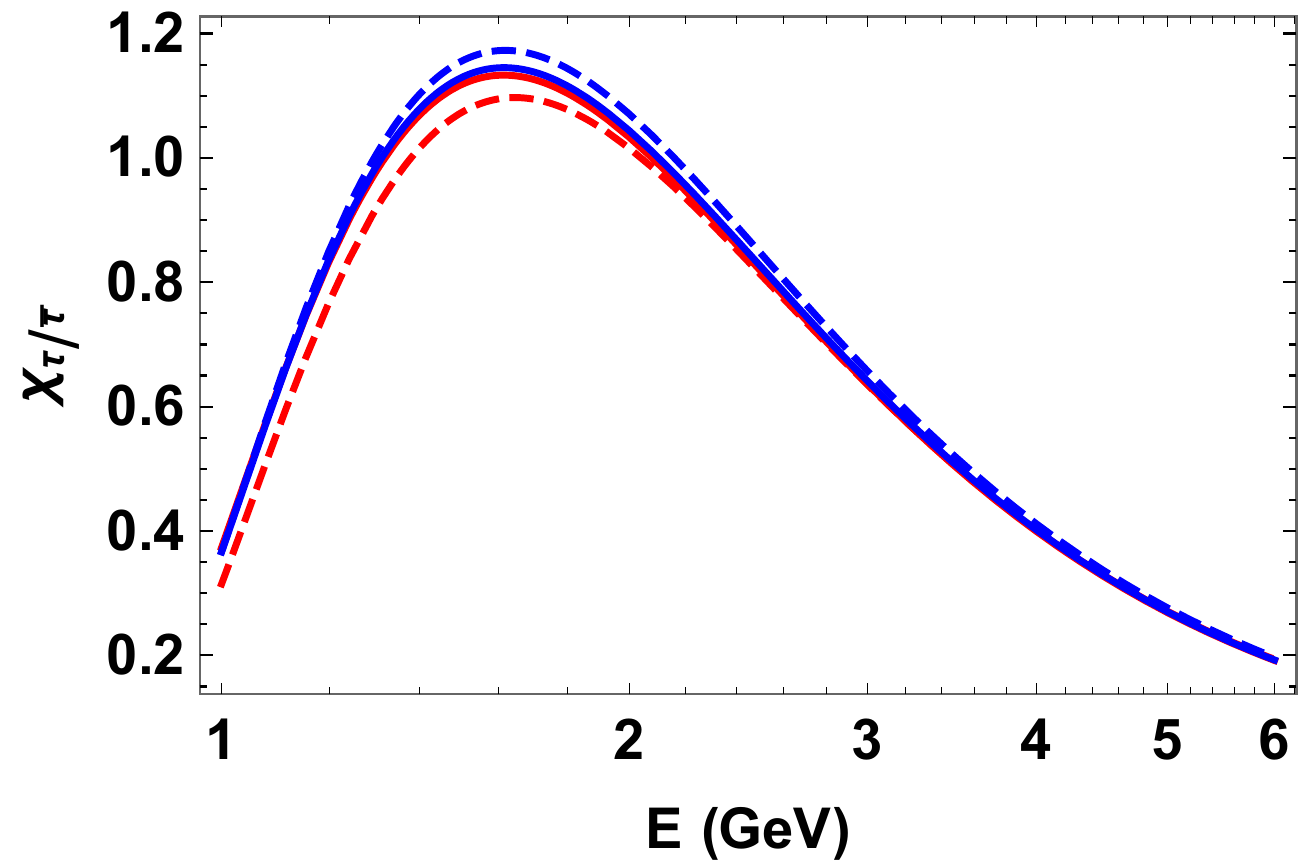}
	\end{tabular}
	\caption{NOvA: Complexities and $P_{\mu e}$ with respect to neutrino-energy $E$ where red and blue curves represent neutrino and antineutrino case, respectively, with solid (normal ordering) and dashed (inverted ordering) lines. Here $L=810$ km and $\delta=-90^o$ are considered.}
	\label{Chi_massoreder_antinu}
\end{figure*}
\begin{itemize}
	\item 
	We have inspected the spread complexity for two-flavor neutrino oscillations. We find that in this case, the Krylov basis is equivalent to the basis spanned by the flavor states of neutrino. Hence, the complexity for both cases of the initial flavor of neutrino comes out to be equal to the oscillation probability, {\it i.e.,} $\chi_e = P_{e\mu}$ and $\chi_{\mu} = P_{\mu e}$ as can be seen in Eqs.~(\ref{twoflavorchie}) and (\ref{twoflavorchimu}). It means that complexity and oscillation probabilities contain the same information. Also, since $P_{e\mu}=P_{\mu e}$ for both vacuum and standard matter oscillations, it implies that $\chi_e=\chi_{\mu}$.\\
	\item  
	In the three-flavor neutrino oscillation framework, we find that the Krylov basis is not equal to the flavor state basis. Forms of the Krylov states for all three cases of initial states ($\nu_e$, $\nu_{\mu}$, $\nu_{\tau}$) are given in Eqs.~(\ref{eq:nue_cost}), (\ref{threeflavorchimu}) and (\ref{threeflavorchitau}) of Sec.~\ref{threeflavor}. We find that the spread complexities have extra cross terms apart from the transition probabilities of the initial neutrino flavor. \\
	\item 
	Complexities show oscillatory patterns (see Fig.~\ref{Chi_L}) driven by the two mass-squared differences ($\Delta m_{21}^2$ and $\Delta m_{31}^2$), similar to the probabilities. The relevant probability to compare with $\chi_\alpha$, however, is $1-P_{\alpha\alpha}$. In vacuum, the complexities are maximized over a large $L/E \approx (10-22)\times 10^3$~km for $\chi_{\mu}$ and $\chi_{\tau}$, depending on the CP-phase $\delta$ and at $L/E \approx 16\times 10^3$~km for  $\chi_e$ (see Fig.~\ref{Cost_L_Exp}). Notably, $\chi_e$ has the highest complexity for this large range of $L/E$ but 
	note that, the maximum $L/E$ value accessible in current long-baseline oscillation experiments is about 1000 km/GeV. Hence, in current experimental conditions, the complexity represented by $\chi_e$ is much lower than the complexities of $\chi_{\mu}$ and $\chi_{\tau}$. \\
	\item We have scrutinized the effects of different oscillation parameters on the complexities. In vacuum, $\chi_\mu$ and $\chi_\tau$ are maximized for the CP-violating phase $\delta \approx \pm 90^o$, respectively, while $\chi_e$ is maximized at $\delta = \pm 90^o$ (see Figs.~\ref{Cost_L_Exp} and \ref{Cost_E_Exp}).  This maximization happens at very large $L/E$ as mentioned above. For $L/E \sim 1000$~km, local maxima for $\chi_\mu$ and $\chi_\tau$ are still at $\delta \approx \pm 90^o$ but can be different for $\chi_e$ depending on the exact $L/E$. We found that sensitivity of complexities to the octant of $\theta_{23}$ is small (see Fig.~\ref{Cost_theta23_Exp}). $\chi_e$ essentially has no sensitivity whereas $\chi_{\mu}$ and $\chi_{\tau}$ show small but non-zero variation with respect to $\theta_{23}$ when varied over its 3$\sigma$ allowed range. \\
	\item We have investigated $\chi_\alpha$ particularly for the setups of the T2K and NOvA experiments, two long-baseline neutrino oscillation experiments currently operating. For the 810~km baseline of NOvA, the matter effect is important and enhances the complexity embedded in the evolution of all three flavors of neutrinos (see Fig.~\ref{Cost_matter}). A detailed examination of $\chi_\alpha$ in the $E-\delta$ plane (see Figs.~\ref{Cost_Edelta_t2k} and \ref{Cost_Edelta_NOvA}) shows that $\chi_e$ is less affected by the variation of $\delta$, whereas, $\chi_{\mu}$ and $\chi_{\tau}$ show stronger variation and the maxima are found around $\delta = -90^o$ and $+90^o$, respectively, which coincide with the relevant $E$ for these experiments. $\delta = -90^o$ for maximum $\chi_{\mu}$ is consistent with results from T2K but is in contradiction with results from NOvA. Even though the T2K result is obtained with  1$\sigma$ confidence only, it is encouraging and looks like quantum information theory is providing a theoretical justification for this preference. The enhancements of $\chi_{e}$ for T2K at around $\delta=135^o$ and $-45^o$, and at $E \sim 0.2$ GeV are outside the current experimental setup and we cannot check their validity. These, however, correspond to local maxima for the particular $L/E$ as mentioned above and the global maximum at $\delta = \pm90^o$ is inaccessible currently (see Fig.~\ref{Cost_L_Exp}). It will be interesting to probe these features with a $\nu_e$ beam in a future experiment. \\
	
	\item Neutrino mass hierarchy, whether normal or inverted, affects the complexity and the matter effect is essential to distinguish between them (see Fig.~\ref{Chi_NOvA_massorder}). Similarly, the effect of neutrino and antineutrino oscillations affected by mass hierarchy is also embedded in the complexity $\chi_\alpha$ (see Fig.~\ref{Chi_massoreder_antinu}). 
	
\end{itemize}

\section{Conclusions}

This study examines the spread complexity of neutrino states in two- and three-flavor oscillation scenarios. In the two-flavor scenario, complexity and transition probabilities yield equivalent information. However, in the case of three-flavor oscillation, a different pattern emerges. An initial flavor state evolves into two mixed final states and the complexity, while compared to the total oscillation probability, contains additional information. In particular, we examined sensitivity of complexity on the yet unknown value of the CP-violating phase angle. Remarkably, when we explored complexity across various phase angles, we found that the complexity is maximized for a value of the phase angle for which CP is also maximally violated. Notably, the T2K experimental data also favors this phase angle, which is obtained from studying the flavor transition. This matching is quite fascinating both from the perspectives of neutrino physics and understanding quantum complexity for natural evolution. 

Another intriguing aspect regarding complexity is its ability to differentiate between the oscillation probabilities of muon and tau neutrinos, unlike the total oscillation probabilities which remain indistinguishable for a given CP-violating phase angle. If we were to set the phase angle at its maximally CP-violating values, the complexities of muon and tau neutrinos would exhibit slight disparities. Consequently, complexity offers a distinguishing factor between these two scenarios that the total probability fails to provide. Although, this and many other features of complexities we have explored are not accessible to current experimental setups, our study may motivate future studies. Similar analysis can be done for the mixing in the quark sector through the $CKM$ matrix which will be the future direction of our project. Moreover, it will be interesting to explore if there exists a correlation between the complexity and other non-classical features embedded in the system. In this context, a comparative analysis of the results shown in \cite{Bittencourt:2023asd} and the ones obtained with complexity perusal will be potentially interesting future work.\\

In conclusion, quantum spread complexity emerges as a potent and novel quantity for investigating neutrino oscillations. Not only does it successfully reproduce existing results, but it also demonstrates the potential to serve as a theoretical tool for predicting new outcomes in future experiments. Its application holds promise in advancing our understanding of neutrino physics and astrophysics in general.

\appendix
\section{Vacuum oscillation amplitudes}
\begin{equation*}
	\begin{split}
		A_{e\mu} = -\frac{1}{2} &\cos \theta_{13} e^{-\frac{1}{2} i \left(2 \delta +\frac{t (\Delta m^2_{21}+\Delta m^2_{31})}{E}\right)} \left(\sin 2 \theta_{12} \cos \theta_{23} \left(-1+e^{\frac{i \Delta m^2_{21} t}{2 E}}\right) e^{\frac{1}{2} i \left(2 \delta +\frac{\Delta m^2_{31} t}{E}\right)}\right.\\
		&\left.-\sin \theta_{13} \sin \theta_{23} \left(\cos 2 \theta_{12} \left(e^{\frac{i \Delta m^2_{31} t}{2 E}}-e^{\frac{i t (\Delta m^2_{21}+\Delta m^2_{31})}{2 E}}\right)-e^{\frac{i t (\Delta m^2_{21}+\Delta m^2_{31})}{2 E}}\right.\right.\\
		&\left.\left.~~~~~~~~~~~~~~~~~~~~~~~~~~~~~~~~+2 e^{\frac{i \Delta m^2_{21} t}{2 E}}-e^{\frac{i \Delta m^2_{31} t}{2 E}}\right)\right)
	\end{split}
\end{equation*}
\begin{equation*}
	\begin{split}
		A_{e\tau} = \frac{1}{2} &\cos \theta_{13} e^{-\frac{1}{2} i \left(2 \delta +\frac{t (\Delta m^2_{21}+\Delta m^2_{31})}{E}\right)} \left(\sin 2\theta_{12} \sin \theta_{23} \left(-1+e^{\frac{i \Delta m^2_{21} t}{2 E}}\right) e^{\frac{1}{2} i \left(2 \delta +\frac{\Delta m^2_{31} t}{\text{E$\nu $}}\right)}\right.\\
		&\left.+\sin \theta_{13} \cos \theta_{23} \left(\cos 2\theta_{12} \left(e^{\frac{i \Delta m^2_{31} t}{2 E}}-e^{\frac{i t (\Delta m^2_{21}+\Delta m^2_{31})}{2 E}}\right)\right.\right.\\
		&\left.\left.~~~~~~~~~~~~~~-e^{\frac{i t (\Delta m^2_{21}+\Delta m^2_{31})}{2 E}}+2 e^{\frac{i \Delta m^2_{21} t}{2 E}}-e^{\frac{i \Delta m^2_{31} t}{2 E}}\right)\right)
	\end{split}
\end{equation*} 
\begin{equation*}
	\begin{split}
		A_{\mu e} = -\frac{1}{2}& \cos \theta_{13} e^{-\frac{i t (\Delta m^2_{21}+\Delta m^2_{31})}{2 E}} \left(-e^{i \delta } \sin \theta_{13} \sin \theta_{23} \left(\sin ^2\theta_{12} \left(-1+e^{\frac{i \Delta m^2_{21} t}{2 E}}\right) e^{\frac{i \Delta m^2_{31} t}{2 E}}\right.\right.\\
		&\left.\left.-e^{\frac{i t (\Delta m^2_{21}+\Delta m^2_{31})}{2 E}}+2 e^{\frac{i \Delta m^2_{21} t}{2 E}}
		-e^{\frac{i \Delta m^2_{31} t}{2 E}}\right)+\cos ^2\theta_{12} \sin \theta_{13} \sin \theta_{23} \left(-1+e^{\frac{i \Delta m^2_{21} t}{2 E}}\right)\right.\\ 
		&\left.e^{\frac{1}{2} i \left(2 \delta +\frac{\Delta m^2_{31} t}{E}\right)}+\sin 2\theta_{12} \cos \theta_{23} e^{\frac{i t (\Delta m^2_{21}+\Delta m^2_{31})}{2 E}}
		-2 \sin \theta_{12} \cos \theta_{12} \cos \theta_{23} e^{\frac{i \Delta m^2_{31} t}{2 E}}\right)
	\end{split}	
\end{equation*}
\begin{equation*}
	\begin{split}
		A_{\mu \tau} = \frac{1}{32} &e^{-\frac{1}{2} i \left(2 \delta +\frac{t (\Delta m^2_{21}+\Delta m^2_{31})}{E}\right)} \left(8 \sin 2\theta_{12} \sin \theta_{13} \left(\left(1+e^{2 i \delta }\right) \cos 2\theta_{23}-e^{2 i \delta }+1\right) \left(-1+e^{\frac{i \Delta m^2_{21} t}{2 E}}\right)\right.\\
		 &\left.e^{\frac{i \Delta m^2_{31} t}{2 E}}
		+2 e^{i \delta } \sin 2\theta_{23} \left(\cos (2 (\theta_{12}-\theta_{13})) \left(e^{\frac{i \Delta m^2_{31} t}{2 E}}-e^{\frac{i t (\Delta m^2_{21}+\Delta m^2_{31})}{2 E}}\right)\right.\right.\\
		&\left.\left.-\cos (2 (\theta_{12}+\theta_{13})) e^{\frac{i t (\Delta m^2_{21}+\Delta m^2_{31})}{2 E}}
		-6 \cos 2\theta_{12} \left(e^{\frac{i \Delta m^2_{31} t}{2 E}}-e^{\frac{i t (\Delta m^2_{21}+\Delta m^2_{31})}{2 E}}\right)\right.\right.\\
		&\left.\left.-2 \cos 2\theta_{13} e^{\frac{i t (\Delta m^2_{21}+\Delta m^2_{31})}{2 E}}-2 e^{\frac{i t (\Delta m^2_{21}+\Delta m^2_{31})}{2 E}}
		+4 \cos 2\theta_{13} e^{\frac{i \Delta m^2_{21} t}{2 E}}+4 e^{\frac{i \Delta m^2_{21} t}{2 E}}\right.\right.\\
		&\left.\left.+e^{\frac{i \Delta m^2_{31} t}{2 E}} \cos (2 (\theta_{12}+\theta_{13}))-2 \cos 2\theta_{13} e^{\frac{i \Delta m^2_{31} t}{2 E}}-2 e^{\frac{i \Delta m^2_{31} t}{2 E}}\right)\right)
	\end{split}
\end{equation*}
\begin{equation*}
	\begin{split}
		A_{\tau e} = \frac{1}{2} &\cos \theta_{13} e^{-\frac{i t (\Delta m^2_{21}+\Delta m^2_{31})}{2 E}} \left(e^{i \delta } \sin \theta_{13} \cos \theta_{23} \left(\cos 2\theta_{12} \left(e^{\frac{i \Delta m^2_{31} t}{2 E}}-e^{\frac{i t (\Delta m^2_{21}+\Delta m^2_{31})}{2 E}}\right)\right.\right.\\
		&\left.\left.-e^{\frac{i t (\Delta m^2_{21}+\Delta m^2_{31})}{2 E}}+2 e^{\frac{i \Delta m^2_{21} t}{2 E}}\right.\right.\\
		&\left.\left.-e^{\frac{i \Delta m^2_{31} t}{2 E}}\right)+\sin 2\theta_{12} \sin \theta_{23} \left(-1+e^{\frac{i \Delta m^2_{21} t}{2 E}}\right) e^{\frac{i \Delta m^2_{31} t}{2 E}}\right)
	\end{split}
\end{equation*}
\begin{equation*}
	\begin{split}
		A_{\tau \mu} = \frac{1}{32} &e^{-\frac{1}{2} i \left(2 \delta +\frac{t (\Delta m^2_{21}+\Delta m^2_{31})}{E}\right)} \left(8 \sin 2\theta_{12} \sin \theta_{13} \left(\left(1+e^{2 i \delta }\right) \cos 2\theta_{23}+e^{2 i \delta }-1\right)\right.\\
		&\left.\left(-1+e^{\frac{i \Delta m^2_{21} t}{2 E}}\right) e^{\frac{i \Delta m^2_{31} t}{2 E}}
		+2 e^{i \delta } \sin 2\theta_{23} \left(\cos (2 (\theta_{12}-\theta_{13})) \left(e^{\frac{i \Delta m^2_{31} t}{2 E}}-e^{\frac{i t (\Delta m^2_{21}+\Delta m^2_{31})}{2 E}}\right)\right.\right.\\
		&\left.\left.-\cos (2 (\theta_{12}+\theta_{13})) e^{\frac{i t (\Delta m^2_{21}+\Delta m^2_{31})}{2 E}}-6 \cos 2\theta_{12} \left(e^{\frac{i \Delta m^2_{31} t}{2 E}}-e^{\frac{i t (\Delta m^2_{21}+\Delta m^2_{31})}{2 E}}\right)\right.\right.\\
		&\left.\left.-2 \cos 2\theta_{13} e^{\frac{i t (\Delta m^2_{21}+\Delta m^2_{31})}{2 E}}-2 e^{\frac{i t (\Delta m^2_{21}+\Delta m^2_{31})}{2 E}}+4 \cos 2\theta_{13} e^{\frac{i \Delta m^2_{21} t}{2 E}}+4 e^{\frac{i \Delta m^2_{21} t}{2 E}}\right.\right.\\
		&\left.\left.+e^{\frac{i \Delta m^2_{31} t}{2 E}} \cos (2 (\theta_{12}+\theta_{13}))-2 \cos 2\theta_{13} e^{\frac{i \Delta m^2_{31} t}{2 E}}-2 e^{\frac{i \Delta m^2_{31} t}{2 E}}\right)\right)
	\end{split}
\end{equation*}

\acknowledgments

We thank Pratik Nandi, Arpan Bhattacharyya, Jaco van Zyl, Ushak Rahaman and Alexei Smirnov for their helpful discussions. This work was partially supported by grants from the National Institute of Theoretical and Computational Sciences (NITheCS) and from the University of Johannesburg Research Council.



\end{document}